\documentclass[10 pt]{article}
\usepackage{style}

\title{\Large \bfseries {A Finite-Sample Analysis of Payoff-Based Independent Learning in Zero-Sum Stochastic Games}}
\author{\normalsize 
	Zaiwei Chen\textsuperscript{1,$*$}, Kaiqing Zhang\textsuperscript{2}, Eric Mazumdar\textsuperscript{1,$\dagger$}, \\ \normalsize   Asuman Ozdaglar\textsuperscript{3}, and Adam Wierman\textsuperscript{1,$\ddagger$}\\
{\footnotesize
\textsuperscript{1}Caltech  \href{mailto:zchen458@caltech.edu}{$^*$\textit{zchen458@caltech.edu}},
\href{mailto:mazumdar@caltech.edu}{$^\dagger$\textit{mazumdar@caltech.edu}},
\href{mailto:mazumdar@caltech.edu}{$^\ddagger$\textit{adamw@caltech.edu}}
}\\
{\footnotesize\textsuperscript{2}University of Maryland, College Park  \href{mailto:kaiqing@umd.edu}{\textit{kaiqing@umd.edu}}}~~
{\footnotesize\textsuperscript{3}MIT   \href{mailto:asuman@mit.edu}{\textit{asuman@mit.edu}}}
}
\date{\vspace{-0.3 in}}

\begin{document}
\maketitle

\pagestyle{fancy}
\fancyhf{} 
\fancyhead[R]{\thepage}

\begin{abstract}
We study two-player zero-sum stochastic games,  
and  propose a form of independent learning dynamics called Doubly Smoothed Best-Response dynamics, which integrates   a discrete and doubly smoothed variant of the best-response dynamics into  temporal-difference (TD)-learning and minimax value iteration. The resulting dynamics are payoff-based,  convergent, rational, and symmetric among players.  Our main results provide finite-sample guarantees. In particular, we prove the first-known $\tilde{\mathcal{O}}(1/\epsilon^2)$ sample complexity bound for payoff-based independent learning dynamics, up to a smoothing bias. In the special case where the stochastic game has only one state (i.e., matrix games), we provide a sharper $\tilde{\mathcal{O}}(1/\epsilon)$ sample complexity.
Our analysis uses a novel coupled Lyapunov drift approach to capture the evolution of multiple sets of coupled and stochastic iterates, which might be of independent interest.
\end{abstract}

\section{Introduction}\label{sec:intro}
Recent years have seen remarkable successes of reinforcement learning (RL) in a variety of applications, such as board games \citep{silver2017mastering}, autonomous driving \citep{shalev2016safe}, city navigation \citep{mirowski2018learning}, and fusion plasma control \citep{degrave2022magnetic}. A common feature of these applications is that there are \emph{multiple} decision-makers interacting with each other in an unknown environment. While empirical successes have shown the potential of multi-agent reinforcement learning (MARL) \citep{bu2008comprehensive,zhang2021multi}, the training of MARL agents largely relies on heuristics and parameter-tuning, and is not always reliable. In particular, many practical MARL algorithms are heuristically extended from their single-agent counterparts and lack theoretical guarantees. 

A growing literature seeks to provide theoretical insights to substantiate the empirical success of MARL and inform the design of efficient, and provably convergent algorithms. Work along these lines can  be broadly categorized  into work on cooperative MARL such as  \cite{arslan2017decentralized,zhang2018fully,qu2020scalable,zhang2022global} where agents seek to reach a common goal, and work on competitive MARL where agents have individual (and possibly misaligned) objectives \citep{littman1994markov,littman2001friend,hu2003nash,daskalakis2020independent,sayin2021decentralized,bai2020provable,xie2020learning,zhang2021gradient,ding2022independent,liu2020sharp,jin2021v,daskalakis2022complexity}. While some earlier work focused on providing guarantees on  asymptotic convergence, the more recent ones share an increasing interest in understanding the  finite-time/sample behavior. This follows from the line of recent successes in understanding the finite-sample behavior of single-agent RL algorithms, see e.g.,  \cite{bhandari2018finite,srikant2019finite,li2020sample,chen2020finite} and many others.

In this paper, we focus on the  benchmark-setting of two-player\footnote{Hereafter, we may use \emph{player} and \emph{agent} interchangeably.} zero-sum matrix and stochastic games, and develop multi-agent learning dynamics with provable finite-sample guarantees. Crucially,  our  dynamics are independent (requiring no coordination between the agents in learning), rational (each agent will converge to the best response to the opponent if the opponent plays an (asymptotically) stationary policy \citep{bowling2001rational}), and hence capture the learning in settings with multiple game-theoretic agents. Indeed, game-theoretic agents are self-interested, and ideal learning dynamics should not enforce any communication of information or coordination among agents. In addition, we focus on the more challenging but practically  relevant settings of \emph{payoff-based learning}, where each agent can only observe the realized payoff of itself during learning, without observing the policy or even the action taken by the opponent. For these learning dynamics, we establish for the first time finite-sample guarantees for both two-player zero-sum matrix and stochastic games. We detail our contributions as follows. 

\subsection{Contributions} 
We take a principled approach to algorithm design: we first construct independent learning dynamics for the special case of zero-sum matrix games. Then, we generalize the dynamics to the setting of Markov games and present the finite-sample guarantees. Further, in both cases, our dynamics are easily implementable and have a simple, intuitive structure that links them to well-known dynamics from the learning in games literature. 

\subsubsection{Independent Learning for Two-Player Zero-Sum Matrix Games}

\paragraph{Algorithm Design.}
We design a new learning dynamics  called \emph{Doubly Smoothed Best-Response} (DSBR) dynamics for solving matrix games.  It maintains two sets of iterates on a single time scale: the policies and the \emph{local} state-action functions (denoted as the $q$-functions). The policy update can be viewed as a  variant of the best-response dynamics, where the best response is constructed by introducing the $q$-function as an estimate of the payoff marginalized by the opponent's current policy. The name of \emph{doubly smoothed} follows from two key algorithmic ideas that enable our finite-sample analysis: (1) we introduce a stepsize to \emph{smooth} the    update of the policy, so that it does not change too abruptly during learning; (2) we use a \emph{smoothed} best-response to the local $q$-function when updating the policy. Idea (1) is an alternative to independent learning dynamics for matrix games \citep{leslie2005individual} that use \emph{adaptive}  stepsizes, enabling us to smooth the variation of the local $q$-function (and thus policy) in a more controlled way (and thus to establish finite-sample guarantees).  Idea (2) has been exploited in the well-known dynamics of smoothed fictitious play \citep{ref:Fudenberg93}, in order to encourage exploration and make the learning dynamics consistent \citep{fudenberg1995consistency}. 

\paragraph{Finite-Sample Analysis.} We establish a finite-sample bound (measured in terms of the  Nash gap) for our DSBR dynamics  when using stepsizes of various decay rates. The best convergence rate is achieved with a stepsize of $\mathcal{O}(1/k)$, in which case the learning dynamics  enjoys an overall $\mathcal{O}(1/k)$ rate of convergence
to a Nash equilibrium up to a smoothing bias. The smoothing bias arises from the use of softmax policies in the update.  

\subsubsection{Independent Learning for Two-Player Zero-Sum Markov Games}\label{subsubsec:contribution_Markov}

\paragraph{Algorithm Design.} Building on the results for matrix games, we design a new  learning dynamics for Markov games called \emph{Doubly Smoothed Best-Response dynamics with Value Iteration} (DSBR-VI), driven by a \emph{single trajectory} of Markovian samples. The dynamics consist of two loops, and can be viewed as a combination of the DSBR dynamics for an induced auxiliary matrix game (conducted in the inner loop) and an independent way of performing minimax value iteration (conducted in the outer loop).  In particular, in the inner loop, the iterate of the outer loop, i.e., the value function, is fixed, and the players learn the approximate Nash equilibrium of an auxiliary matrix game induced by the value function; then the outer loop is updated by approximating the minimax value iteration updates for Markov games, with only local information. 

\paragraph{Finite-Sample Analysis.}
We then establish finite-sample bounds for our DSBR-VI dynamics when using either constant stepsize or diminishing stepsizes, and under  weaker assumptions compared to existing work. Our dynamics achieves an overall $\tilde{\mathcal{O}}(1/\epsilon^2)$ sample complexity up to a smoothing bias. To the best of our knowledge, this is the first finite-sample analysis of best-response type independent learning dynamics that are convergent and rational for Markov games. Most existing MARL algorithms are either symmetric across players  but not payoff-based, e.g., \cite{cen2021fast,cen2022faster,zhang2022policy,zengregularized,erez2022regret},  or not symmetric and thus not rational, e.g.,  \cite{daskalakis2020independent,zhao2021provably,zhang2021derivative,alacaoglu2022natural}, or do not have finite-sample guarantees, e.g.,  \cite{leslie2020best,sayin2021decentralized,baudinsmooth}.

\subsection{Challenges \& Our Techniques} \label{subsec:challenge}
At a high level, we develop a novel \textit{coupled Lyapunov drift argument} to establish the finite-sample bounds. Specifically, we design a Lyapunov function for each set of the iterates (i.e., value functions, policies, and $q$-functions) and establish coupled Lyapunov drift inequalities for each. We then carefully combine the coupled Lyapunov drift inequalities to establish the  finite-sample bounds. While a more detailed analysis is provided in Section \ref{sec:sketch}, we briefly give an overview of the main challenges encountered in analyzing the  payoff-based independent learning dynamics in Markov games --- as well as how we overcome them in our analysis and algorithm design.

\paragraph{Time-Inhomogeneous Markovian Noise.}
The fact that our algorithm is \emph{payoff-based} imposes additional challenges in handling the stochastic errors in the update. In particular, due to the best-response nature of the dynamics, the behavior policy for sampling becomes \emph{time-varying}. In fact, the samples used for learning form a \textit{time-inhomogeneous} Markov chain. This makes it challenging to establish finite-sample guarantees, as time-inhomogeneity prevents us from directly exploiting the uniqueness of stationary distributions and the fast mixing of Markov chains. Building on existing work \cite{bhandari2018finite,srikant2019finite}, we overcome this challenge by tuning the algorithm design and developing a refined conditioning argument. 

\paragraph{Possible Non-Smoothness of the Lyapunov Function.} To use a Lyapunov argument to study the convergence rate of discrete and stochastic dynamics, existing work, e.g., \cite{chen2020finite} shows that the \emph{smoothness}\footnote{Here ``smooth'' is the counter-part of ``strongly convex'' in optimization. See for example \cite{beck2017first} for the definition of smoothness.} of the Lyapunov function plays an important role. However, the Lyapunov function proposed to study the continuous-time smoothed best-response dynamics is not a smooth function on the joint probability simplex. Our smoothed best-response update comes to the rescue: we show that it ensures the policies generated by our learning dynamics are naturally uniformly bounded away from zero. By restricting our analysis to the interior of the joint probability simplex (which does not contain any extreme points), we are able to establish the smoothness of the Lyapunov function, thereby making way for our Lyapunov approach.

\paragraph{Non-Zero-Sum Payoffs  due to  Independent Learning.} As illustrated in Section \ref{subsubsec:contribution_Markov}, the inner loop of DSBR-VI is designed to learn the Nash equilibrium of an auxiliary matrix game induced by the value functions $v_t^i$ and $v_t^{-i}$, where $t$ is the outer-loop iteration index. Importantly, $v_t^i$ and $v_t^{-i}$ are  maintained individually by players $i$ and $-i$, and hence do not necessarily satisfy $v_t^i+v_t^{-i}=0$ due to independent learning. As a result, the auxiliary matrix game from the inner loop does not necessarily admit a zero-sum structure. See Section \ref{subsec:Markov_Algorithm} for more details. The error induced from such non-zero-sum structure appears in existing work \cite{sayin2021decentralized,sayin2022fictitious}, and was handled by designing a novel truncated Lyapunov function. However, the truncated Lyapunov function was sufficient to establish the asymptotic convergence, but did not provide the explicit rate at which the induced error goes to zero. To facilitate finite-sample analysis, in addition to the standard Lyapunov functions used to analyze the $q$-functions, the policies, and the $v$-functions, we introduce $\|v_t^i+v_t^{-i}\|_\infty$ as an additional Lyapunov function to capture the behavior of the induced error from the non-zero-sum structure of the inner-loop auxiliary matrix game. 

\paragraph{Coupled Lyapunov Drift Inequalities.} When using Lyapunov arguments for finite-sample analysis, once the Lyapunov drift inequality is established, the finite-sample bound follows   straightforwardly  by repeatedly invoking the result. However, since our learning dynamics  maintains multiple sets of iterates (the value functions, the policies, and the $q$-functions) and updates them in a coupled manner, the Lyapunov drift inequalities we establish are also highly coupled. Decoupling the Lyapunov drift inequalities without compromising the convergence rate is a major challenge. We develop a systematic strategy for decoupling, which crucially relies on a bootstrapping argument where we first establish a crude bound of the Lyapunov function and then substitute the bound back into the Lyapunov drift inequalities to obtain a tighter one.

\subsection{Related Work}\label{sec:related_work}
Before presenting our problem formulations and analysis, we first briefly summarize related and prior work in single-agent RL, MARL, and learning in games.

\paragraph{Single-Agent RL.} The most related works (in single-agent RL) to our paper are those that perform finite-sample analysis for RL in infinite-horizon discounted  Markov decision processes following a single trajectory of Markovian samples  \citep{even2003learning,bhandari2018finite,zou2019finite,srikant2019finite,li2020sample,chen2020finite,qu2020finite,chen2021finite,lan2021policy,yan2022efficacy}. In particular, \cite{bhandari2018finite,srikant2019finite} establish finite-sample bounds for TD-learning (with linear function approximation), and \cite{li2020sample,qu2020finite,chen2021finite} establish finite-sample bounds for $Q$-learning. In both cases, the behavior policy for sampling is some \emph{stationary} policy. For \emph{non-stationary} behavior policies as we  consider, 
\cite{zou2019finite} establishes finite-sample bounds for SARSA, an on-policy RL algorithm, with additional assumptions that control the varying rate of the non-stationary policy.

\paragraph{Sample-Efficient MARL.} 
There has been increasing study of MARL with sample efficiency guarantees recently \citep{bai2020provable,bai2020near,liu2020sharp,xie2020learning,jin2021v,songcan,mao2022improving,daskalakis2022complexity,cui2023break}. Most of them focus on the finite-horizon episodic setting with online explorations and perform regret analysis, which differs from our finite-sample analysis. Additionally, these algorithms are episodic due to the finite-horizon nature of the setting, and are not best-response type independent learning dynamics that are repeatedly run for infinitely long, which can be viewed as a non-equilibrating adaptation process. In fact, the primary focus of this line of work is a \emph{self-play} setting where all the players can be controlled to perform centralized learning \citep{wei2017online,bai2020provable,bai2020near,liu2020sharp,xie2020learning}.  Beyond the online setting, finite-sample efficiency has also been established for MARL using a generative model \citep{zhang2020model,li2022minimax} or offline datasets  \citep{cuioffline,cui2022provably,zhong2022pessimistic,yan2022efficacy}.  These algorithms tend to be centralized in nature and focus on \emph{equilibrium computation}, and thus do not perform independent learning.

Finite-sample complexity has also been established for \emph{policy gradient}  methods, a popular RL approach, when applied to solving zero-sum stochastic games \citep{daskalakis2020independent,zhao2021provably,zhang2021derivative,alacaoglu2022natural}. However, to ensure convergence, these methods are \emph{asymmetric} in that the players update their policies at  \emph{different} timescales,  e.g., one player updates faster than the other with larger stepsizes; or one player fixes its policy while waiting for the other to update. Such asymmetric policy gradient methods are not completely independent, as some implicit coordination is required to enable such a timescale separation across agents. This style of implicit coordination is also required for the finite-sample analysis of decentralized learning in certain general-sum stochastic games, e.g., \cite{gao2021finite}, which improves the asymptotic convergence in  \cite{arslan2017decentralized}. 

\paragraph{Independent Learning in Games.} Independent learning has been well-studied in the literature on  learning in matrix games. \emph{Fictitious play} (FP) \citep{brown1951iterative} may be viewed as the earliest of this kind, and its convergence analysis  for the zero-sum setting is provided in \cite{robinson1951iterative}. In FP, each player chooses the \emph{best response}  to its estimate of the opponent's strategy via the history of the play, an idea we also follow. Smoothed versions of FP have been developed \citep{ref:Fudenberg93,hofbauer2002global} to make the learning dynamics consistent \citep{fudenberg1995consistency,fudenberg1998theory}. Moreover, no-regret learning algorithms, extensively studied  in online learning, can also be used as independent learning dynamics for matrix games \citep{cesa2006prediction}. It is known that they are both convergent and rational by the definition of \cite{bowling2001rational}, and are usually implemented in a symmetric way.  See \cite{cesa2006prediction} for a detailed introduction to no-regret learning in games.

For stochastic games, independent and symmetric policy gradient methods have been developed in recent years, mostly for the case of \emph{potential} games \citep{zhang2021gradient,ding2022independent,leonardos2022global}. The zero-sum case is more challenging since there is no off-the-shelf Lyapunov function, which the potential function in the potential game case serves as. For non-potential game settings, symmetric variants of policy gradient methods  have been proposed, but have only been studied under the \emph{full-information} setting without finite-sample guarantees \citep{cen2021fast,cen2022faster,pattathil2022symmetric,zhang2022policy,zengregularized,erez2022regret}, with the exception of \cite{wei2021last,chenzy2021sample}. However, the learning algorithm in \cite{wei2021last} requires some coordination between the players when sampling, and is thus not completely independent; that in 
\cite{chenzy2021sample} is extragradient-based and not best-response-type, and needs some stage-based sampling process that also requires coordination across players.  

Best-response type independent learning  for stochastic games has attracted increasing attention lately  \citep{leslie2020best,sayin2021decentralized,sayin2022fictitious,sayin2022fictitious2,baudinsmooth,baudin2022fictitious,maheshwari2022independent}, with \cite{sayin2021decentralized,sayin2022fictitious,baudinsmooth,baudin2022fictitious}  tackling  the zero-sum setting. However, only asymptotic convergence  was  established in these works. 

\section{Independent Learning for Zero-Sum Matrix Games}\label{sec:bandit}

As a warm-up, we begin by considering zero-sum matrix games.  
This setting introduces both algorithmic and technical ideas that are important for the stochastic game setting, which may be of independent interest. 

Let $\mathcal{A}^1$ (respectively, $\mathcal{A}^2$) be the finite action-space of player $1$ (respectively, player $2$), and let $\mathcal{R}^1\in\mathbb{R}^{|\mathcal{A}^1|\times |\mathcal{A}^2|}$ (respectively, $\mathcal{R}^2=-(\mathcal{R}^1)^\top$) be the payoff matrix of player $1$ (respectively, player $2$). The decision variables here are the policies $\pi^i\in \Delta^{|\mathcal{A}^i|}$, $i\in \{1,2\}$, where $\Delta^{|\mathcal{A}^i|}$ denotes the $|\mathcal{A}^i|$-dimensional probability simplex. We assume without loss of generality that $\max_{a^i,a^{-i}}|\mathcal{R}^i(a^i,a^{-i})|\leq 1$, and denote  $A_{\max}=\max(|\mathcal{A}^1|,| \mathcal{A}^2|)$. In what follows, we use $-i$ as the index of player $i$'s opponent. 

\begin{definition}[Nash Gap in Matrix Games]\label{def:Nash_gap_matrix}
Given a joint policy $\pi=(\pi^i,\pi^{-i})$, the \emph{Nash gap} is defined as 
\begin{align*}
	\text{NG}(\pi^i,\pi^{-i}):=\sum_{i=1,2}\max_{\hat{\pi}^i\in\Delta^{|\mathcal{A}^i|}}(\hat{\pi}^i-\pi^i)^\top \mathcal{R}^i\pi^{-i}.
\end{align*}
\end{definition}

\subsection{Algorithm: Doubly Smoothed Best-Response Dynamics}\label{subsec:bandit_algorithm}
The high-level idea behind our proposed dynamics is to use a discrete and smoothed variant of the best-response dynamics, where the players  construct approximations of the best response to the opponent's policy using a local $q$-function update that is in the spirit of temporal-difference (TD)-learning in RL \citep{sutton1988learning}.  Importantly, while the learning dynamics  maintains two sets of iterates (the policies and the $q$-functions), they are updated on a \emph{single} time scale with only a multiplicative constant difference in their stepsizes.  The details of the learning dynamics are summarized in Algorithm \ref{algo:bandit}, where $\sigma_\tau:\mathbb{R}^{|\mathcal{A}^i|}\mapsto\mathbb{R}^{|\mathcal{A}^i|}$  stands for the softmax function with temperature $\tau>0$. Specifically, we define $[\sigma_\tau(q^i)](a^i)=\exp(q^i(a^i)/\tau)/\sum_{\tilde{a}^i}\exp(q^i(\tilde{a}^i)/\tau)$ for all $a^i\in\mathcal{A}^i$ and $q^i\in\mathbb{R}^{|\mathcal{A}^i|}$.

\begin{algorithm}[!t]\caption{Doubly Smoothed Best-Response Dynamics}\label{algo:bandit}
	\begin{algorithmic}[1]
		\STATE \textbf{Input:} Integer $K$, initializations $q_0^i=\bm{0}\in\mathbb{R}^{|\mathcal{S}||\mathcal{A}^i|}$ and $\pi_0^i \sim \text{Unif}(\mathcal{A}^i)$.
		\FOR{$k=0,1,\cdots,K-1$}
		\STATE $\pi_{k+1}^i=\pi_k^i+\beta_k(\sigma_\tau(q_k^i)-\pi_k^i)$
		\STATE Play
		$A_k^i\sim \pi_{k+1}^i(\cdot)$ (against $A_k^{-i}$), and receive reward $\mathcal{R}^i(A_k^i,A_k^{-i})$
		\STATE $q_{k+1}^i(a^i)=q_k^i(a^i)+\alpha_k\mathds{1}_{\{a^i=A_k^i\}} \left(\mathcal{R}^i(A_k^i,A_k^{-i})-q_k^i(A_k^i)\right)$ for all $a^i\in\mathcal{A}^i$
		\ENDFOR
		\STATE \textbf{Output:} $\pi_K^i$
	\end{algorithmic}
\end{algorithm}

To motivate the algorithm design,
we start with the discrete best-response dynamics:
\begin{equation}\label{eq:FP}
	\begin{aligned}
		\pi_{k+1}^i=\;&\pi_k^i+\frac{1}{k+1}(\text{br}(\pi_k^{-i})-\pi_k^i),&\text{br}(\pi_k^{-i})\in {\arg\max}_{a^i}[\mathcal{R}^i\pi_k^{-i}](a^i),\;i=1,2,
	\end{aligned}
\end{equation}
where $e(a^i)$ is the $a^i$-th unit vector in $\mathbb{R}^{|\mathcal{A}^i|}$. In Eq. (\ref{eq:FP}), each player updates its \textit{randomized} policy $\pi_k^i$ incrementally towards the best response to its opponent's current policy, and chooses an action $A_{k+1}^i\sim \pi_{k+1}^i(\cdot)$. While the dynamics in Eq. (\ref{eq:FP}) provably converges for zero-sum matrix games, see e.g.,   \citep{hofbauer2006best}, implementing it requires player $i$ to compute ${\arg\max}_{a^i}[\mathcal{R}^i\pi_k^{-i}](a^i)$. Note that ${\arg\max}_{a^i}[\mathcal{R}^i\pi_k^{-i}](a^i)$ involves the exact knowledge of the opponent's policy, which cannot be accessed in independent learning.

To tackle this issue, suppose for now that we are given a \textit{stationary} joint policy $\pi=(\pi^i,\pi^{-i})$. The problem of player $i$ estimating $\mathcal{R}^i\pi^{-i}$ can be viewed as a \emph{policy evaluation} problem, which is usually solved with TD-learning in reinforcement learning. 
Specifically, the two players repeatedly play the matrix game with the joint policy $\pi=(\pi^i,\pi^{-i})$ and produce a sequence of joint actions $\{(A_k^i,A_k^{-i})\}_{k\geq 0}$. Then,  player $i$ forms an \emph{estimate} of $\mathcal{R}^i\pi^{-i}$ through the following iterative algorithm: 
\begin{align}\label{eq:q_bandit}
	q_{k+1}^i(a^i)=q_k^i(a^i)+\alpha_k \mathds{1}_{\{a^i=A_k^i\}}(\mathcal{R}^i(A_k^i,A_k^{-i})-q_k^i(A_k^i)),\quad \forall\;a^i\in\mathcal{A}^i,
\end{align}
with an arbitrary initialization $q_0^i\in\mathbb{R}^{|\mathcal{A}^i|}$, where $\alpha_k>0$ is the stepsize. To understand (\ref{eq:q_bandit}), suppose that $q_k^i$ converges to some $\Bar{q}^i$. Then the update equation (\ref{eq:q_bandit}) should be ``stationary'' at the limit point $\Bar{q}^i$ in the sense that 
\begin{align*}
    \mathbb{E}_{A^i\sim \pi^i(\cdot),A^{-i}\sim \pi^{-i}(\cdot)}[\mathds{1}_{\{a^i=A^i\}}(\mathcal{R}^i(A^i,A^{-i})-\Bar{q}^i(A^i))]=0
\end{align*}
for all $a^i$, which would imply that $\Bar{q}^i=\mathcal{R}^i\pi^{-i}$, as desired. The update equation (\ref{eq:q_bandit}), which can be viewed as a simplification of TD-learning in RL to the stateless case, is promising; however, to use $q_k^i$ as an estimate of $\mathcal{R}^i\pi_k^{-i}$ in Eq. (\ref{eq:FP}), we need to overcome two challenges as below, which inspire us to develop the ``double smoothing'' technique in Algorithm \ref{algo:bandit}. 

\begin{itemize}
    \item While we motivated the use of the update equation (\ref{eq:q_bandit}) in the case when the joint policy $(\pi^i,\pi^{-i})$ is stationary, the joint policy 
$\pi_k=(\pi_k^i,\pi_k^{-i})$ from the discrete best-response dynamics (\ref{eq:FP}) is time-varying. To make TD-learning (\ref{eq:q_bandit}) work for time-varying target policies, a natural approach is to make sure that the policies evolve at a slower rate compared to that of the $q$-functions, so that $\pi_k$ is close to being \emph{stationary} from the perspectives of $q_k^i$. This represents the first form of \emph{smoothing} in our algorithm. To implement this smoothing, we view $1/(k+1)$ in Eq. (\ref{eq:FP}) as a stepsize and replace it with a more flexible $\beta_k$, which is chosen to be smaller (but only by a constant multiplicative factor) than the stepsize $\alpha_k$ (cf. Lines  $3$ and $5$ of Algorithm \ref{algo:bandit}). 
\item Now we can view $\pi_k$ as if it is stationary in updating the $q$-functions. In order for TD-learning (\ref{eq:q_bandit}) to converge, a necessary condition is that the target policy (the value of which we want to estimate) should ensure exploration \citep{sutton2018reinforcement}. To see this, suppose that we are evaluating a deterministic policy, which has no exploration components. Then $q_k^i$ generated by TD-learning (\ref{eq:q_bandit}) clearly cannot converge because essentially only one entry of the vector-valued iterate $q_k^i$ is updated. To overcome this challenge, we \emph{smooth} the update by using a softmax instead of a hardmax (cf. Line $3$ of Algorithm \ref{algo:bandit}), which prevents the policies we want to evaluate from ever being deterministic. We term this as our second form of \emph{smoothing}.
\end{itemize}

Given the two algorithmic ideas described above, we arrive at Algorithm \ref{algo:bandit} -- a payoff-based independent learning dynamics for zero-sum matrix games. While TD-learning \citep{sutton1988learning,tsitsiklis1997analysis} and best-response dynamics \citep{hofbauer2006best,harris1998rate,leslie2020best}  are both extensively studied in isolation, the combined use of them to form independent learning dynamics is less studied. The most related work is  \cite{leslie2005individual}, in which an individual $Q$-learning algorithm is proposed. Compared with Algorithm \ref{algo:bandit}, the algorithm in \cite{leslie2005individual} uses stochastic stepsizes (which are adaptively updated based on the algorithm trajectory), and a rapidly time-varying behavior policy. In addition, only asymptotic convergence was shown in \cite{leslie2005individual}.

As an aside, the continuous version of Eq. (\ref{eq:FP}), i.e., the best-response dynamics 
\begin{align*}
    \dot{\pi}^i\in \arg\max_{\hat{\pi}^i\in\Delta^i}(\hat{\pi}^i)^\top \mathcal{R}^i\pi^{-i}-\pi^i,
\end{align*}
is frequently used to analyze the convergence behavior of the celebrated FP dynamics for solving zero-sum matrix games; see \cite{leslie2020best} for more details.

\subsection{Finite-Sample Analysis}\label{sec:bandit_analysis}

We now present a finite-sample analysis of Algorithm \ref{algo:bandit}, deferring its  proofs to Appendix \ref{pf:thm:bandit}.  We consider stepsizes of the form $\alpha_k=\alpha/(k+h)^z$, where $\alpha,h>0$ and $z\in [0,1]$. Note that $z=0$ corresponds to the  \emph{constant} stepsize case. The stepsize $\beta_k$ satisfies $\beta_k=c_{\alpha,\beta}\alpha_k$ for any $k\geq 0$, where $c_{\alpha,\beta}\in (0,1)$ is a tunable constant. Importantly, the stepsizes $\alpha_k$ and $\beta_k$ differ only by a multiplicative constant, which makes Algorithm \ref{algo:bandit} a \emph{single} time-scale algorithm that is easier to implement than a two time-scale one.

In the following theorem, the parameters $\{c_j\}_{0\leq j\leq 3}$ are numerical constants, and the parameter $\ell_\tau$ (the explicit expression of which is presented in Appendix \ref{sec:analysis}) depends only on the temperature $\tau$ and $A_{\max}$. 

\begin{theorem}\label{thm:bandit}
	Suppose that both players follow Algorithm \ref{algo:bandit} and $c_{\alpha,\beta}\leq \frac{\ell_\tau^3\tau^3}{c_0A_{\max}^2}$.
	\begin{enumerate}[(1)]
		\item When $\alpha_k\equiv \alpha$, we have
		\begin{align}\label{eq:bound_bandit_constant}
			\mathbb{E}[\text{NG}(\pi_K^i,\pi_K^{-i})]\leq 3\left(1-\frac{c_{\alpha,\beta}\alpha}{2}\right)^K+\frac{c_1A_{\max}^{3/2}}{c_{\alpha,\beta}}\alpha+2\tau\log(A_{\max}).
		\end{align}
		\item When $\alpha_k=\alpha/(k+h)$ 
		with $\alpha >2/c_{\alpha,\beta}$ and $h>\alpha$, we have 
		\begin{align*}
			\mathbb{E}[\text{NG}(\pi_K^i,\pi_K^{-i})]\leq 3\left(\frac{h}{K+h}\right)^{ c_{\alpha,\beta}\alpha/2}+\frac{c_2A_{\max}^{3/2}\alpha}{c_{\alpha,\beta}\alpha-2}\frac{\alpha}{K+h}+2\tau\log(A_{\max}).
		\end{align*}
		\item When $\alpha_k=\alpha/(k+h)^z$ with $\alpha>0$, $z\in (0,1)$, and $h\geq (\frac{4z}{c_{\alpha,\beta}\alpha})^{\frac{1}{1-z}}$, we have 
		\begin{align*}
			\mathbb{E}[\text{NG}(\pi_K^i,\pi_K^{-i})]\leq\;& 3\exp\left(-\frac{\alpha((K+h)^{1-z}-h^{1-z})}{2c_{\alpha,\beta}(1-z)}\right)+\frac{c_3A_{\max}^{3/2}}{c_{\alpha,\beta}}\frac{\alpha}{(K+h)^z}\\
			&+2\tau\log(A_{\max}).
		\end{align*}
	\end{enumerate}
\end{theorem}

In all the three cases in Theorem \ref{thm:bandit}, the bound is a combination of \textit{convergence bias}, \textit{variance}, and \textit{smoothing bias}. The behavior of the convergence bias and the variance agrees with existing literature on stochastic approximation algorithms \citep{srikant2019finite,chen2021finite,bhandari2018finite}. 
In particular, large stepsizes result in smaller convergence bias but larger variance. When using $\mathcal{O}(1/k)$ stepsizes, we achieve the best convergence rate of $\mathcal{O}(1/K)$. The smoothing bias $2\tau\log(A_{\max})$ arises from  using softmax instead of hardmax in Algorithm \ref{algo:bandit}, which can also be viewed as the difference between the Nash distribution \citep{leslie2005individual} and a Nash equilibrium.

Importantly, with only bandit feedback, we achieve an $\mathcal{O}(1/K)$ rate of convergence to a Nash equilibrium up to a smoothing bias. In general, for smooth and strongly monotone games, the lower bound for the rate of convergence of payoff-based or zeroth-order algorithms is $\mathcal{O}(1/\sqrt{K})$ \citep{lin2021optimal}. We have an improved $\mathcal{O}(1/K)$. convergence rate because our learning dynamics can exploit the bilinear structure of the game. In particular, we are able to use only the bandit feedback to construct an efficient estimator (using the $q$-functions) of the marginalized payoff $\mathcal{R}^i\pi^{-i}_k$ (which can also be interpreted as the  gradient), thereby enjoy the fast $\mathcal{O}(1/K)$ rate of convergence that is comparable to first-order method \citep{beznosikov2022stochastic}.

Based on Theorem \ref{thm:bandit}, we next derive the sample complexity in the following corollary.

\begin{corollary}[Sample Complexity]
	Given $\epsilon>0$, to achieve $\mathbb{E}[\text{NG}(\pi_K^i,\pi_K^{-i})]\leq \epsilon+2\tau\log(A_{\max})$, the sample complexity is $\mathcal{O}(\epsilon^{-1})$.
\end{corollary}

Notably, we achieve $\tilde{\mathcal{O}}(1/\epsilon)$ sample complexity up to a smoothing bias. The stepsize ratio appears only as a multiplicative constant in the bound, and does not impact the rate, which is the advantage of using a single time-scale algorithm.

When the opponent does not follow Algorithm \ref{algo:bandit}, but plays with a stationary policy, the following corollary states that we have the same $\mathcal{O}(1/\epsilon)$ sample complexity for the player to find an optimal policy against its opponent.

\begin{corollary}[Rationality]\label{co:convergent_opponent}
	Suppose that player $i$ follows the learning dynamics presented in Algorithm \ref{algo:bandit}, but its opponent follows a stationary policy $\pi^{-i}$. Then, given $\epsilon>0$, to achieve $\mathbb{E}[\max_{\hat{\pi}^i}(\hat{\pi}^i-\pi_K^i)^\top \mathcal{R}^i\pi^{-i}]\leq \epsilon+2\tau\log(A_{\max})$, the sample complexity is $\mathcal{O}(1/\epsilon)$.
\end{corollary}

According to the definition in \cite{bowling2001rational}, a dynamics being rational means that the player following this dynamics will converge to the best response to its opponent when the opponent uses an \textit{asymptotically} stationary policy. Since we are performing finite-sample analysis, we assume the opponent's policy \emph{is} stationary, because otherwise, the convergence rate  (which may be arbitrary) of the opponent's policy will also impact the bound.

\section{Independent Learning for Zero-Sum Markov Games}\label{sec:two_time_scale}

This section presents our main technical and algorithmic contributions. We introduce a payoff-based, single-trajectory, convergent, rational, and independent learning dynamics for zero-sum Markov games. Consider an infinite-horizon two-player zero-sum Markov game $\mathcal{M}=(\mathcal{S},\mathcal{A}^1,\mathcal{A}^2,p,\mathcal{R}^1,\mathcal{R}^2,\gamma)$, where $\mathcal{S}$ is the finite state-space, $\mathcal{A}^1$ (respectively, $\mathcal{A}^2$) is the finite action-space for player $1$ (respectively, player 2), $p$ represents the transition probabilities, in particular, $p(s'\mid s,a^1,a^2)$ is the probability of transitioning  to state $s'$ after player $1$ taking action $a^1$ and player $2$ taking action $a^2$ simultaneously  at state $s$, $\mathcal{R}^1:\mathcal{S}\times\mathcal{A}^1\times\mathcal{A}^2\mapsto \mathbb{R}$ (respectively, $\mathcal{R}^2:\mathcal{S}\times\mathcal{A}^2\times \mathcal{A}^1\mapsto \mathbb{R}$) is player $1$'s (respectively, player $2$'s) reward function, and $\gamma\in [0,1)$ is the discount factor. Note that we have $\mathcal{R}^1(s,a^1,a^2)+\mathcal{R}^2(s,a^2,a^1)=0$ for all $(s,a^1,a^2)$. 
We assume without loss of generality that $\max_{s,a^1,a^2}|\mathcal{R}^1(s,a^1,a^2)|\leq 1$, and denote $A_{\max}=\max(|\mathcal{A}^1|,| \mathcal{A}^2|)$.

Given a joint stationary policy $\pi=(\pi^i,\pi^{-i})$, where $\pi^i:\mathcal{S}\mapsto\Delta^{|\mathcal{A}^i|}$ and $\pi^{-i}:\mathcal{S}\mapsto\Delta^{|\mathcal{A}^{-i}|}$, we define the \emph{local} $q$-function $q_\pi^i\in\mathbb{R}^{|\mathcal{S}||\mathcal{A}^i|}$ of player $i$ as 
\begin{align*}
    q_\pi^i(s,a^i)=\mathbb{E}_\pi\left[\sum_{k=0}^\infty\gamma^i\mathcal{R}^i(S_k,A_k^i,A_k^{-i})\;\middle|\; S_0=s,A_0^i=a^i\right]
\end{align*}
for all $(s,a^i)$, where we use the notation $\mathbb{E}_\pi[\,\cdot\,]$ to indicate  that the actions are chosen according to the joint policy $\pi$. In addition, we define the $v$-function $v_\pi^i\in\mathbb{R}^{|\mathcal{S}|}$ as $v_\pi^i(s)=\mathbb{E}_{A^i\sim \pi^i(\cdot|s)}[q_\pi^i(s,A^i)]$ for all $s$, and the utility function $U^i(\pi^i,\pi^{-i})\in\mathbb{R}$ as $U^i(\pi^i,\pi^{-i})=\mathbb{E}_{S\sim p_o}[v^i_\pi(S)]$, where $p_o\in\Delta^{|\mathcal{S}|}$ is an arbitrary initial distribution on the states.
\begin{definition}[Nash Gap in Markov Games]\label{Nash_gap_markov}
Given a joint policy $\pi=(\pi^i,\pi^{-i})$, the Nash gap is defined as
\begin{align*}
	\textit{NG}(\pi^i,\pi^{-i})=\sum_{i=1,2}\left(\max_{\hat{\pi}^i}U^i(\hat{\pi}^i,\pi^{-i})- U^i(\pi^i,\pi^{-i})\right).
\end{align*}
\end{definition}

In what follows, we will frequently work with the real vectors  in $\mathbb{R}^{|\mathcal{S}||\mathcal{A}^i|}$, $\mathbb{R}^{|\mathcal{S}||\mathcal{A}^{-i}|}$, and $\mathbb{R}^{|\mathcal{S}||\mathcal{A}^i||\mathcal{A}^{-i}|}$. To simplify the notation, for any $Q\in\mathbb{R}^{|\mathcal{S}||\mathcal{A}^i||\mathcal{A}^{-i}|}$, we use $Q(s)$ to denote the $|\mathcal{A}^i|\times|\mathcal{A}^{-i}|$ matrix with the $(a^i,a^{-i})$-th entry being $Q(s,a^i,a^{-i})$. Similarly, for any $q\in\mathbb{R}^{|\mathcal{S}||\mathcal{A}^i|}$, we use $q(s)$ to denote the $|\mathcal{A}^i|$-dimensional vector with its $a^i$-th entry being $q(s,a^i)$. 

\subsection{Algorithm: Doubly Smoothed Best-Response Dynamics with Value Iteration}\label{subsec:Markov_Algorithm}

Our learning dynamics for Markov games (cf. Algorithm \ref{algorithm:tabular}) builds on the ideas presented in our algorithm design for matrix games in Section \ref{subsec:bandit_algorithm},  with the additional incorporation  of minimax value iteration,  a well-known approach for zero-sum stochastic  games \citep{shapley1953stochastic}.  

\paragraph{Algorithmic Ideas.} 
To motivate the algorithm design, we need to introduce the following notation. For $i\in \{1,2\}$, let $\mathcal{T}^i:\mathbb{R}^{|\mathcal{S}|}\mapsto\mathbb{R}^{|\mathcal{S}||\mathcal{A}^i||\mathcal{A}^{-i}|}$ be an operator defined as
\begin{align*}
	\mathcal{T}^i(v)(s,a^i,a^{-i})= \mathcal{R}^i(a,a^i,a^{-i})+\gamma\mathbb{E}\left[ v(S_1)\mid S_0=s,A_0^i=a^i,A_0^{-i}=a^{-i}\right]
\end{align*}
for all $(s,a^i,a^{-i})$ and $v\in\mathbb{R}^{|\mathcal{S}|}$. We also define  $\textit{val}^i:\mathbb{R}^{|\mathcal{A}^i|\times |\mathcal{A}^{-i}|}\mapsto\mathbb{R}$ to be the following  operator  
\begin{align*} 
	\textit{val}^i(X)=\max_{\mu^i\in\Delta^{|\mathcal{A}^i|}}\min_{\mu^{-i}\in\Delta^{|\mathcal{A}^{-i}|}}\{(\mu^i)^\top X\mu^{-i}\}
	=\min_{\mu^{-i}\in\Delta^{|\mathcal{A}^{-i}|}}\max_{\mu^i\in\Delta^{|\mathcal{A}^i|}}\{(\mu^i)^\top X\mu^{-i}\}
\end{align*}
for all $X\in \mathbb{R}^{|\mathcal{A}^i|\times |\mathcal{A}^{-i}|}$. Then, the minimax Bellman operator $\mathcal{B}^i:\mathbb{R}^{|\mathcal{S}|}\mapsto\mathbb{R}^{|\mathcal{S}|}$ is defined as 
\begin{align*}
    \mathcal{B}^i(v)(s)=\textit{val}^i(\mathcal{T}^i(v)(s))
\end{align*}
for all $s\in\mathcal{S}$,
where $\mathcal{T}^i(v)(s)$ is an $|\mathcal{A}^i|\times|\mathcal{A}^{-i}|$ matrix according to our notation. It is known that the operator $\mathcal{B}^i(\cdot)$ is a $\gamma$ -- contraction mapping with respect to the $\ell_\infty$-norm \citep{shapley1953stochastic}, hence admits a unique fixed-point, which we denote by $v_*^i$.

A common approach for solving Markov games is to first implement the minimax value iteration $v^i_{t+1}=\mathcal{B}^i(v_t^i)$
until (approximate) convergence to $v_*^i$, and then solve the matrix game 
\begin{align*}
    \max_{\mu^i\in\Delta^{|\mathcal{A}^i|}}\min_{\mu^{-i}\in\Delta^{|\mathcal{A}^{-i}|}}(\mu^i)^\top\mathcal{T}^i(v_*^i)(s)\mu^{-i}
\end{align*}for each state $s$ to obtain an  (approximate) Nash equilibrium policy. However, implementing this  algorithm  requires complete knowledge of the underlying transition probabilities. Moreover, since it is an off-policy algorithm, the output is independent of the opponent's policy. Thus, it is not rational by the definition in \cite{bowling2001rational}. To design a model-free and rational learning dynamics, let us first rewrite the minimax value iteration in the following equivalent way:
\begin{align}
	\hat{v}(s)=\;&\max_{\mu^i}\min_{\mu^{-i}}(\mu^i)^\top \mathcal{T}^i(v_t^i)(s)\mu^{-i},\; \forall\;s\in\mathcal{S},\label{eq:minimax_reformulate2}\\
	v_{t+1}^i=\;&\hat{v}.\label{eq:minimax_reformulate3}
\end{align}
In view of Eqs. (\ref{eq:minimax_reformulate2}) and (\ref{eq:minimax_reformulate3}), we need to solve a matrix game with payoff matrix $\mathcal{T}^i(v_t^i)(s)$ for each state $s$ and then update  the value of the game to $v_{t+1}^i(s)$. In light of Algorithm \ref{algo:bandit}, we already know how to solve matrix games with independent learning. Thus, what remains is to combine Algorithm \ref{algo:bandit} with value iteration, i.e.,  Eq. (\ref{eq:minimax_reformulate3}).  This combination yields Algorithm \ref{algorithm:tabular}.

\begin{algorithm}[!t]\caption{Doubly Smoothed Best-Response Dynamics with Value Iteration}\label{algorithm:tabular} 
	\begin{algorithmic}[1]
		\STATE \textbf{Input:} Integers $K$ and $T$, initializations $v_0^i=\bm{0}\in\mathbb{R}^{|\mathcal{S}|}$, $q_{t,0}^i=\bm{0}\in\mathbb{R}^{|\mathcal{S}||\mathcal{A}^i|}$ for all $t$, $\pi_{t,0}^i(a^i|s)=1/|\mathcal{A}^i|$ for all $(s,a^i)$ and $t$, and $S_0$ arbitrarily.
		\FOR{$t=0,1,\cdots,T$}
		\FOR{$k=0,1,\cdots,K-1$}
		\STATE $\pi_{t,k+1}^i(s)=\pi_{t,k}^i(s)+\beta_k(\sigma_\tau(q_{t,k}^i(s))-\pi_{t,k}^i(s))$ for all $s\in\mathcal{S}$
		\STATE Play
		$A_k^i\sim \pi_{t,k+1}^i(\cdot|S_k)$ (against $A_k^{-i}$), and observe $S_{k+1}\sim p(\cdot\mid S_k,A_k^i,A_k^{-i})$
		\STATE $q_{t,k+1}^i(s,a^i)=q_{t,k}^i(s,a^i)+\alpha_k\mathds{1}_{\{(s,a^i)=(S_k,A_k^i)\}}(\mathcal{R}^i(S_k,A_k^i,A_k^{-i})+\gamma v_t^i(S_{k+1})-q_{t,k}^i(S_k,A_k^i))$ for all $(s,a^i)$
		\ENDFOR
		\STATE $v_{t+1}^i(s)=\pi_{t,K}^i(s)^\top q_{t,K}^i(s)$ for all $s\in\mathcal{S}$ and set $S_0=S_K$
		\ENDFOR
		\STATE \textbf{Output:} $\pi_{T,K}^i$
	\end{algorithmic}
\end{algorithm} 

\paragraph{Algorithm Details.} 
For each state $s$, the inner-loop of Algorithm \ref{algorithm:tabular} is designed to solve a matrix game with payoff matrices $\mathcal{T}^i(v_t^i)(s)$ and $\mathcal{T}^{-i}(v_t^{-i})(s)$, which reduces to Algorithm \ref{algo:bandit} when (1) the Markov game has only one state, and (2) $v_t^i=v_t^{-i}=\bm{0}$. However, since $v_t^i$ and $v_t^{-i}$ are \emph{independently} maintained by player $i$ and its opponent, the quantity 
\begin{align*}
    \mathcal{T}^i(v_t^i)(s,a^i,a^{-i})+\mathcal{T}^{-i}(v_t^{-i})(s,a^{-i},a^i)= \gamma \sum_{s'}p(s'\mid s,a^i,a^{-i})(v_t^i(s)+v_t^{-i}(s))
\end{align*}
is in general \emph{non-zero} during learning. As a result, the auxiliary matrix game (with payoff matrices $\mathcal{T}^i(v_t^i)(s)$ and $\mathcal{T}^{-i}(v_t^{-i})(s)$) that the inner loop of Algorithm \ref{algorithm:tabular} is designed to solve is not necessarily zero-sum, which presents a major challenge in the finite-sample analysis, as illustrated previously in Section \ref{subsec:challenge}.

The outer loop of Algorithm \ref{algorithm:tabular} is an ``on-policy'' variant of minimax value iteration. To see this, note that ideally we would synchronize $v_{t+1}^i(s)$ with $\pi_{t,K}^i(s)^\top \mathcal{T}^i(v_t^i)(s)\pi_{t,K}^{-i}(s)$, which is an approximation of $\textit{val}^i(\mathcal{T}^i(v_t^i)(s))$ by design of our inner loop. However, player $i$ has no access to $\pi_K^{-i}$ in independent learning. Fortunately, the $q$-function $q_{t,K}^i$ is precisely constructed as an \emph{estimate} of $\mathcal{T}^i(v_t^i)(s)\pi_{t,K}^{-i}(s)$, as illustrated in Section \ref{subsec:bandit_algorithm}, which leads to the outer loop of Algorithm \ref{algorithm:tabular}. In Line $8$ of Algorithm \ref{algorithm:tabular}, we set $S_0=S_K$ to ensure that the initial state of the next inner-loop is the last state of the previous inner-loop, hence Algorithm \ref{algorithm:tabular} is driven by a single trajectory of Markovian samples.

\subsection{Finite-Sample Analysis}\label{subsec:finite-sample-analysis}

We now state our main results, which provide the \emph{first} finite-sample bounds for best-response type independent learning dynamics in zero-sum Markov games. 
A detailed analysis is provided in Section \ref{sec:sketch} and the complete proof are provided in Appendix \ref{sec:analysis}. Our results rely on one assumption.

\begin{assumption}\label{as:MC}
	There \emph{exists} a joint policy $\pi_b=(\pi_b^i,\pi_b^{-i})$ such that the Markov chain $\{S_k\}_{k\geq 0}$ induced by $\pi_b$ is irreducible and aperiodic.
\end{assumption}

Most, if not all, analyses of RL algorithms driven by \emph{time-varying}  behavior policies assume that the induced Markov chain of \emph{any} policy, or any policy from the algorithm trajectory, is uniformly geometrically ergodic \citep{zou2019finite,khodadadian2021finite,chenziyi2022sample,chenzy2021sample,xu2021sample,wu2020finite,qiu2019finite}.  Assumption \ref{as:MC} is weaker, since it assumes only the existence of \emph{one} policy that induces an irreducible and aperiodic Markov chain. 

In the following theorems, we consider using either constant stepsize $\alpha_k\equiv \alpha$, or diminishing stepsize $\alpha_k=\alpha/(k+h)$. In either case, $\beta_k=c_{\alpha,\beta}\alpha_k$ with $c_{\alpha,\beta}\in (0,1)$ being a tunable constant. The parameters $\{\hat{c}_j\}_{0\leq j\leq 4}$ used to state the following theorem are numerical constants, and $c_{\tau}$, $\ell_\tau$, and $\hat{L}_{\tau}$ are constants that depend on the temperature $\tau$ and $A_{\max}$. See Appendix \ref{sec:analysis} for the explicit expressions of the quantities.

\begin{theorem}[Constant Stepsize Bound]\label{thm:tabular}
	Suppose that both players follow Algorithm \ref{algorithm:tabular}, Assumption \ref{as:MC} is satisfied, and the stepsize ratio satisfies $c_{\alpha,\beta}\leq  \frac{c_{\tau}\tau^3\ell_{\tau}^2(1-\gamma)^2}{\hat{c}_0|\mathcal{S}|A_{\max}^2}$. Then, there exists a threshold $z_\beta=\mathcal{O}(\log(1/\beta))$ such that the following inequality holds as long as $K\geq z_\beta$:
	\begin{align}\label{eq:Markov_bound_constant}
		\mathbb{E}[\textit{NG}(\pi_{T,K}^i,\pi_{T,K}^{-i})]\!\leq  &\underbrace{\frac{\hat{c}_1|\mathcal{S}|A_{\max}T}{\tau(1-\gamma)^3}\left(\frac{1+\gamma}{2}\right)^{T-1}}_{\mathcal{E}_1:\text{ Value Iteration Bias}}\nonumber\\
		&+\underbrace{\frac{\hat{c}_2(|\mathcal{S}|A_{\max})^{3/2}(K-z_\beta)^{1/2}}{\tau(1-\gamma)^5}\left(1\!-\!\frac{c_{\alpha,\beta}\alpha}{2}\right)^{\frac{K-z_\beta-1}{2}}}_{\mathcal{E}_2: \text{Convergence Bias in the Inner-Loop}}\nonumber\\
		&+\underbrace{\frac{\hat{c}_3|\mathcal{S}|^2A_{\max}^2\hat{L}_\tau}{c_{\alpha,\beta}(1-\gamma)^5}z_\beta^2\alpha^{1/2}}_{\mathcal{E}_3: \text{ Variance in the Inner-Loop}}+\underbrace{\frac{\hat{c}_4\tau \log(A_{\max})}{(1-\gamma)^2}}_{\mathcal{E}_4: \text{ Smoothing Bias}}.
	\end{align}
\end{theorem}

As in Theorem \ref{thm:bandit} for matrix games, the bound includes terms for the convergence bias, variance, and smoothing bias. However, now there is an additional term capturing the value iteration bias.  More specifically, the first term $\mathcal{E}_1$ on the right-hand side of  Eq. (\ref{eq:Markov_bound_constant}) is referred to as  the value iteration bias, and would be the only error term if we were able to perform minimax value iteration to solve the game. The terms $\mathcal{E}_2$ and $\mathcal{E}_3$ are the counterparts of the first two terms  on the right-hand side of Eq. (\ref{eq:bound_bandit_constant}) in Theorem \ref{thm:bandit}, and capture the convergence bias and the variance in the inner-loop. The term $\mathcal{E}_4$ represents the smoothing bias resulted from using softmax instead of hardmax in the learning dynamics. Since a Markov game is a sequential decision making problem, the smoothing bias is accumulated over time, and hence is multiplied by a factor depending on the effective horizon of the problem compared to its counterpart in matrix games. 

Notably, the terms $\mathcal{E}_2$ and $\mathcal{E}_3$ are order-wise larger compared to their matrix game counterparts, which is the (mathematical) reason that Algorithm \ref{algorithm:tabular} has a slower convergence rate (or larger sample complexity) compared to that of Algorithm \ref{algo:bandit}. Intuitively, the reason is that the induced auxiliary matrix game (with payoff matrices $\mathcal{T}^i(v_t^i)(s)$ and $\mathcal{T}^{-i}(v_t^{-i})(s)$) that the inner-loop of Algorithm \ref{algorithm:tabular} is designed to solve does not necessarily have a zero-sum structure (see the discussion after Algorithm \ref{algorithm:tabular}). Consequently, the error due to such ``non-zero-sum'' structure propagates through the algorithm and eventually undermines the rate of convergence.

We next consider using diminishing stepsizes, i.e., $\alpha_k=\alpha/(k+h)$ and $\beta_k=c_{\alpha,\beta}\alpha_k$. The requirement for choosing $\alpha$ and $h$ are presented in Appendix \ref{sec:analysis}. The parameters $\{\hat{c}'_j\}_{0\leq j\leq 3}$ used in presenting the following theorem are numerical constants.

\begin{theorem}[Diminishing Stepsizes Bound]\label{thm:diminishing}
	Suppose that both players follow the learning dynamics in Algorithm \ref{algorithm:tabular}, Assumption \ref{as:MC} is satisfied, and the stepsize ratio satisfies $c_{\alpha,\beta}\leq  \frac{c_{\tau}\tau^3\ell_{\tau}^2(1-\gamma)^2}{\hat{c}_0'|\mathcal{S}|A_{\max}^2}$. Then there exists a threshold $k_0>0$ such that the following inequality holds as long as $K\geq k_0$:
	\begin{align*}
		\mathbb{E}[\textit{NG}(\pi_{T,K}^i,\pi_{T,K}^{-i})]\leq\;&\underbrace{\frac{\hat{c}_1'|\mathcal{S}|A_{\max}T}{\tau(1-\gamma)^3}\left(\frac{\gamma+1}{2}\right)^{T-1}}_{\mathcal{E}_1'}+\underbrace{\frac{\hat{c}_2'|\mathcal{S}|^2A_{\max}^2\hat{L}_\tau}{\alpha_{k_0} c_{\alpha,\beta}(1-\gamma)^5}\frac{z_K^2\alpha^{1/2}}{(K+h)^{1/2}}}_{\mathcal{E}_{2,3}'}\\
		&+\underbrace{\frac{\hat{c}_3'\tau \log(A_{\max})}{(1-\gamma)^2}}_{\mathcal{E}_4'},\;\text{where }z_K=\mathcal{O}(\log(K)).
	\end{align*}
\end{theorem}

The terms $\mathcal{E}_1'$ and $\mathcal{E}_4'$ are quantitatively similar to the terms $\mathcal{E}_1$ and $\mathcal{E}_4$ in Eq. (\ref{eq:Markov_bound_constant}), and represent the value iteration bias and the smoothing bias. The term $\mathcal{E}_{2,3}'$ corresponds to $\mathcal{E}_2+\mathcal{E}_3$ in Theorem \ref{thm:tabular}, and captures the combined error of the convergence bias and the variance in the inner loop. Since we are using diminishing stepsizes, unlike Eq. (\ref{eq:Markov_bound_constant}), the convergence bias and the variance are balanced, and are both converging at the same rate.

We next present the sample complexity of Algorithm \ref{algorithm:tabular}, which does not depend on whether constant  or diminishing stepsizes are used.

\begin{corollary}[Sample Complexity]
	To achieve $\mathbb{E}[\textit{NG}(\pi_{T,K}^i,\pi_{T,K}^{-i})]\leq \epsilon+\frac{\hat{c}_4\tau \log(A_{\max})}{(1-\gamma)^2}$ for some $\epsilon>0$, the sample complexity is 
	$\tilde{\mathcal{O}}\left(\epsilon^{-2}\right)$.
\end{corollary}

Notably, we achieve an $\tilde{\mathcal{O}}(1/\epsilon^2)$ sample complexity to find a Nash equilibrium up to a smoothing bias, which is order-wise the same compared with the sample complexity of popular RL algorithms in the single agent setting, such as $Q$-learning \citep{qu2020finite,li2020sample,chen2021finite}. We want to emphasize that there are no asymptotic bias terms in those single-agent RL algorithms while we have a smoothing bias. An interesting future direction of this work is to investigate the use of a time-varying temperature $\tau_k$ and establish a sharp rate of convergence with an asymptotically vanishing smoothing bias. We suspect that this is a much more challenging task as even in single-agent $Q$-learning (which is arguably one of the most popular and well-studied algorithms), finite-sample analysis under $\epsilon$-greedy policy with a time-varying $\epsilon$ (or softmax exploration policy with a time-varying temperature) was not performed in the literature.

Finally, we consider the case where the opponent plays with a stationary policy, and provide a sample complexity bound for the player to find the best-response. 

\begin{corollary}[Rationality]\label{co:convergent_opponent_Markov}
	Suppose that player $i$ follows the learning dynamics presented in Algorithm \ref{algorithm:tabular}, but its opponent follows a stationary policy $\pi^{-i}$. Then, given  $\epsilon>0$, to achieve $\max_{\hat{\pi}^i}U^i(\hat{\pi}^i,\pi^{-i})-\mathbb{E}[U^i(\pi_{T,K}^i,\pi^{-i})]\leq \epsilon+\frac{\hat{c}_4\tau \log(A_{\max})}{(1-\gamma)^2}$, the sample complexity is $\tilde{\mathcal{O}}(1/\epsilon^2)$.
\end{corollary}

Intuitively, the reason that our algorithm is rational is that it performs the so-called \emph{on-policy} update in RL. In contrast to an off-policy update, where the behavior policy can be arbitrarily different from the policy being generated during learning (such as in $Q$-learning and off-policy TD-learning), in the on-policy update for games, each player is actually playing with the policy that is moving towards the best-response to its opponent. As a result, when the opponent's policy is stationary, it reduces to a single-agent problem and the player naturally finds the best response (also up to a smoothing bias). This is also exactly the advantage of  symmetric and independent learning dynamics. 

\section{Analyzing the Learning Dynamics in Algorithm \ref{algorithm:tabular}}\label{sec:sketch} 

In this section, we present the key steps and technical ideas used to prove Theorem \ref{thm:tabular} and Theorem \ref{thm:diminishing}.  The core challenge here is that Algorithm \ref{algorithm:tabular} maintains $3$ sets of iterates ($\{q_{t,k}^i\}$, $\{\pi_{t,k}^i\}$, and $\{v_t^i\}$), which are coupled. The coupling of their update equations means that it is not possible to separately analyze them. Instead, we develop a \textit{coupled Lyapunov drift argument} to establish the finite-sample bounds of Algorithm \ref{algorithm:tabular}. Specifically, we first show that the expected Nash gap can be upper bounded by a sum of properly defined Lyapunov functions, one for each set of the iterates (i.e., the $v$-functions, the policies, and the $q$-functions). Then, we establish a set of coupled Lyapunov drift inequalities -- one for each Lyapunov function. Finally, we decouple the Lyapunov drift inequalities to establish the overall finite-sample bounds. We outline the key steps in the argument below.

To begin with, we show that the $q$-functions $\{q_{t,k}^i\}$ and the $v$-functions $\{v_t^i\}$ generated by Algorithm \ref{algorithm:tabular} are uniformly bounded from above in $\ell_\infty$-norm by $1/(1-\gamma)$ (cf. Lemma \ref{le:boundedness}), and the entries of the policies $\{\pi_{t,k}^i\}$ are uniformly bounded below by $\ell_\tau>0$ (cf. Lemma \ref{le:margin}). These two results are frequently used in our analysis.

At the core of our argument is the following inequality:
\begin{align}\label{eq:sketch:overall}
	\textit{NG}(\pi_{T,K}^i,\pi_{T,K}^{-i})\!\leq\! C_0\bigg(2\|v_T^i+v_T^{-i}\|_\infty\!+\!\sum_{i=1,2}\|v^i_T-v^i_{*}\|_\infty\!+\!\mathcal{L}_\pi(T,K)\!+\!\tau \log(A_{\max})\bigg),
\end{align}
where $C_0$ is a constant, and $\mathcal{L}_\pi(\cdot)$ stands for the Lyapunov function we use to analyze the policies (the explicit expression of which is presented in Eq. (\ref{eq:Lyaunov_short}). Eq. (\ref{eq:sketch:overall}) follows from Lemma \ref{le:Nash_to_v} and Lemma \ref{le:Nash_Gap}.

\subsection{Analysis of the Outer Loop: $v$-Function Update}
Motivated by Eq. (\ref{eq:sketch:overall}), we need to bound $\|v_T^i+v_T^{-i}\|_\infty$ and $\|v^i_T-v^i_{*}\|_\infty$. To do so, we view them as Lyapunov functions and establish Lyapunov drift inequalities for them. Specifically, we show in Lemma \ref{le:outer-loop} and Lemma \ref{le:outer-sum} that
\begin{align}
	\|v_{t+1}^i-v_*^i\|_\infty
	\leq \;&\underbrace{\gamma  \|v_t^i-v_*^i\|_\infty}_{\text{Negative Drift}}\nonumber\\
	&+\underbrace{C_1(\|v_t^i+v_t^{-i}\|_\infty+\mathcal{L}_\pi(t,K)+\mathcal{L}_q^{1/2}(t,K)+\tau \log(A_{\max}))}_{\text{Additive Errors}},\label{eq:sketchv}\\
	\|v_{t+1}^i+v_{t+1}^{-i}\|_\infty\leq\;& \underbrace{\gamma\|v_t^i+v_t^{-i}\|_\infty}_{\text{Negative Drift}}+\underbrace{C_2\mathcal{L}_q^{1/2}(t,K)}_{\text{Additive Errors}}\label{eq:sketchvsum}
\end{align}
for all $t\geq 0$,
where $C_1$ and $C_2$ are constants, and $\mathcal{L}_q(\cdot)$ stands for the  Lyapunov function we use to analyze the $q$-functions (the  expression of which is presented in Eq. (\ref{eq:Lyaunov_short})). If the \textit{Additive Errors} in the previous two inequalities were only functions of $v_t^i$ and $v_t^{-i}$, then these two Lyapunov drift inequalities can be repeatedly used to obtain a convergence bound for $\|v_T^i+v_T^{-i}\|_\infty$ and $\|v^i_T-v^i_{*}\|_\infty$. However, the coupled nature of Eqs. (\ref{eq:sketchv}) and (\ref{eq:sketchvsum}) requires us to analyze the policies and the $q$-functions in the inner loop, and establish their Lyapunov drift inequalities.

\subsection{Analysis of the Inner Loop: Policy Update}  As illustrated in Section \ref{subsec:bandit_algorithm} and Section \ref{subsec:Markov_Algorithm}, for each state $s$, the update equation of the policies can be viewed as a discrete and stochastic variant of the smoothed best-response dynamics for solving matrix games \citep{leslie2005individual}. Typically, the following Lyapunov function is used to study such dynamics:
\begin{align}\label{eq:sketchV} 
	V_X(\mu^i,\mu^{-i})=\sum_{i=1,2}\max_{\hat{\mu}^i\in\Delta^{|\mathcal{A}^i|}}\{(\hat{\mu}^i-\mu^i)^\top X_i\mu^{-i}+\tau \nu(\hat{\mu}^i)-\tau\nu(\mu^i)\},
\end{align}
where $X_i$ and $X_{-i}$ are the payoff matrices for player $i$ and player $-i$, respectively, and $\nu(\cdot)$ is the entropy function defined as $\nu(\mu^i)=-\sum_{a^i}\mu^i(a^i)\log(\mu^i({a^i}))$. 
Specialized to our case, given a joint $v$-function $v=(v^i,v^{-i})$ from the outer loop\footnote{Due to the nested-loop structure of Algorithm \ref{algorithm:tabular}, conditioned on the history up to the beginning of the $t$-th outer loop, the $v$-functions $v_t^i$ and $v_t^{-i}$ are constants. Thus, when focusing on the inner loop we omit the subscript $t$.}, and a state $s\in\mathcal{S}$, we would like to use 
\begin{align*}
	V_{v,s}(\pi^i(s),\pi^{-i}(s)):=\sum_{i=1,2}\max_{\hat{\mu}^i\in\Delta^{|\mathcal{A}^i|}}\{(\hat{\mu}^i-\pi^i(s))^\top \mathcal{T}^i(v^i)(s)\pi^{-i}(s)+\tau \nu(\hat{\mu}^i)-\tau\nu(\pi^i(s))\}
\end{align*}
as our Lyapunov function. Unlike the continuous-time smoothed best-response dynamics\footnote{The continuous-time best-response dynamics is an ordinary differential equation (ODE) use to study matrix games, and is defined as $\dot{\pi}^i=\sigma_\tau(\mathcal{R}^i\pi^{-i})-\pi^i$.}, our policy update equation in Algorithm \ref{algorithm:tabular} Line $4$ is discrete and stochastic. To use $V_{v,s}(\pi^i(s),\pi^{-i}(s))$ as our Lyapunov function to study the policy convergence, we need to show that $V_{v,s}(\cdot,\cdot)$ is a smooth function.  However, since the entropy $\nu(\cdot)$ is not a smooth function, the function $V_{v,s}(\cdot,\cdot)$ is in general not smooth on the joint simplex $\Delta^{|\mathcal{A}^i|}\times \Delta^{|\mathcal{A}^{-i}|}$. 

To overcome this difficulty, recall that we have shown that all the policies from the algorithm trajectory have uniformly lower-bounded entries, with the lower bound being $\ell_\tau$ (cf. Lemma \ref{le:margin}). Therefore, it is enough to only consider $V_{v,s}(\cdot,\cdot)$ on the following proper subset of the joint probability simplex $\Pi_{\ell_\tau}:=\{\mu=(\mu^i,\mu^{-i})\in \Delta^{|\mathcal{A}^i|}\times \Delta^{|\mathcal{A}^{-i}|} \mid  \min_{a^i}\mu^i(a^i)>\ell_\tau,\min_{a^{-i}}\mu^{-i}(a^{-i})>\ell_\tau \}$.
Since the extreme points of the joint simplex are excluded, we are able to establish the smoothness of $V_{v,s}(\cdot,\cdot)$ on $\Pi_{\ell_\tau}$, which is key in our Lyapunov approach for analyzing the policies.  Eventually, we obtain the following Lyapunov drift inequality for $V_{v,s}(\cdot,\cdot)$: 
\begin{align}
	\sum_{s}\mathbb{E}[V_{v,s}(\pi_{k+1}^i(s),\pi_{k+1}^{-i}(s))]\leq \;& \underbrace{(1-C_1'\beta_k)\sum_{s}\mathbb{E}[V_{v,s}(\pi_k^i(s),\pi_k^{-i}(s))]}_{\text{Negative Drift}}\nonumber\\
	&+\underbrace{C_2'(\beta_k^2+\beta_k\mathbb{E}[\mathcal{L}_q(k)]+\beta_k\|v^i+v^{-i}\|_\infty^2)}_{\text{Additive Errors}}\label{eq:sketchpi},
\end{align}
where $C_1'$ and $C_2'$ are (problem-dependent) constants.
To interpret the above, suppose that we were considering the continuous-time smoothed best-response dynamics (which is an ODE). Then, the additive error term would disappear in the sense that the time-derivative of the Lyapunov function $V_{v,s}(\cdot)$ along the trajectory of the ODE is strictly negative. Thus, the three terms in the \textit{Additive Errors} can be interpreted as (1) the discretization error in the update equation, (2) the stochastic error in the $q$-function estimate, and (3) the error due to the non-zero-sum structure of the inner-loop auxiliary matrix game; see Section \ref{subsec:Markov_Algorithm}.

\subsection{Analysis of the Inner Loop: $q$-Function Update}

Our next focus is the $q$-function update. The $q$-function update equation is in the same spirit as TD-learning, and a  necessary condition for the convergence of TD-learning is that the behavior policy (i.e., the policy used to collect samples) should enable the agent to sufficiently explore the environment. To achieve this goal, since we have shown that all joint policies from the algorithm trajectory have uniformly lower-bounded entries (with lower bound $\ell_\tau>0$), it is enough to restrict our attention to a ``soft'' policy class $\Pi_\delta:=\{\pi=(\pi^i,\pi^{-i})\mid \min_{s,a^i}\pi^i(a^i|s)>\delta_i,\min_{s,a^{-i}}\pi^{-i}(a^{-i}|s)>\delta_{-i} \}$, where $(\delta_i,\delta_{-i})$ represent the margins. The following lemma, which is an extension of \cite[Lemma 4]{zhang2022global}, establishes a uniform exploration property under Assumption \ref{as:MC}.

To present the result, we need the following notation. Under Assumption \ref{as:MC}, the Markov chain induced by the joint policy $\pi_b$ has a unique stationary distribution $\mu_b\in\Delta^{|\mathcal{S}|}$, the minimum component of which is denoted by $\mu_{b,\min}$. In addition, there exists $\rho_b\in (0,1)$ such that $\max_{s\in\mathcal{S}}\left\|P_{\pi_b}^k(s,\cdot)-\mu_b(\cdot)\right\|_{\text{TV}}\leq 2\rho_b^k$ for all $k\geq 0$ \citep{levin2017markov}, where $P_{\pi_b}$ is the transition probability matrix of the Markov chain $\{S_k\}$ under $\pi_b$.
We also define the mixing time in the following. Given a joint policy $\pi=(\pi^i,\pi^{-i})$ and an accuracy level $\eta>0$, the $\eta$ -- mixing time of the Markov chain $\{S_k\}$ induced by $\pi$ is defined as 
\begin{align*}
    t_{\pi,\eta}=\min\left\{k\geq 0\;:\;\max_{s\in\mathcal{S}}\|P_\pi^k(s,\cdot)-\mu_\pi(\cdot)\|_{\text{TV}}\leq \eta\right\},
\end{align*}
where $P_\pi$ is the $\pi$-induced transition probability matrix and $\mu_\pi$ is the stationary distribution of $\{S_k\}$ under $\pi$, provided that it exists and is unique. When the induced Markov chain mixes at a geometric rate, it is easy to see that $t_{\pi,\eta}=\mathcal{O}(\log(1/\eta))$.

\begin{lemma}[An Extension of Lemma 4 in \cite{zhang2022global}]\label{thm:exploration}
	Suppose that Assumption \ref{as:MC} is satisfied. Then we have the following results.
	\begin{enumerate}[(1)]
		\item For any $\pi=(\pi^i,\pi^{-i})\in \Pi_\delta$, the Markov chain $\{S_k\}$ induced by the joint policy $\pi$ is irreducible and aperiodic, hence admits a unique stationary distribution $\mu_\pi\in\Delta^{|\mathcal{S}|}$.
		\item It holds that $\sup_{\pi\in \Pi_\delta}\max_{s\in\mathcal{S}}\|P_\pi^k(s,\cdot)-\mu_\pi(\cdot)\|_{\text{TV}}\leq 2\rho_\delta^k$ for any $k\geq 0$, where  $\rho_\delta=\rho_b^{(\delta_i\delta_{-i})^{r_b}\mu_{b,\min}}$ and $r_b:=\min\{k\geq 0\;:\; P_{\pi_b}^k(s,s')>0,\;\forall\;(s,s')\}$. As a result, we have
		\begin{align}\label{eq:mixing_time_definition}
			t(\delta,\eta):=\sup_{\pi\in \Pi_\delta}t_{\pi,\eta}\leq \frac{t_{\pi_b,\eta}}{(\delta_i\delta_{-i})^{r_b}\mu_{b,\min}},
		\end{align}
		where we recall that $t_{\pi,\eta}$ is the $\eta$ -- mixing time of the Markov chain $\{S_k\}$ induced by $\pi$.
		\item 
		Let $G:\mathbb{R}^{|\mathcal{S}|A_{\max}}\mapsto\mathbb{R}^{|\mathcal{S}|}$ be the mapping from a policy $\pi\in \Pi_\delta$ to the unique stationary distribution $\mu_\pi$ of the Markov chain $\{S_k\}$ induced by $\pi$. Then $G(\cdot)$ is Lipschitz continuous with respect to $\|\cdot\|_\infty$, with Lipschitz constant $\hat{L}_\delta:=\frac{2\log(8|\mathcal{S}|/\rho_\delta)}{\log(1/\rho_\delta)}$.
		\item $\mu_\delta:=\inf_{\pi\in \Pi_\delta}\min_{s\in\mathcal{S}}\mu_\pi(s)>0$.
	\end{enumerate}
\end{lemma}
\begin{remark}
Lemma \ref{thm:exploration} (1), (3), and (4) were previous established in \cite[Lemma 4]{zhang2022global}. Lemma \ref{thm:exploration} (2) enables us to see the explicit dependence of the ``uniform mixing time'' on the margins $\delta_i$, $\delta_{-i}$ and the mixing time of the benchmark exploration policy $\pi_b$.
\end{remark}

In view of Lemma \ref{thm:exploration} (2), we have fast mixing for all policies in $\Pi_\delta$ if \textit{(i)} the margins $\delta_i,\delta_{-i}$ are large, and \textit{(ii)} the Markov chain $\{S_k\}$ induced by the benchmark exploration policy $\pi_b$ is well-behaved. By ``well-behaved'' we mean the mixing time is small (i.e., small $t_{\pi_b,\eta}$) and the stationary distribution is relatively well-balanced (i.e., large $\mu_{b,\min}$). Point \textit{(i)} agrees with our intuition as large margins encourage more exploration. To make sense of point \textit{(ii)}, since $\pi(a|s)\geq \delta_i\delta_{-i}\pi_b(a|s)$ for all $s$ and $a=(a^i,a^{-i})$, we can write $\pi$ as a convex combination between $\pi_b$ and some residual policy $\tilde{\pi}$: $\pi=\delta_i\delta_{-i}\pi_b+(1-\delta_i\delta_{-i})\tilde{\pi}$. Therefore, since any $\pi\in \Pi_\delta$ has a portion of the benchmark exploration policy $\pi_b$ in it, it makes intuitive sense that fast mixing of $\{S_k\}$ under $\pi_b$ implies, to some extent, fast mixing of $\{S_k\}$ under $\pi\in \Pi_\delta$. Note that, as the margins $\delta_i,\delta_{-i}$ approach zero, the uniform mixing time in Lemma \ref{thm:exploration} (2) goes to infinity. This is not avoidable in general, as demonstrated by a simple MDP example constructed in Appendix \ref{ap:mixing_example}. 

When $\Pi_\delta=\Pi_{\ell_\tau}$, we denote $\rho_\tau:=\rho_\delta$, $\mu_\tau:=\mu_\delta$, and $\hat{L}_\tau:=\hat{L}_\delta$. We also define $c_\tau:=\mu_\tau\ell_\tau$. With Lemma \ref{thm:exploration} in hand, we are now able to analyze the behavior of the $q$-functions. We model the $q$-function update as a stochastic approximation algorithm driven by \emph{time-inhomogeneous}  Markovian noise, and use the norm-square function 
\begin{align*}
	\sum_{i=1,2}\sum_{s}\|q^i(s)-\mathcal{T}^i(v^i)(s)\pi_k^{-i}(s)\|_2^2
\end{align*}
as the Lyapunov function to study its behavior. The key challenge to establishing a Lyapunov drift inequality is to control a difference of the form
\begin{align}\label{eq:Markov_term}
	\mathbb{E}[F^i(q^i,S_k,A_k^i,A_k^{-i},S_{k+1})]-\mathbb{E}[F^i(q^i,\hat{S},\hat{A}^i,\hat{A}^{-i},\hat{S}')]
\end{align}
for any $q^i\in\mathbb{R}^{|\mathcal{S}||\mathcal{A}^i|}$, where $F^i(\cdot)$ is some appropriately defined operator that captures the dynamics of the update equation; see Appendix \ref{subsec:q-function} for its definition. In the term (\ref{eq:Markov_term}), the random tuple $(S_k,A_k^i,A_k^{-i},S_{k+1})$ is the $k$-th sample from the time-inhomogeneous Markov chain $\{(S_n,A_n^i,A_n^{-i},S_{n+1})\}_{n\geq 0}$ generated by the time-varying joint policies $\{\pi_n\}_{n\geq 0}$, and $(\hat{S},\hat{A}^i,\hat{A}^{-i},\hat{S}')$ is a random tuple such that $S\sim  \mu_k(\cdot)$, $A^i\sim \pi_k^i$, $A^{-i}\sim \pi_k^{-i}$, and $S'\sim p(\cdot|S,A^i,A^{-i})$, where $\mu_k(\cdot)$ is the unique stationary distribution of the Markov chain $\{S_n\}$ induced by the joint policy $\pi_k$. Due to Lemma \ref{thm:exploration}, $\mu_k$ exists and is unique.

In existing literature, when $\{(S_k,A_k^i,A_k^{-i},S_{k+1})\}$ is sampled either in an i.i.d. manner or forms an ergodic time-homogeneous Markov chain, there are techniques that successfully handle (\ref{eq:Markov_term}) \citep{bertsekas1996neuro,srikant2019finite,bhandari2018finite}. To deal with time-inhomogeneous Markovian noise, building upon existing conditioning results \citep{bhandari2018finite,srikant2019finite,zou2019finite,khodadadian2021finite} and also Lemma \ref{thm:exploration}, we develop a refined conditioning argument to show that
\begin{align*}
	(\ref{eq:Markov_term})=\mathcal{O}\left(z_k\sum_{n=k-z_k}^{k-1}\beta_n\right),
\end{align*}
where $z_k=t(\ell_\tau,\beta_k)$ is a uniform upper bound on the $\beta_k$  -- the mixing time (i.e., the uniform mixing time with accuracy $\beta_k$, see Eq. (\ref{eq:mixing_time_definition})) of the Markov chain $\{S_n\}$ induced by an arbitrary joint policy from the algorithm trajectory. Suppose we are using diminishing stepsizes $\beta_k=\beta/(k+h)$ (similar results hold for constant stepsize). Then, the uniform mixing property from Lemma \ref{thm:exploration} (2) implies that $z_k=\mathcal{O}(\log(1/k))$. As a result, we have $\lim_{k\rightarrow \infty} (\ref{eq:Markov_term})\leq \lim_{k\rightarrow \infty} z_k\sum_{n=k-z_k}^{k-1}\beta_n=0$, which provides us a way to control (\ref{eq:Markov_term}). After successfully handling (\ref{eq:Markov_term}), we are able to establish a Lyapunov drift inequality of the following form:
\begin{align}\label{eq:sketchq}
	\sum_{i=1,2}\mathbb{E}[\|q_{k+1}^i-\Bar{q}_{k+1}^i\|_2^2]\leq \;&\underbrace{\left(1-C_1''\alpha_k\right)\sum_{i=1,2}\mathbb{E}[\|q_k^i-\Bar{q}_k^i\|_2^2]}_{\text{Negative Drift}}\nonumber\\
	&+\underbrace{C_2''(\alpha_k^2+\beta_k\sum_{s}\mathbb{E}[V_{v,s}(\pi_k^i(s),\pi_k^{-i}(s))])}_{\text{Additive Error}}
\end{align}
where $C_1''$ and $C_2''$ are (problem-dependent) constants, $\Bar{q}_k^i(s):=\mathcal{T}^i(v^i)(s)\pi_k^{-i}(s)$ for all $s\in\mathcal{S}$, and $V_{v,s}(\cdot,\cdot)$ is the Lyapunov function we used to study the policy convergence.

\subsection{Solving Coupled Lyapunov Drift Inequalities}
Until this point, we have established the Lyapunov drift inequalities for the individual $v$-functions, the sum of the $v$-functions, the policies, and the $q$-functions in Eqs. (\ref{eq:sketchv}), (\ref{eq:sketchvsum}), (\ref{eq:sketchpi}), and (\ref{eq:sketchq}), respectively. The last challenge is to find a strategic way of using these coupled inequalities to derive the finite-sample bound. To elaborate, we first restate all the Lyapunov drift inequalities in the following. For simplicity of notation, we denote
\begin{align}
	\mathcal{L}_{v}(t)=\;&\sum_{i=1,2}\|v_t^i-v_*^i\|_\infty,\;\qquad\mathcal{L}_{\text{sum}}(t)=\|v_t^i+v_t^{-i}\|_\infty,\nonumber\\
	\mathcal{L}_{\pi}(t,k)=\;&\sum_{s}V_{v_t,s}(\pi_{t,k}^i(s),\pi_{t,k}^{-i}(s)),\;\text{ and }\;\mathcal{L}_q(t,k)=\sum_{i=1,2}\sum_{s}\|q_{t,k}^i(s)-\Bar{q}_{t,k}^i(s)\|_2^2.\label{eq:Lyaunov_short}
\end{align}
Then Eqs. (\ref{eq:sketchv}), (\ref{eq:sketchvsum}), (\ref{eq:sketchpi}), and (\ref{eq:sketchq}) can be compactly written as
\begin{align}
	\mathcal{L}_{v}(t+1)\leq \;&\gamma\mathcal{L}_{v}(t)+C_1(\mathcal{L}_{\text{sum}}(t)+\mathcal{L}_{\pi}(t,K)+\mathcal{L}_q^{1/2}(t,K)+\tau\log(A_{\max})),\label{eq:sketch1}\\
	\mathcal{L}_{\text{sum}}(t+1)\leq\;& \gamma \mathcal{L}_{\text{sum}}(t)+C_2\mathcal{L}_q^{1/2}(t,K),\label{eq:sketch2}\\
	\mathbb{E}_t[\mathcal{L}_{\pi}(t,k+1)]\leq\;& (1-C_1'\beta_k)\mathbb{E}_t[\mathcal{L}_{\pi}(t,k)]+C_2'(\beta_k^2+\beta_k\mathbb{E}_t[\mathcal{L}_{q}(t,k)]+\beta_k \mathcal{L}_{\text{sum}}^2(t)),\label{eq:sketch3}\\
	\mathbb{E}_t[\mathcal{L}_{q}(t,k+1)]\leq\;& (1-C_1''\alpha_k)\mathbb{E}_t[\mathcal{L}_{q}(t,k)]+C_2''(\alpha_k^2+\beta_k\mathbb{E}_t[\mathcal{L}_{\pi}(t,k)]).\label{eq:sketch4}
\end{align}
where $\mathbb{E}_t[\,\cdot\,]$ stands for conditional expectation conditioned on the history up to the beginning of the $t$-th outer loop.

\paragraph{A Vanilla Approach.}
Recall that we have shown that the iterates $\{v_t^i\}$ and $\{q_{t,k}^i\}$ are uniformly bounded (cf. Lemma \ref{le:boundedness}). As a result, all the Lyapunov functions $\mathcal{L}_v(\cdot)$, $\mathcal{L}_{\text{sum}}(\cdot)$, $\mathcal{L}_\pi(\cdot)$, and $\mathcal{L}_q(\cdot)$ are uniformly bounded too, which provides us a handle to decouple the inequalities. As a clear example, observe that
\begin{align*}
    \mathcal{L}_{\pi}(t,k)=\;&\sum_{s}V_{v_t,s}(\pi_{t,k}^i(s),\pi_{t,k}^{-i}(s))\\
    =\;&\sum_{s}\sum_{i=1,2}\max_{\hat{\mu}^i\in\Delta^{|\mathcal{A}^i|}}\{(\hat{\mu}^i-\pi_{t,k}^i(s))^\top \mathcal{T}^i(v_t^i)(s)\pi_{t,k}^{-i}(s)+\tau \nu(\hat{\mu}^i)-\tau\nu(\pi_{t,k}^i(s))\}\\
    \leq \;&\sum_{s}\sum_{i=1,2}\left(2 \max_{s,a^i,a^{-i}}|\mathcal{T}^i(v_t^i)(s,a^i,a^{-i})|+\tau \log(A_{\max})\right)\\
    \leq \;&\sum_{s}\sum_{i=1,2}(2+2\gamma \|v_t^i\|_\infty+\tau \log(A_{\max}))\tag{Definition of $\mathcal{T}^i(\cdot)$}\\
    \leq \;&4|\mathcal{S}|\left(\frac{1}{1-\gamma}+\tau \log(A_{\max})\right):= L_{\text{bound}},
\end{align*}
where the last line follows from the boundedness of the $v$-functions (cf. Lemma \ref{le:boundedness}). Therefore, we can replace $\mathcal{L}_\pi(t,k)$ in Eq. (\ref{eq:sketch4}) by its uniform upper bound established in the previous inequality to obtain
\begin{align}\label{eq:sketch4'}
    \mathbb{E}_t[\mathcal{L}_{q}(t,k+1)]
    \leq (1-C_1''\alpha_k)\mathbb{E}_t[\mathcal{L}_{q}(t,k)]+C_2''\alpha_k^2+C_2'' L_{\text{bound}}\beta_k.
\end{align}
Note that Eq. (\ref{eq:sketch4'}) is now a \textit{decoupled} Lyapunov drift inequality solely for $\mathcal{L}_{q}(\cdot)$, which can be repeatedly used to derive a finite-sample bound for $\mathbb{E}_t[\mathcal{L}_q(\cdot)]$. In particular, when using $\alpha_k=\alpha/(k+h)$ with properly chosen $\alpha$ and $h$, we have
\begin{align}\label{eq:sketch4''}
    \mathbb{E}_t[\mathcal{L}_{q}(t,K)]\leq \;&\mathcal{O}\left(\frac{1}{K+h}\right)+\mathcal{O}\left(c_{\alpha,\beta}\right).
\end{align}
With the same decoupling technique, we can establish finite-sample bounds of $\mathcal{L}_v(\cdot)$, $\mathcal{L}_{\text{sum}}(\cdot)$, and $\mathcal{L}_\pi(\cdot)$. However, in view of Eq. (\ref{eq:sketch4''}), even when using diminishing stepsizes, due to the presence of $\mathcal{O}(c_{\alpha,\beta})$, we cannot make $\mathbb{E}_t[\mathcal{L}_{q}(t,K)]$ arbitrarily small by just increasing the iteration number $K$. In other words, to make $\mathbb{E}_t[\mathcal{L}_{q}(t,K)]$ arbitrarily small, it is necessary to use a diminishing stepsize ratio, which implies $\beta_k=o(\alpha_k)$ and hence making Algorithm \ref{algorithm:tabular} a \emph{two} time-scale learning dynamics. The fact that $\beta_k$ has to be order-wise smaller than $\alpha_k$ will also undermine the convergence rate. Specifically, with this approach (i.e., using the uniform upper bounds to decouple the Lyapunov drift inequalities and enforcing convergence by introducing another time-scale), the overall sample complexity will be order-wise larger than $\tilde{\mathcal{O}}(\epsilon^{-2})$. This is not surprising as we essentially use constants (i.e., the uniform upper bounds) to bound quantities that are actually converging to zero.

In general, we observe from existing literature that once an iterative algorithm has multiple time scales, oftentimes  the convergence rate is downgraded  \citep{khodadadian2021finite,zhang2021globalsoft}.

\paragraph{Our Decoupling Approach.}
To establish a sharper rate without introducing another time-scale, the high-level ideas are (1) using the Lyapunov drift inequalities in a combined way instead of in a separate manner, and (2) a bootstrapping procedure where we first derive a crude bound and then substitute the crude bound back into the Lyapunov drift inequalities to derive a tighter bound. We next present our approach. 

For simplicity of notation, for a scalar-valued quantity $W$ that is a function of $k$ and/or $t$, we say $W=o_k(1)$ if $\lim_{k\rightarrow\infty} W=0$ and $W=o_t(1)$ if $\lim_{t\rightarrow\infty} W=0$. The explicit convergence rates of the $o_k(1)$ term and the $o_t(1)$ term will be revealed in the complete proof in Appendix \ref{sec:strategy}, but is not important for the illustration here.

\paragraph{Step 1.} Adding up Eq. (\ref{eq:sketch3}) and (\ref{eq:sketch4}) and then repeatedly using the resulting inequality, and we obtain:
\begin{align}\label{eq:sketch5}
	\mathbb{E}_t[\mathcal{L}_{\pi}(t,k)]\leq \mathbb{E}_t[\mathcal{L}_{\pi}(t,k)+\mathcal{L}_{q}(t,k)]= o_k(1)+\mathcal{O}(1)\mathcal{L}_{\text{sum}}^2(t),\;\forall\;t,k.
\end{align}

\paragraph{Step 2.} Substituting the bound for $\mathbb{E}_t[\mathcal{L}_{\pi}(t,k)]$ in Eq. (\ref{eq:sketch5}) into Eq. (\ref{eq:sketch4}) and repeatedly using the resulting inequality, and we obtain:
\begin{align*}
	\mathbb{E}_t[\mathcal{L}_{q}(t,K)]=o_K(1)+\mathcal{O}(c_{\alpha,\beta})\mathcal{L}_{\text{sum}}^2(t),\;\forall\;t,
\end{align*}
which in turn implies (by first using Jensen's inequality and then taking total expectation) that:
\begin{align}\label{eq:sketch6}
    \mathbb{E}[\mathcal{L}_{q}^{1/2}(t,K)]=o_K(1)+\mathcal{O}(c^{1/2}_{\alpha,\beta})\mathbb{E}[\mathcal{L}_{\text{sum}}(t)],\;\forall\;t,
\end{align}
where we recall that $c_{\alpha,\beta}=\beta_k/\alpha_k$ is the stepsize ratio. The fact that we are able to get a factor of $\mathcal{O}(c^{1/2}_{\alpha,\beta})$ in front of $\mathbb{E}[\mathcal{L}_{\text{sum}}(t)]$ is crucial for the decoupling procedure.

\paragraph{Step 3.} Taking total expectation on both sides of Eq. (\ref{eq:sketch2}) and then using the upper bound of $\mathbb{E}[\mathcal{L}^{1/2}_{q}(t,K)]$ we obtained in Eq. (\ref{eq:sketch6}), and we obtain
\begin{align*}
	\mathbb{E}[\mathcal{L}_{\text{sum}}(t+1)]\leq (\gamma +\mathcal{O}(c^{1/2}_{\alpha,\beta}))\mathbb{E}[\mathcal{L}_{\text{sum}}(t)]+o_K(1),\;\forall\;t.
\end{align*}
By choosing $c_{\alpha,\beta}$ so that $\mathcal{O}(c^{1/2}_{\alpha,\beta})\leq (1-\gamma)/2$, the previous inequality implies
\begin{align}\label{eq:obsorb}
	\mathbb{E}[\mathcal{L}_{\text{sum}}(t+1)]\leq \left(1-\frac{1-\gamma}{2}\right)\mathbb{E}[\mathcal{L}_{\text{sum}}(t)]+o_K(1),\;\forall\;t,
\end{align}
which can be repeatedly used to obtain
\begin{align}\label{eq:sketch7}
	\mathbb{E}[\mathcal{L}_{\text{sum}}(t)]= o_t(1)+o_K(1).
\end{align}
Substituting the previous bound on $\mathbb{E}[\mathcal{L}_{\text{sum}}(t)]$ into Eq. (\ref{eq:sketch5}) and we have
\begin{align}\label{eq:sketch8}
	\max(\mathbb{E}[\mathcal{L}_{\pi}(t,K)],\mathbb{E}[\mathcal{L}_{q}(t,K)])= o_t(1)+o_K(1).
\end{align}

\paragraph{Step 4.}
Substituting the bounds we obtained for $\mathbb{E}[\mathcal{L}_{\pi}(t,K)]$,  $\mathbb{E}[\mathcal{L}_{q}(t,K)]$, and $\mathbb{E}[\mathcal{L}_{\text{sum}}(t)]$ in Eqs. (\ref{eq:sketch7}), and (\ref{eq:sketch8}) into Eq. (\ref{eq:sketch1}), and then repeatedly using the resulting inequality from $t=0$ to $t=T$, we have
\begin{align*}
	\mathbb{E}[\mathcal{L}_{v}(T)]=o_T(1)+o_K(1)+\mathcal{O}(\tau).
\end{align*}
Now that we have obtained finite-sample bounds for $\mathbb{E}[\mathcal{L}_{v}(T)]$, $\mathbb{E}[\mathcal{L}_{\text{sum}}(T)]$, $\mathbb{E}[\mathcal{L}_{\pi}(T,K)]$, and $\mathbb{E}[\mathcal{L}_{q}(T,K)]$, using them in Eq. (\ref{eq:sketch:overall}) and we finally obtain the desired finite-sample bound for the expected Nash gap.

Looking back at the decoupling procedure, Steps $2$ and $3$ are crucial. In fact, in Step $1$ we already obtain a bound on $\mathbb{E}_t[\mathcal{L}_q(t,k)]$, where the additive error is $\mathcal{O}(1)\mathbb{E}[\mathcal{L}_{\text{sum}}(t)]$. However, directly using this bound on $\mathbb{E}_t[\mathcal{L}_q(t,k)]$ in Eq. (\ref{eq:sketch2}) would result in an expansive inequality for $\mathbb{E}[\mathcal{L}_{\text{sum}}(t)]$. By performing Step $2$, we are able to obtain a tighter bound for $\mathbb{E}_t[\mathcal{L}_q(t,k)]$, with the additive error being $\mathcal{O}(\sqrt{c_{\alpha,\beta}})\mathbb{E}[\mathcal{L}_{\text{sum}}(t)]$. Furthermore, we can choose $c_{\alpha,\beta}$ so that after using the bound from Eq. (\ref{eq:sketch6}) in Eq. (\ref{eq:sketch2}), the additive error $\mathcal{O}(\sqrt{c_{\alpha,\beta}})\mathbb{E}[\mathcal{L}_{\text{sum}}(t)]$ is dominated by the negative drift in Eq. (\ref{eq:obsorb}).

\section{Conclusion}
In this work, we consider solving zero-sum matrix games and Markov games with independent learning dynamics. In both  settings, we design learning dynamics  that are payoff-based, convergent, and rational. In addition, both learning dynamics are intuitive and natural to implement. Our main results provide finite-sample bounds on both learning dynamics, establishing an $\tilde{\mathcal{O}}(1/\epsilon)$ sample complexity in the matrix game setting and an  $\tilde{\mathcal{O}}(1/\epsilon^2)$ sample complexity in the Markov game setting.  Our analysis provides a number of new tools that are likely to be of interest more broadly, such as our strategy to handle coupled Lyapunov drift inequalities. 

As mentioned in Section \ref{subsec:finite-sample-analysis}, an immediate future direction is to investigate using a time-varying temperature $\tau_k$ and establish a sharp rate of convergence with an asymptotically vanishing smoothing bias. In long term, we are interested to see if the algorithmic ideas and the analysis techniques developed in this work can be used to study other classes of games beyond zero-sum stochastic games.

\bibliographystyle{apalike}
{\small \bibliography{references}}

\newpage

\begin{center}
    {\Large \bfseries{Appendices}}
\end{center}

\appendix

\section{Proof of Theorem \ref{thm:tabular} and Theorem \ref{thm:diminishing}}\label{sec:analysis}

We first explicitly state the requirement for choosing the stepsizes. For simplicity of notation, given $k_1\leq k_2$, we denote $\beta_{k_1,k_2}=\sum_{k=k_1}^{k_2}\beta_k$ and  $\alpha_{k_1,k_2}=\sum_{k=k_1}^{k_2}\alpha_k$. For any $k\geq 0$, let $z_k=t(\ell_\tau,\beta_k)$, where $t(\cdot,\cdot)$ is the uniform mixing time defined in Lemma \ref{thm:exploration} (2), and $\ell_\tau$ is the uniform lower bound of the policies derived in Lemma \ref{le:margin}. When using constant stepsize, $z_k$ is not a function of $k$, and is simply denoted by $z_\beta$. Observe that $z_k=\mathcal{O}(\log(k))$ (when using diminishing stepsizes) and $z_\beta=\mathcal{O}(\log(1/\beta))$ due to the uniform geometric mixing property established in Lemma \ref{thm:exploration} (2).

\begin{condition}\label{con:stepsize}
	It holds that $\alpha_{k-z_k,k-1}\leq 1/4$ for all $k\geq z_k$ and $c_{\alpha,\beta}\leq \frac{c_\tau\ell_{\tau}^2\tau^3(1-\gamma)^2}{512|\mathcal{S}|A_{\max}^2}$. When using diminishing stepsizes $\alpha_k=\frac{\alpha}{k+h}$ and $\beta_k=\frac{\beta}{k+h}$, we additionally require $\beta>2$.
\end{condition}

Condition \ref{con:stepsize} is easy to satisfy as (1) $z_k=\mathcal{O}(\log(1/k))$ while $\alpha_k=\mathcal{O}(1/k)$ when using diminishing stepsizes and (2) $z_k=\mathcal{O}(\log(1/\alpha))$ and $\alpha_k=\alpha$ when using constant stepsize. The parameter $k_0$ is defined to be $\min\{k\geq 0\mid k\geq z_k \}$. Note that $k_0=z_\beta$ when using constant stepsize.

\subsection{Notation}
We begin with a summary of some notation that is used in the proof. 

\begin{enumerate}[(1)]
	\item Given a pair of matrices $X_i\in\mathbb{R}^{|\mathcal{A}^i|\times |\mathcal{A}^{-i}|}$, $X_{-i}\in\mathbb{R}^{ |\mathcal{A}^{-i}|\times |\mathcal{A}^i|}$ and a pair of distributions $\mu^i\in\Delta^{|\mathcal{A}^i|},\mu^{-i}\in \Delta^{|\mathcal{A}^{-i}|}$, we define
	\begin{align}\label{def:Nash_Gap}
		V_{X}(\mu^i,\mu^{-i})=\sum_{i=1,2}\max_{\hat{\mu}^i\in\Delta^{|\mathcal{A}^i|}}\left\{(\hat{\mu}^i-\mu^i)^\top X_i\mu^{-i}+\tau \nu(\hat{\mu}^i)-\tau\nu(\mu^i)\right\},
	\end{align}
	where $\nu(\mu^i)=-\sum_{a^i}\mu^i(a^i)\log(\mu^i(a^i))$ is the entropy function.
	\item Given a pair of $v$-functions $(v^i,v^{-i})$ and a state $s\in\mathcal{S}$, when $X_i=\mathcal{T}^i(v^i)(s)$ and $X_{-i}=\mathcal{T}^{-i}(v^{-i})(s)$, we write $V_{v,s}(\cdot,\cdot)$ for $V_X(\cdot,\cdot)$. 
	\item For any $(\pi^i,\pi^{-i})$ and $s$, define  $v^i_{*,\pi^{-i}}(s)=\max_{\hat{\pi}^i}v^i_{\hat{\pi}^i,\pi^{-i}}(s)$, $v^i_{\pi^i,*}=\min_{\hat{\pi}^{-i}}v^i_{\pi^i,\hat{\pi}^{-i}}(s)$, $v^{-i}_{\pi^{-i},*}(s)=\min_{\hat{\pi}^i}v^{-i}_{\pi^{-i},\hat{\pi}^i}(s)$, and $v^{-i}_{*,\pi^i}(s)=\max_{\hat{\pi}^{-i}}v^{-i}_{\hat{\pi}^{-i},\hat{\pi}^i}(s)$. Note that we have $v^i_{*,\pi^{-i}}+v^{-i}_{\pi^{-i},*}=0$ and $v^i_{\pi^i,*}+v^{-i}_{*,\pi^i}=0$ because of the zero-sum structure.
	\item Denote $v_*^i$ (respectively, $v_*^{-i}$) as the unique fixed-point of the equation $\mathcal{B}^i(v^i)=v^i$ (respectively, $\mathcal{B}^{-i}(v^{-i})=v^{-i}$). Note that we have $v_*^i+v_*^{-i}=0$.
\end{enumerate}

\subsection{Boundedness of the Iterates}\label{subsec:boundedness}

We first show in the following two lemmas that all the $q$-functions and $v$-functions generated by Algorithm \ref{algorithm:tabular} are uniformly bounded from above, and the policies are uniformly bounded from below.

\begin{lemma}[Proof in Appendix \ref{pf:le:boundedness}]\label{le:boundedness}
	It holds for all $t,k\geq 0$ and $i\in \{1,2\}$ that 
	\begin{enumerate}[(1)]
		\item $\|v_t^i\|_\infty\leq \frac{1}{1-\gamma}$,
		\item $\|q_{t,k}^i\|_\infty\leq \frac{1}{1-\gamma}$.
	\end{enumerate} 
\end{lemma}

\begin{lemma}[Proof in Appendix \ref{pf:le:margin}]\label{le:margin}
	It holds for all $t,k\geq 0$ and $(s,a^i,a^{-i})$ that
	\begin{enumerate}[(1)]
		\item $\pi_{t,k}^i(a^i|s)\geq\ell_{\tau}$,
		\item $\pi_{t,k}^{-i}(a^{-i}|s)\geq\ell_{\tau}$,
	\end{enumerate}
	where $\ell_\tau=[1+(A_{\max}-1)\exp(2/[(1-\gamma)\tau])]^{-1}$.
\end{lemma}

\subsection{Analysis of the Outer-Loop: $v$-Function Update}\label{subsec:outer-loop}

Our ultimate goal is to bound the expected Nash gap
\begin{align*}
	\mathbb{E}[\textit{NG}(\pi_{T,K}^i,\pi_{T,K}^{-i})]=\mathbb{E}\left[\sum_{i=1,2}\left(\max_{\pi^i}U^i(\pi^i,\pi_{T,K}^{-i})-U^i(\pi_{T,K}^i,\pi_{T,K}^{-i})\right)\right].
\end{align*}

We first bound the Nash gap using the value functions of the output policies of Algorithm \ref{algorithm:tabular}. 

\begin{lemma}[Proof in Appendix \ref{pf:le:Nash_to_v}]\label{le:Nash_to_v}
	The following inequality holds:
	\begin{align}\label{eq:Nash_to_v_policy}
		\sum_{i=1,2}\left(\max_{\pi^i}U^i(\pi^i,\pi_{T,K}^{-i})-U^i(\pi_{T,K}^i,\pi_{T,K}^{-i})\right)\leq \sum_{i=1,2}\left\|v^i_{*,\pi_{T,K}^{-i}}-v^i_{\pi_{T,K}^i,\pi_{T,K}^{-i}}\right\|_\infty.
	\end{align}
\end{lemma}

The next lemmas connects the RHS of Eq. (\ref{eq:Nash_to_v_policy}) to the $v$-function iterates $\{(v_t^i,v_t^{-i})\}_{t\geq 0}$ of Algorithm \ref{algorithm:tabular}. 

\begin{lemma}[Proof in Appendix \ref{pf:le:Nash_Gap}]\label{le:Nash_Gap}
	It holds for all $t\geq 0$ and $i=1,2$ that
	\begin{align*}
		\left\|v^i_{*,\pi_{t,K}^{-i}}-v^i_{\pi_{t,K}^i,\pi_{t,K}^{-i}}\right\|_\infty\leq \;&\frac{2}{1-\gamma}\bigg(2\|v_t^i+v_t^{-i}\|_\infty+2\|v^i_t-v^i_{*}\|_\infty\\
		&+\max_{s}V_{v_t,s}(\pi_{t,K}^i(s),\pi_{t,K}^{-i}(s))+2\tau \log(A_{\max})\bigg).
	\end{align*}
\end{lemma}

In view of Lemma \ref{le:Nash_Gap}, we need to further bound the terms $\|v_t^i+v_t^{-i}\|_\infty$, $\|v^i_t-v^i_{*}\|_\infty$, and
$\max_{s}V_{v_t,s}(\pi_{t,K}^i(s),\pi_{t,K}^{-i}(s))$. We first consider $\|v^i_t-v^i_{*}\|_\infty$, and establish a one-step Lyapunov drift inequality for it.

\begin{lemma}[Proof in Appendix \ref{pf:le:outer-loop}]\label{le:outer-loop}
	It holds for all $t\geq 0$ and $i=1,2$ that
	\begin{align}\label{eq:result:le:outer-loop}
		\|v_{t+1}^i-v_*^i\|_\infty
		\leq \;&\gamma  \|v_t^i-v_*^i\|_\infty+2\max_{s\in\mathcal{S}}V_{v_t,s}(\pi_{t,K}^i(s),\pi_{t,K}^{-i}(s))+4\tau \log(A_{\max})\nonumber\\
		&+\max_{s\in\mathcal{S}}\|\mathcal{T}^i(v^i_t)(s) \pi_{t,K}^{-i}(s)- q_{t,K}^i(s)\|_\infty+2\gamma\|v_t^i+v_t^{-i}\|_\infty.
	\end{align}
\end{lemma}

Our next step is to control $\|v_t^i+v_t^{-i}\|_\infty$. Similar to $\|v_t^i-v_*^i\|_\infty$, we also establish a one-step Lyapunov drift inequality for $\|v_t^i+v_t^{-i}\|_\infty$ in the following lemma. 
\begin{lemma}[Proof in Appendix \ref{pf:le:outer-sum}]\label{le:outer-sum}
	It holds for all $t\geq 0$ that
	\begin{align*}
		\|v_{t+1}^i+v_{t+1}^{-i}\|_\infty\leq \gamma\|v_t^i+v_t^{-i}\|_\infty+\sum_{i=1,2}\max_{s\in\mathcal{S}}\| q_{t,K}^i(s)-\mathcal{T}^i(v_t^i)(s)\pi_{t,K}^{-i}(s)\|_\infty.
	\end{align*}
\end{lemma}

In view of Lemma \ref{le:outer-loop} and Lemma \ref{le:outer-sum}, our next task is to control the following two terms: $\max_{s\in\mathcal{S}}\| q_{t,K}^i(s)-\mathcal{T}^i(v_t^i)(s)\pi_{t,K}^{-i}(s)\|_\infty$, and $\max_{s\in\mathcal{S}}V_{v_t,s}(\pi_{t,K}^i(s),\pi_{t,K}^{-i}(s))$. For ease of exposition, we write down only the inner-loop of Algorithm \ref{algorithm:tabular} in the following. All results derived for the $q$-functions and policies of Algorithm \ref{algorithm:inner-loop} can be directly combined with the outer-loop of Algorithm \ref{algorithm:tabular} using a simple conditioning argument together with the Markov property.

\begin{algorithm}[H]\caption{Inner-Loop of Algorithm \ref{algorithm:tabular}}\label{algorithm:inner-loop}
	\begin{algorithmic}[1]
		\STATE \textbf{Input:} Integer $K$, initializations $q_0^i=\bm{0}\in\mathbb{R}^{|\mathcal{S}| |\mathcal{A}^i|}$ and $\pi_0^i(a^i|s)=1/|\mathcal{A}^i|$ for all $s\in\mathcal{S}$, and a joint $v$-function $v=(v^i,v^{-i})$ from the outer-loop satisfying $
		\max(\|v^{-i}\|_\infty,\|v^i\|_\infty)\leq 1/(1-\gamma)$.
		\FOR{$k=0,1,\cdots,K-1$}
		\STATE $\pi_{k+1}^i(s)=\pi_k^i(s)+\beta_k(\sigma_\tau(q_k^i(s))-\pi_k^i(s))$ for all $s\in\mathcal{S}$
		\STATE Sample
		$A_k^i\sim \pi_{k+1}^i(\cdot\mid S_k)$, and observe $S_{k+1}\sim p(\cdot\mid S_k,A_k^i,A_k^{-i})$
		\STATE $q_{k+1}^i(S_k,A_k^i)=q_k^i(S_k,A_k^i)+\alpha_k \left(\mathcal{R}^i(S_k,A_k^i,A_k^{-i})+\gamma v^i(S_{k+1})-q_k^i(S_k,A_k^i)\right)$
		\ENDFOR
		\STATE \textbf{Output:} $q_K^i$ and $\pi_K^i$
	\end{algorithmic}
\end{algorithm} 

\subsection{Analysis of the Inner-Loop: Policy Update}\label{subsec:policy}

We consider $\{(\pi_k^i,\pi_k^{-i})\}_{k\geq 0}$ generated by Algorithm \ref{algorithm:inner-loop}, and use $V_X(\cdot,\cdot)$ defined in Eq. (\ref{def:Nash_Gap}) as the Lyapunov function to study them. For simplicity of notation, we use $\nabla_1V_X(\cdot,\cdot)$ (respectively, $\nabla_2V_X(\cdot,\cdot)$) to denote the gradient with respect to the first argument (respectively, the second argument). The following lemma establishes the strongly convexity and the smoothness of $V_X(\mu^i,\mu^{-i})$. We only state the results regarding the argument $\mu^i$. Similar results also hold for the argument $\mu^{-i}$.

\begin{lemma}[Proof in Appendix \ref{pf:le:properties_Lyapunov}]\label{le:properties_Lyapunov}
	The function $V_X(\cdot,\cdot)$ has the following properties.
	\begin{enumerate}[(1)]
		\item For any $\mu^{-i}\in\Delta^{|\mathcal{A}^{-i}|}$, $V_X(\mu^i,\mu^{-i})$ as a function of $\mu^i$ is $\tau$ -- strongly convex with respect to $\|\cdot\|_2$.
		\item For any $\delta_i>0$ and $\mu^{-i}\in\Delta^{|\mathcal{A}^{-i}|}$, $V_X(\mu^i,\mu^{-i})$ as a function of $\mu^i$ is $L_\tau$ -- smooth on $\{\mu^i\in\Delta^{|\mathcal{A}^i|}\mid \min_{a^i}\mu^i(a^i)\geq \delta_i\}$ with respect to $\|\cdot\|_2$, where $L_\tau=\frac{\sigma^2_{\max}(X_{-i})}{\tau}+\frac{\tau}{\delta_i}$.
		\item It holds for any $(\mu^i,\mu^{-i})$ that
		\begin{align*}
			&\langle \nabla_1V_X(\mu^i,\mu^{-i}),\sigma_\tau(X_i\mu^{-i})-\mu^i \rangle+\langle \nabla_2V_X(\mu^i,\mu^{-i}),\sigma_\tau(X_{-i}\mu^i)-\mu^{-i} \rangle\\
			\leq\;& -\frac{7}{8}V_X(\mu^i,\mu^{-i})+\frac{16}{\tau}\|X_i+X_{-i}^\top \|_2^2.
		\end{align*}
		\item For any $u^i\in\mathbb{R}^{|\mathcal{A}^i|},u^{-i}\in\mathbb{R}^{|\mathcal{A}^{-i}|}$ , we have for all $(\mu^i,\mu^{-i})\in \{\mu^i\in\Delta^{|\mathcal{A}^i|},\mu^{-i}\in\Delta^{|\mathcal{A}^{-i}|}\mid \min_{a^i}\mu^i(a^i)\geq \delta_i,\min_{a^{-i}}\mu^{-i}(a^{-i})\geq \delta_{-i}\}$ (where $\delta_i,\delta_{-i}>0$) that
		\begin{align*}
			&\langle \nabla_1V_X(\mu^i,\mu^{-i}),\sigma_\tau(u^i)-\sigma_\tau(X_i\mu^{-i})\rangle+\langle \nabla_2V_X(\mu^i,\mu^{-i}),\sigma_\tau(u^{-i})-\sigma_\tau(X_{-i}\mu^i)\rangle\\
			\leq \;&\left(\frac{\tau}{\delta_i}+\frac{\tau}{\delta_{-i}}+\|X_i\|_2+\|X_{-i}\|_2\right)\bigg[\frac{2\Bar{c}}{\tau}V_X(\mu^i,\mu^{-i})+\frac{1}{\Bar{c}\tau^2}\|u^i-X_i\mu^{-i} \|_2^2\\
			&+\frac{1}{\Bar{c}\tau^2}\| u^{-i}-X_i\mu^i\|_2^2\bigg]
		\end{align*}
		where $\Bar{c}$ is any positive real number.
	\end{enumerate}
\end{lemma}

With the properties of $V_X(\cdot,\cdot)$ established above, we can now use it as a Lyapunov function to study $\pi_k^i$ and $\pi_k^{-i}$. Specifically, using the smoothness of $V_X(\cdot,\cdot)$, the update equation in Algorithm \ref{algorithm:inner-loop} Line $3$, and Lemma \ref{le:properties_Lyapunov} (3) and (4), we have the desired one-step Lyapunov drift inequality for $\sum_{s}V_{v,s}(\pi_k^i(s),\pi_k^{-i}(s))$, which is presented in the following.

\begin{lemma}[Proof in Appendix \ref{pf:le:policy_drift}]\label{le:policy_drift}
The following inequality holds for all $k\geq 0$:
\begin{align*}
	\sum_{s}\mathbb{E}[V_{v,s}(\pi_{k+1}^i(s),\pi_{k+1}^{-i}(s))]
	\leq \;&\left(1-\frac{3\beta_k}{4}\right)\sum_{s}\mathbb{E}[V_{v,s}(\pi_k^i(s),\pi_k^{-i}(s))]+\frac{4|\mathcal{S}|A_{\max}^2}{\ell_{\tau}(1-\gamma)^2}\beta_k^2\\
	&+\frac{256A_{\max}^2\beta_k}{\ell_{\tau}^2\tau^3(1-\gamma)^2}\sum_{i=1,2}\sum_{s}\mathbb{E}[\|q_k^i(s)-\mathcal{T}^i(v^i)(s)\pi_k^{-i}(s) \|_2^2]\\
	&+\frac{16|\mathcal{S}|A_{\max} \beta_k}{\tau}\|v^i+v^{-i}\|_\infty^2.
\end{align*}
\end{lemma}

\subsection{Analysis of the Inner-Loop: $q$-Function Update}\label{subsec:q-function}

In this section, we consider $q_k^i$ generated by Algorithm \ref{algorithm:inner-loop}. 
We begin by reformulating the update of the $q$-function as a stochastic approximation algorithm for estimating a time-varying target. Let $F^i:\mathbb{R}^{|\mathcal{S}| |\mathcal{A}^i|}\times \mathcal{S}\times \mathcal{A}^i\times \mathcal{A}^{-i}\times \mathcal{S}\mapsto \mathbb{R}^{|\mathcal{S}| |\mathcal{A}^i|}$ be an operator defined as
\begin{align*}
	[F^i(q^i,s_0,a_0^i,a_0^{-i},s_1)](s,a^i)=\mathds{1}_{\{(s,a^i)=(s_0,a_0^i)\}}\left(\mathcal{R}^i(s_0,a_0^i,a_0^{-i})+\gamma v^i(s_1)-q^i(s_0,a_0^i)\right)
\end{align*}
for all $(q^i,s_0,a_0^i,a_0^{-i},s_1)$ and $(s,a^i)$. Then Algorithm \ref{algorithm:inner-loop} Line $5$ can be compactly written as
\begin{align}\label{sa:reformulation}
	q_{k+1}^i=q_k^i+\alpha_k F^i(q_k^i,S_k,A_k^i,A_k^{-i},S_{k+1}).
\end{align}
Denote the stationary distribution of the Markov chain $\{S_k\}$ induced by the joint policy $\pi_k=(\pi_k^i,\pi_k^{-i})$ by $\mu_k\in\Delta^{|\mathcal{S}|}$, the existence and uniqueness of which is guaranteed by Lemma \ref{le:margin} and Lemma \ref{thm:exploration} (1).
Let $\Bar{F}_k^i:\mathbb{R}^{|\mathcal{S}| |\mathcal{A}^i|}\mapsto \mathbb{R}^{|\mathcal{S}| |\mathcal{A}^i|}$ be defined as
\begin{align*}
	\Bar{F}_k^i(q^i)=\mathbb{E}_{S_0\sim \mu_k(\cdot),A_0^i\sim \pi_k^i(\cdot|S_0), A_k^{-i}\sim \pi_k^{-i}(\cdot|S_0), S_1\sim p(\cdot|S_0,A_0^i,A_0^{-i})}\left[F^i(q^i,S_0,A_0^i,A_0^{-i},S_1)\right]
\end{align*}
for all $q^i\in\mathbb{R}^{|\mathcal{S}||\mathcal{A}^i|}$.
Then Eq. (\ref{sa:reformulation}) can be viewed as a stochastic approximation algorithm for solving the (time-varying) equation $\Bar{F}_k^i(q^i)=0$ with time-inhomogeneous Markovian noise $\{(S_k,A_k^i,A_k^{-i},S_{k+1})\}_{k\geq 0}$. We next establish the properties of the operators $F^i(\cdot)$ and $\Bar{F}_k^i(\cdot)$ in the following lemma.

\begin{lemma}[Proof in Appendix \ref{pf:le:operators}]\label{le:operators}
	The following inequalities hold:
	\begin{enumerate}[(1)]
		\item $\|F^i(q_1^i,s_0,a_0^i,a_0^{-i},s_1)-F^i(q_2^i,s_0,a_0^i,a_0^{-i},s_1)\|_2\leq \|q_1^i-q_2^i\|_2$ for any $(q_1^i,q_2^i)$ and $(s_0,a_0^i,a_0^{-i},s_1)$.
		\item $\|F^i(\bm{0},s_0,a_0^i,a_0^{-i},s_1)\|_2\leq \frac{1}{1-\gamma}$ for all $ (s_0,a_0^i,a_0^{-i},s_1)$.
		\item $\bar{F}_k^i(q^i)=0$ has a unique solution $\bar{q}_k^i$, which is explicitly given as $\bar{q}_k^i(s)=\mathcal{T}^i(v^i)(s)\pi_k^{-i}(s)$ for all $s$.
		\item $\langle \Bar{F}_k^i(q_1^i)-\Bar{F}_k^i(q_2^i),q_1^i-q_2^i\rangle\leq   -c_\tau\|q_1^i-q_2^i\|_2^2$ for all $(q_1^i,q_2^i)$.
	\end{enumerate}
\end{lemma}

Using $\|\cdot\|_2^2$ as a Lyapunov function and we have by the equivalent update equation (\ref{sa:reformulation}) that
\begin{align}
	&\mathbb{E}[\|q_{k+1}^i-\bar{q}_{k+1}^i\|_2^2]\nonumber\\
	=\;&\mathbb{E}[\|q_{k+1}^i-q_k^i+q_k^i-\bar{q}_k^i+\bar{q}_k^i-\bar{q}_{k+1}^i\|_2^2]\nonumber\\
	=\;&\mathbb{E}[\|q_k^i-\bar{q}_k^i\|_2^2]+\mathbb{E}[\|q_{k+1}^i-q_k^i\|_2^2]+\mathbb{E}[\|\bar{q}_k^i-\bar{q}_{k+1}^i\|_2^2]\nonumber\\
	&+\alpha_k\mathbb{E}[\langle F^i(q_k^i,S_k,A_k^i,A_k^{-i},S_{k+1}),q_k^i-\bar{q}_k^i\rangle]+\mathbb{E}[\langle q_{k+1}^i-q_k^i,\bar{q}_k^i-\bar{q}_{k+1}^i\rangle]\nonumber\\
	&+\mathbb{E}[\langle q_k^i-\bar{q}_k^i,\bar{q}_k^i-\bar{q}_{k+1}^i\rangle]\nonumber\\
	=\;&\mathbb{E}[\|q_k^i-\bar{q}_k^i\|_2^2]+\alpha_k\underbrace{\mathbb{E}[\langle \bar{F}_k^i(q_k^i),q_k^i-\bar{q}_k^i\rangle]}_{N_1}\nonumber\\
	&+\alpha_k\underbrace{\mathbb{E}[\langle F^i(q_k^i,S_k,A_k^i,A_k^{-i},S_{k+1})-\bar{F}_k^i(q_k^i),q_k^i-\bar{q}_k^i\rangle]}_{N_2}\nonumber\\
	&+\mathbb{E}[\|q_{k+1}^i-q_k^i\|_2^2]+\mathbb{E}[\|\bar{q}_k^i-\bar{q}_{k+1}^i\|_2^2]\nonumber\\
	&+\mathbb{E}[\langle q_{k+1}^i-q_k^i,\bar{q}_k^i-\bar{q}_{k+1}^i\rangle]+\mathbb{E}[\langle q_k^i-\bar{q}_k^i,\bar{q}_k^i-\bar{q}_{k+1}^i\rangle].\label{eq:all_terms}
\end{align}
What remains to do is to bound the terms on the RHS of the previous inequality. Among them, we want to highlight the two terms $N_1$ and $N_2$. For the term $N_1$, using Lemma \ref{le:operators} (4) and we have
\begin{align}\label{eq:q_drift}
	N_1=\mathbb{E}[\langle \bar{F}_k^i(q_k^i),q_k^i-\bar{q}_k^i\rangle]
	=\mathbb{E}[\langle \bar{F}_k^i(q_k^i)-\bar{F}_k^i(\bar{q}_k^i),q_k^i-\bar{q}_k^i\rangle]
	\leq -c_\tau\mathbb{E}[\|q_k^i-\bar{q}_k^i\|_2^2],
\end{align}
which provides us the desired negative drift.

The term $N_2$ involves the difference between the operator $F^i(q_k^i,S_k,A_k^i,A_k^{-i},S_{k+1})$ and its expected version $\bar{F}_k^i(q_k^i)$, and hence can be viewed as the stochastic error due to sampling. The fact that the Markov chain $\{(S_k,A_k^i,A_k^{-i},S_{k+1})\}$ is time-inhomogeneous presents major theoretical challenges in our analysis. To overcome this challenge, observe that: (1) the policy (hence the transition probability matrix of the induced Markov chain) is changing slowly compared to the $q$-function; see Algorithm \ref{algorithm:inner-loop} Line $3$, and (2) the stationary distribution as a function of the policy is Lipschitz (cf. Lemma \ref{thm:exploration} (3)). These two observations together enable us to develop a refined conditioning argument to handle the time-inhomogeneous Markovian noise. The result is presented in following. Similar ideas were previous used in \cite{bhandari2018finite,srikant2019finite,chen2021finite,zou2019finite,khodadadian2021finite} for finite-sample analysis of single-agent RL algorithms.

\begin{lemma}[Proof in Appendix \ref{pf:le:noise}]\label{le:noise}
	When $\alpha_{k-z_k,k-1}\leq 1/4$ for all $k\geq z_k$, we have for all $k\geq z_k$ that
	\begin{align*}
		N_2\leq \frac{340|\mathcal{S}|^{3/2}A_{\max}^{3/2}\hat{L}_\tau}{(1-\gamma)^2}z_k\alpha_{k-z_k,k-1}.
	\end{align*}
\end{lemma}

When using constant stepsize, we have $z_k\alpha_{k-z_k,k-1}=z_\beta^2 \alpha=\mathcal{O}(\alpha\log^2(1/\beta))$. Since the two stepsizes $\alpha$ and $\beta$ differ only by a multiplicative constant $c_{\alpha,\beta}$, we have $\lim_{\alpha\rightarrow 0}z_\beta^2 \alpha=0$. Similarly, we also have $\lim_{k\rightarrow \infty}z_k \alpha_{k-z_k,k-1}=0$ when using diminishing stepsizes. Therefore, Lemma \ref{le:noise} implies $N_2=o(1)$.

We next bound the rest of terms on the RHS of Eq. (\ref{eq:all_terms}) in the following lemma.

\begin{lemma}[Proof in Appendix \ref{pf:le:other_terms}]\label{le:other_terms}
	The following inequalities
	hold for all $k\geq 0$.
	\begin{enumerate}[(1)]
		\item $\mathbb{E}[\|q_{k+1}^i-q_k^i\|_2^2]\leq \frac{4|\mathcal{S}|A_{\max}\alpha_k^2}{(1-\gamma)^2}$.
		\item $\mathbb{E}[\|\bar{q}_k^i-\bar{q}_{k+1}^i\|_2^2]\leq \frac{4|\mathcal{S}|A_{\max}\beta_k^2}{(1-\gamma)^2}$.
		\item $\mathbb{E}[\langle q_{k+1}^i-q_k^i,\bar{q}_k^i-\bar{q}_{k+1}^i\rangle]\leq \frac{4|\mathcal{S}|A_{\max}\alpha_k\beta_k}{(1-\gamma)^2}$.
		\item $
		\mathbb{E}[\langle q_k^i-\bar{q}_k^i,\bar{q}_k^i-\bar{q}_{k+1}^i\rangle]
		\leq \frac{17A_{\max}^2\beta_k}{\tau(1-\gamma)^2}\mathbb{E}[\| q_k^i-\bar{q}_k^i\|_2^2]+\frac{\beta_k}{16}\sum_{s}\mathbb{E}[V_{v,s}(\pi_k^i(s),\pi_k^{-i}(s))]$.
	\end{enumerate}
\end{lemma}

Using the upper bounds we obtained for all the terms on the RHS of Eq. (\ref{eq:all_terms}) and we have the one-step Lyapunov drift inequality for $q_k^i$. Following the same line of analysis and we also obtain the one-step inequality for $q_k^{-i}$. Both results are presented in the following lemma.

\begin{lemma}[Proof in Appendix \ref{pf:le:q-function-drift}]\label{le:q-function-drift}
	The following inequality holds for all $k\geq z_k$ and $i\in \{1,2\}$:
	\begin{align*}
		\mathbb{E}[\|q_{k+1}^i-\bar{q}_{k+1}^i\|_2^2]\leq \;&\left(1-c_\tau\alpha_k+\frac{17A_{\max}^2\beta_k}{\tau(1-\gamma)^2}\right)\mathbb{E}[\|q_k^i-\bar{q}_k^i\|_2^2]\\
		&+\frac{352|\mathcal{S}|^{3/2}A_{\max}^{3/2}\hat{L}_\tau}{(1-\gamma)^2}z_k\alpha_k\alpha_{k-z_k,k-1}+\frac{\beta_k}{16}\sum_{s}\mathbb{E}[V_{v,s}(\pi_k^i(s),\pi_k^{-i}(s))].
	\end{align*}
\end{lemma}

\subsection{Solving Coupled Lyapunov Drift Inequalities}\label{sec:strategy}

We first restate the Lyapunov drift inequalities from previous sections. For simplicity of notation, we denote $\mathcal{L}_q(t,k)=\sum_{i=1,2}\|q_{t,k}^i-\bar{q}_{t,k}^i\|_2^2$, $\mathcal{L}_\pi(t,k)=\sum_{s}V_{v_t,s}(\pi_{t,k}^i(s),\pi_{t,k}^{-i}(s))$, and $\mathcal{F}_t$ as the history of Algorithm \ref{algorithm:tabular} right before the $t$-th outer-loop iteration. Note that $v_t^i$ and $v_t^{-i}$ are both measurable with respect to $\mathcal{F}_t$. In what follows, we denote $\mathbb{E}_t[\;\cdot\;]$ for $\mathbb{E}[\;\cdot\;\mid \mathcal{F}_t]$. 

\begin{itemize}
	\item \textbf{Lemma \ref{le:outer-loop}:} It holds for all $t\geq 0$ that 
	\begin{align}\label{eq:Lyapunov_v}
		\|v_{t+1}^i-v_*^i\|_\infty
		\leq \;&\gamma  \|v_t^i-v_*^i\|_\infty+2\|v_t^i+v_t^{-i}\|_\infty+4\tau \log(A_{\max})\nonumber\\
		&+2\mathcal{L}_\pi(t,K)+\sum_{i=1,2}\|q_{t,K}^i-\bar{q}_{t,K}^i\|_2.
	\end{align}
	\item \textbf{Lemma \ref{le:outer-sum}:} It holds for all $t\geq 0$ that
	\begin{align}\label{eq:Lyapunov_v+}
		\|v_{t+1}^i+v_{t+1}^{-i}\|_\infty\leq\;& \gamma\|v_t^i+v_t^{-i}\|_\infty+\sum_{i=1,2}\|q_{t,K}^i-\bar{q}_{t,K}^i\|_2.
	\end{align}
	\item \textbf{Lemma \ref{le:policy_drift}: } It holds for all $t,k\geq 0$ that
	\begin{align}\label{eq:Lyapunov_pi}
		\mathbb{E}_t[\mathcal{L}_\pi(t,k+1)]
		\leq &\left(1-\frac{3\beta_k}{4}\right)\mathbb{E}_t[\mathcal{L}_\pi(t,k)]+\frac{256A_{\max}^2\beta_k}{\ell_{\tau}^2\tau^3(1-\gamma)^2}\mathbb{E}_t[\mathcal{L}_q(t,k)]\nonumber\\
		&+\frac{16|\mathcal{S}|A_{\max} \beta_k}{\tau}\|v_t^i+v_t^{-i}\|_\infty^2+\frac{4|\mathcal{S}|A_{\max}^2\beta_k^2}{\ell_{\tau}(1-\gamma)^2}.
	\end{align}
	\item \textbf{Lemma \ref{le:q-function-drift}:} It holds for all $t\geq 0$ and $k\geq z_k$ that
	\begin{align}\label{eq:Lyapunov_q}
		\mathbb{E}_t[\mathcal{L}_q(t,k+1)]\leq \;&\left(1-c_\tau\alpha_k+\frac{17A_{\max}^2\beta_k}{\tau(1-\gamma)^2}\right)\mathbb{E}_t[\mathcal{L}_q(t,k)]\nonumber\\
		&+\frac{\beta_k}{16}\mathbb{E}_t[\mathcal{L}_\pi(t,k)]+\frac{352|\mathcal{S}|^{3/2}A_{\max}^{3/2}\hat{L}_\tau}{(1-\gamma)^2}z_k\alpha_k\alpha_{k-z_k,k-1}.
	\end{align}
\end{itemize}
Adding up Eqs. (\ref{eq:Lyapunov_pi}) and (\ref{eq:Lyapunov_q}) and we have
\begin{align*}
	&\mathbb{E}_t[\mathcal{L}_\pi(t,k+1)+\mathcal{L}_q(t,k+1)]\\
	\leq \;&\left(1-\frac{\beta_k}{2}\right)\mathbb{E}_t[\mathcal{L}_\pi(t,k)]+\left(1-c_\tau\alpha_k+\frac{256A_{\max}^2\beta_k}{\ell_{\tau}^2\tau^3(1-\gamma)^2}\right)\mathbb{E}_t[\mathcal{L}_q(t,k)]\nonumber\\
	&\frac{16|\mathcal{S}|A_{\max} \beta_k}{\tau}\|v_t^i+v_t^{-i}\|_\infty^2+\frac{4|\mathcal{S}|A_{\max}^2\beta_k^2}{\ell_{\tau}(1-\gamma)^2}+\frac{352|\mathcal{S}|^{3/2}A_{\max}^{3/2}\hat{L}_\tau}{(1-\gamma)^2}z_k\alpha_k\alpha_{k-z_k,k-1}\\
	= \;&\left(1-\frac{c_{\alpha,\beta}\alpha_k}{2}\right)\mathbb{E}_t[\mathcal{L}_\pi(t,k)]+\left(1-c_\tau\alpha_k+\frac{256A_{\max}^2c_{\alpha,\beta}\alpha_k}{\ell_{\tau}^2\tau^3(1-\gamma)^2}\right)\mathbb{E}_t[\mathcal{L}_q(t,k)]\nonumber\\
	&\frac{16|\mathcal{S}|A_{\max} \beta_k}{\tau}\|v_t^i+v_t^{-i}\|_\infty^2+\frac{4|\mathcal{S}|A_{\max}^2\beta_k^2}{\ell_{\tau}(1-\gamma)^2}+\frac{352|\mathcal{S}|^{3/2}A_{\max}^{3/2}\hat{L}_\tau}{(1-\gamma)^2}z_k\alpha_k\alpha_{k-z_k,k-1}.
\end{align*}
Note that Condition \ref{con:stepsize} implies that
\begin{align*}
    \frac{256A_{\max}^2c_{\alpha,\beta}\alpha_k}{\ell_{\tau}^2\tau^3(1-\gamma)^2}\leq \frac{c_\tau}{2}.
\end{align*}
Therefore, we have
\begin{align}\label{eq:before_stepsize}
	&\mathbb{E}_t[\mathcal{L}_\pi(t,k+1)+\mathcal{L}_q(t,k+1)]\nonumber\\
	\leq  \;&\left(1-\frac{c_{\alpha,\beta}\alpha_k}{2}\right)\mathbb{E}_t[\mathcal{L}_\pi(t,k)+\mathcal{L}_q(t,k)]+\frac{16|\mathcal{S}|A_{\max} c_{\alpha,\beta}\alpha_k}{\tau}\|v_t^i+v_t^{-i}\|_\infty^2\nonumber\\
	&+\frac{4|\mathcal{S}|A_{\max}^2c_{\alpha,\beta}^2\alpha_k^2}{\ell_{\tau}(1-\gamma)^2}+\frac{352|\mathcal{S}|^{3/2}A_{\max}^{3/2}\hat{L}_\tau}{(1-\gamma)^2}z_k\alpha_k\alpha_{k-z_k,k-1}.
\end{align}

\subsubsection{Constant Stepsize}
When using constant stepsizes, i.e., $\alpha_k\equiv \alpha$, $\beta_k\equiv \beta$, and $\beta=c_{\alpha,\beta}\alpha$, repeatedly using Eq. (\ref{eq:before_stepsize}) from $z_\beta$ to $k$ and we have
\begin{align}
	&\mathbb{E}_t[\mathcal{L}_\pi(t,k)+\mathcal{L}_q(t,k)]\nonumber\\
	\leq  \;&\left(1-\frac{c_{\alpha,\beta}\alpha}{2}\right)^{k-z_\beta}(\mathcal{L}_\pi(t,0)+\mathcal{L}_q(t,0))\nonumber\\
	&+\frac{32|\mathcal{S}|A_{\max}}{\tau}\|v_t^i+v_t^{-i}\|_\infty^2+\frac{8|\mathcal{S}|A_{\max}^2c_{\alpha,\beta}\alpha}{\ell_{\tau}(1-\gamma)^2}+\frac{704|\mathcal{S}|^{3/2}A_{\max}^{3/2}\hat{L}_\tau}{(1-\gamma)^2c_{\alpha,\beta}}z_\beta^2\alpha.\label{eq:polish1}
\end{align}
We next bound $\mathcal{L}_\pi(t,0)+\mathcal{L}_q(t,0)$.
For $i\in \{1,2\}$, since $\pi_{t,0}^i$ is initialized at a uniformly random policy and $q_{t,0}^i=\bm{0}$, we have
\begin{align*}
    \mathcal{L}_\pi(t,0)=\;&\sum_{s}V_{v_t,s}(\pi_{t,0}^i(s),\pi_{t,0}^{-i}(s))\\
    =\;&\sum_{s}\sum_{i=1,2}\max_{\mu^i}\{(\mu^i-\pi_{t,0}^i(s))^\top \mathcal{T}^i(v_t^i)(s)\pi_{t,0}^{-i}(s)+\tau \nu(\mu^i)-\tau\nu(\pi_{t,0}^i(s))\}\\
    \leq \;&2\sum_{s}\sum_{i=1,2}\max_{s,a^i,a^{-i}}|\mathcal{T}^i(v_t^i)(s,a^i,a^{-i})|\\
    \leq\;& \frac{4|\mathcal{S}|}{(1-\gamma)},
\end{align*}
and
\begin{align*}
    \mathcal{L}_q(t,0)=\sum_{i=1,2}\|\bar{q}_{t,0}^i\|_2^2\leq \frac{2|\mathcal{S}|A_{\max}}{(1-\gamma)^2}.
\end{align*}
Using the previous two bounds in Eq. (\ref{eq:polish1}) and we have
\begin{align}\label{eq:cici}
	&\mathbb{E}_t[\mathcal{L}_\pi(t,k)+\mathcal{L}_q(t,k)]\nonumber\\
	\leq  \;&\frac{4|\mathcal{S}|A_{\max}}{(1-\gamma)^2}\left(1-\frac{c_{\alpha,\beta}\alpha}{2}\right)^{k-z_\beta}+\frac{32|\mathcal{S}|A_{\max}}{\tau}\|v_t^i+v_t^{-i}\|_\infty^2\nonumber\\
	&+\frac{8|\mathcal{S}|A_{\max}^2c_{\alpha,\beta}\alpha}{\ell_{\tau}(1-\gamma)^2}+\frac{704|\mathcal{S}|^{3/2}A_{\max}^{3/2}\hat{L}_\tau}{(1-\gamma)^2c_{\alpha,\beta}}z_\beta^2\alpha,
\end{align}
which implies
\begin{align*}
	\mathbb{E}_t[\mathcal{L}_\pi(t,k)]
	\leq  \;&\frac{4|\mathcal{S}|A_{\max}}{(1-\gamma)^2}\left(1-\frac{c_{\alpha,\beta}\alpha}{2}\right)^{k-z_\beta}+\frac{32|\mathcal{S}|A_{\max}}{\tau}\|v_t^i+v_t^{-i}\|_\infty^2\\
	&+\frac{8|\mathcal{S}|A_{\max}^2c_{\alpha,\beta}\alpha}{\ell_{\tau}(1-\gamma)^2}+\frac{704|\mathcal{S}|^{3/2}A_{\max}^{3/2}\hat{L}_\tau}{(1-\gamma)^2c_{\alpha,\beta}}z_\beta^2\alpha\\
	\leq  \;&\frac{4|\mathcal{S}|A_{\max}}{(1-\gamma)^2}\left(1-\frac{c_{\alpha,\beta}\alpha}{2}\right)^{k-z_\beta}+\frac{32|\mathcal{S}|A_{\max}}{\tau}\|v_t^i+v_t^{-i}\|_\infty^2\\
	&+\frac{712|\mathcal{S}|^{3/2}A_{\max}^{3/2}\hat{L}_\tau}{(1-\gamma)^2c_{\alpha,\beta}}z_\beta^2\alpha.
\end{align*}
Substituting the previous inequality on $\mathbb{E}_t[\mathcal{L}_\pi(t,k)]$ into Eq. (\ref{eq:Lyapunov_q}) and we have
\begin{align*}
	\mathbb{E}_t[\mathcal{L}_q(t,k+1)]\leq \;&\left(1-c_\tau\alpha+\frac{17A_{\max}^2\beta}{\tau(1-\gamma)^2}\right)\mathbb{E}_t[\mathcal{L}_q(t,k)]+\frac{352|\mathcal{S}|^{3/2}A_{\max}^{3/2}}{(1-\gamma)^2}z_\beta^2\alpha^2\\
	&+\frac{c_{\alpha,\beta}\alpha}{16}\left(\frac{4|\mathcal{S}|A_{\max}}{(1-\gamma)^2}\left(1-\frac{c_{\alpha,\beta}\alpha}{2}\right)^{k-z_\beta}+\frac{32|\mathcal{S}|A_{\max}}{\tau}\|v_t^i+v_t^{-i}\|_\infty^2\right.\\
	&+\left.\frac{712|\mathcal{S}|^{3/2}A_{\max}^{3/2}\hat{L}_\tau}{(1-\gamma)^2c_{\alpha,\beta}}z_\beta^2\alpha\right)\\
	\leq  \;&\left(1-\frac{c_\tau\alpha}{2}\right)\mathbb{E}_t[\mathcal{L}_q(t,k)]+\frac{|\mathcal{S}|A_{\max}c_{\alpha,\beta}\alpha}{4(1-\gamma)^2}\left(1-\frac{c_{\alpha,\beta}\alpha}{2}\right)^{k-z_\beta}\\
	&+\frac{2|\mathcal{S}|A_{\max}c_{\alpha,\beta}\alpha}{\tau}\|v_t^i+v_t^{-i}\|_\infty^2+\frac{45|\mathcal{S}|^{3/2}A_{\max}^{3/2}\hat{L}_\tau}{(1-\gamma)^2}z_\beta^2\alpha^2,
\end{align*}
where the last line follows from Condition \ref{con:stepsize}. Repeatedly using the previous inequality from $z_\beta$ to $k$ and we have
\begin{align*}
	\mathbb{E}_t[\mathcal{L}_q(t,k)]
	\leq  \;&\frac{2|\mathcal{S}|A_{\max}}{(1-\gamma)^2}\left(1-\frac{c_\tau\alpha}{2}\right)^{k-z_\beta}+\frac{|\mathcal{S}|A_{\max}c_{\alpha,\beta}\alpha(k-z_\beta)}{4(1-\gamma)^2}\left(1-\frac{c_{\alpha,\beta}\alpha}{2}\right)^{k-z_\beta-1}\\
	&+\frac{4|\mathcal{S}|A_{\max}c_{\alpha,\beta}}{c_\tau\tau}\|v_t^i+v_t^{-i}\|_\infty^2+\frac{90|\mathcal{S}|^{3/2}A_{\max}^{3/2}\hat{L}_\tau}{c_\tau(1-\gamma)^2}z_\beta^2\alpha.
\end{align*}
The next step is to substitute the previous bound on $\mathbb{E}_t[\mathcal{L}_q(t,k)]$ into Eq. (\ref{eq:Lyapunov_v+}). To achieve that, first note that
\begin{align*}
    \sum_{i=1,2}\mathbb{E}_t\left[\|q_{t,K}^i-\bar{q}_{t,K}^i\|_2\right]\leq \;&\sum_{i=1,2}\left(\mathbb{E}_t\left[\|q_{t,K}^i-\bar{q}_{t,K}^i\|_2^2\right]\right)^{1/2}\tag{Jensen's inequality}\\
    \leq \;&2\left(\sum_{i=1,2}\mathbb{E}_t\left[\|q_{t,K}^i-\bar{q}_{t,K}^i\|_2^2\right]\right)^{1/2}\tag{$\sqrt{a}+\sqrt{b}\leq 2\sqrt{a+b}$}\\
    \leq \;&2\mathbb{E}_t^{1/2}[\mathcal{L}_q(t,K)].
\end{align*}
Therefore, we have
\begin{align}\label{eq:cicipp}
	&\sum_{i=1,2}\mathbb{E}_t\left[\|q_{t,K}^i-\bar{q}_{t,K}^i\|_2\right]\nonumber\\
	\leq \;&2\mathbb{E}_t[\mathcal{L}_q(t,K)]^{1/2}\nonumber\\
	\leq \;&\frac{3\sqrt{|\mathcal{S}|A_{\max}}}{(1-\gamma)}\left(1-\frac{c_\tau\alpha}{2}\right)^{\frac{K-z_\beta}{2}}+\frac{\sqrt{|\mathcal{S}|A_{\max}}c^{1/2}_{\alpha,\beta}\alpha^{1/2}(K-z_\beta)^{1/2}}{(1-\gamma)}\left(1-\frac{c_{\alpha,\beta}\alpha}{2}\right)^{\frac{K-z_\beta-1}{2}}\nonumber\\
	&+\frac{4\sqrt{|\mathcal{S}|A_{\max}}c^{1/2}_{\alpha,\beta}}{c_{\tau}^{1/2}\tau^{1/2}}\|v_t^i+v_t^{-i}\|_\infty+\frac{20|\mathcal{S}|^{3/4}A_{\max}^{3/4}\hat{L}_\tau^{1/2}}{c_{\tau}^{1/2}(1-\gamma)}z_\beta\alpha^{1/2}.
\end{align}
Taking the total expectation on both sides of the previous inequality then using the result in Eq. (\ref{eq:Lyapunov_v+}), and we obtain
\begin{align*}
	\mathbb{E}[\|v_{t+1}^i+v_{t+1}^{-i}\|_\infty]\leq\;& \left(\gamma+\frac{4\sqrt{|\mathcal{S}|A_{\max}}c^{1/2}_{\alpha,\beta}}{c_{\tau}^{1/2}\tau^{1/2}}\right)\mathbb{E}[\|v_t^i+v_t^{-i}\|_\infty]\\
	&+\frac{4\sqrt{|\mathcal{S}|A_{\max}}(K-z_\beta)^{1/2}}{(1-\gamma)}\left(1-\frac{c_{\alpha,\beta}\alpha}{2}\right)^{\frac{K-z_\beta-1}{2}}\\
	&+\frac{20|\mathcal{S}|^{3/4}A_{\max}^{3/4}\hat{L}_\tau^{1/2}}{c_{\tau}^{1/2}(1-\gamma)}z_\beta\alpha^{1/2}\\
	\leq\;& \left(\frac{1+\gamma}{2}\right)\mathbb{E}[\|v_t^i+v_t^{-i}\|_\infty]\\
	&+\frac{4\sqrt{|\mathcal{S}|A_{\max}}(K-z_\beta)^{1/2}}{(1-\gamma)}\left(1-\frac{c_{\alpha,\beta}\alpha}{2}\right)^{\frac{K-z_\beta-1}{2}}\\
	&+\frac{20|\mathcal{S}|^{3/4}A_{\max}^{3/4}\hat{L}_\tau^{1/2}}{c_{\tau}^{1/2}(1-\gamma)}z_\beta\alpha^{1/2},
\end{align*}
where the last line follows from Condition \ref{con:stepsize}. Since $\|v_0^i+v_0^{-i}\|_\infty\leq \frac{2}{1-\gamma}$, repeatedly using the previous inequality starting from $0$ and we have for all $t\geq 0$ that
\begin{align}\label{eq:cipa}
	\mathbb{E}[\|v_t^i+v_t^{-i}\|_\infty]\leq\;& \frac{2}{1-\gamma}\left(\frac{1+\gamma}{2}\right)^t+\frac{8\sqrt{|\mathcal{S}|A_{\max}}(K-z_\beta)^{1/2}}{(1-\gamma)^2}\left(1-\frac{c_{\alpha,\beta}\alpha}{2}\right)^{\frac{K-z_\beta-1}{2}}\nonumber\\
	&+\frac{40|\mathcal{S}|^{3/4}A_{\max}^{3/4}\hat{L}_\tau^{1/2}}{c_{\tau}^{1/2}(1-\gamma)^2}z_\beta\alpha^{1/2}.
\end{align}
Now we have obtained finite-sample bounds for $\mathcal{L}_q(t,k)$, $\mathcal{L}_\pi(t,k)$, and $\|v_t^i+v_t^{-i}\|_\infty$. The next step is to use them in Eq. (\ref{eq:Lyapunov_v}) to obtain finite-sample bounds for $\|v_t^i-v_*^i\|_\infty$. Specifically, we have by Eq. (\ref{eq:Lyapunov_v}), Eq. (\ref{eq:cici}), and Eq. (\ref{eq:cicipp}) that
\begin{align*}
	\mathbb{E}[\|v_{t+1}^i-v_*^i\|_\infty]
	\leq \;&\gamma  \mathbb{E}[\|v_t^i-v_*^i\|_\infty]+2\mathbb{E}[\|v_t^i+v_t^{-i}\|_\infty]+4\tau \log(A_{\max})\\
	&+2\mathbb{E}[\mathcal{L}_\pi(t,K)]+\sum_{i=1,2}\mathbb{E}\|q_{t,K}^i-\bar{q}_{t,K}^i\|_2]\\
	\leq \;&\gamma  \mathbb{E}[\|v_t^i-v_*^i\|_\infty]+2\mathbb{E}[\|v_t^i+v_t^{-i}\|_\infty]+4\tau \log(A_{\max})\\
	&+2\mathbb{E}[\mathcal{L}_\pi(t,K)]+2\mathbb{E}_t[\mathcal{L}_q(t,K)]^{1/2}\\
	\leq \;&\gamma  \mathbb{E}[\|v_t^i-v_*^i\|_\infty]+2\mathbb{E}[\|v_t^i+v_t^{-i}\|_\infty]+4\tau \log(A_{\max})\\
	&+\frac{8|\mathcal{S}|A_{\max}}{(1-\gamma)^2}\left(1-\frac{c_{\alpha,\beta}\alpha}{2}\right)^{K-z_\beta}+\frac{64|\mathcal{S}|A_{\max}}{\tau}\mathbb{E}[\|v_t^i+v_t^{-i}\|_\infty^2]\\
    &+\frac{1424|\mathcal{S}|^{3/2}A_{\max}^{3/2}\hat{L}_\tau}{(1-\gamma)^2c_{\alpha,\beta}}z_\beta^2\alpha+\frac{3\sqrt{|\mathcal{S}|A_{\max}}}{(1-\gamma)}\left(1-\frac{c_\tau\alpha}{2}\right)^{\frac{K-z_\beta}{2}}\\
    &+\frac{\sqrt{|\mathcal{S}|A_{\max}}c^{1/2}_{\alpha,\beta}\alpha^{1/2}(K-z_\beta)^{1/2}}{(1-\gamma)}\left(1-\frac{c_{\alpha,\beta}\alpha}{2}\right)^{\frac{K-z_\beta-1}{2}}\\
    &+\frac{4\sqrt{|\mathcal{S}|A_{\max}}c^{1/2}_{\alpha,\beta}}{c_{\tau}^{1/2}\tau^{1/2}}\mathbb{E}[\|v_t^i+v_t^{-i}\|_\infty]+\frac{20|\mathcal{S}|^{3/4}A_{\max}^{3/4}\hat{L}_\tau^{1/2}}{c_{\tau}^{1/2}(1-\gamma)}z_\beta\alpha^{1/2}\\
    \leq \;&\gamma  \mathbb{E}[\|v_t^i-v_*^i\|_\infty]+\frac{134|\mathcal{S}|A_{\max}}{\tau(1-\gamma)}\mathbb{E}[\|v_t^i+v_t^{-i}\|_\infty]+4\tau \log(A_{\max})\\
	&+\frac{12|\mathcal{S}|A_{\max}(K-z_\beta)^{1/2}}{(1-\gamma)^2}\left(1-\frac{c_{\alpha,\beta}\alpha}{2}\right)^{\frac{K-z_\beta-1}{2}}\\
    &+\frac{1444|\mathcal{S}|^{3/2}A_{\max}^{3/2}\hat{L}_\tau}{(1-\gamma)^2c_{\alpha,\beta}}z_\beta^2\alpha^{1/2}\\
    \leq \;&\gamma  \mathbb{E}[\|v_t^i-v_*^i\|_\infty]+4\tau \log(A_{\max})+\frac{1444|\mathcal{S}|^{3/2}A_{\max}^{3/2}\hat{L}_\tau}{(1-\gamma)^2c_{\alpha,\beta}}z_\beta^2\alpha^{1/2}\\
	&+\frac{12|\mathcal{S}|A_{\max}(K-z_\beta)^{1/2}}{(1-\gamma)^2}\left(1-\frac{c_{\alpha,\beta}\alpha}{2}\right)^{\frac{K-z_\beta-1}{2}}\\
    &+\frac{134|\mathcal{S}|A_{\max}}{\tau(1-\gamma)}\left(\frac{2}{1-\gamma}\left(\frac{1+\gamma}{2}\right)^t+\frac{40|\mathcal{S}|^{3/4}A_{\max}^{3/4}\hat{L}_\tau^{1/2}}{c_{\tau}^{1/2}(1-\gamma)^2}z_\beta\alpha^{1/2}\right.\\
    &\left.+\frac{8\sqrt{|\mathcal{S}|A_{\max}}(K-z_\beta)^{1/2}}{(1-\gamma)^2}\left(1-\frac{c_{\alpha,\beta}\alpha}{2}\right)^{\frac{K-z_\beta-1}{2}}\right)\\
    \leq \;&\gamma  \mathbb{E}[\|v_t^i-v_*^i\|_\infty]+4\tau \log(A_{\max})+\frac{268|\mathcal{S}|A_{\max}}{\tau(1-\gamma)^2}\left(\frac{1+\gamma}{2}\right)^t\\
    &+\frac{1084|\mathcal{S}|^{3/2}A_{\max}^{3/2}(K-z_\beta)^{1/2}}{\tau (1-\gamma)^3}\left(1-\frac{c_{\alpha,\beta}\alpha}{2}\right)^{\frac{K-z_\beta-1}{2}}\\
    &+\frac{6804|\mathcal{S}|^2A_{\max}^2\hat{L}_\tau}{c_{\alpha,\beta}(1-\gamma)^3}z_\beta^2\alpha^{1/2}.
\end{align*}
Repeatedly using the previous inequality from $0$ to $T-1$ and we have
\begin{align*}
    \mathbb{E}[\|v_T^i-v_*^i\|_\infty]\leq\;& \frac{4\tau \log(A_{\max})}{1-\gamma}+\frac{270|\mathcal{S}|A_{\max}T}{\tau(1-\gamma)^2}\left(\frac{1+\gamma}{2}\right)^{T-1}\\
    &+\frac{1084|\mathcal{S}|^{3/2}A_{\max}^{3/2}(K-z_\beta)^{1/2}}{\tau (1-\gamma)^4}\left(1-\frac{c_{\alpha,\beta}\alpha}{2}\right)^{\frac{K-z_\beta-1}{2}}\\
    &+\frac{6804|\mathcal{S}|^2A_{\max}^2\hat{L}_\tau}{c_{\alpha,\beta}(1-\gamma)^4}z_\beta^2\alpha^{1/2},
\end{align*}
where we used $\|v_0^i-v_*^i\|_\infty\leq 2/(1-\gamma)$.

Our next step is to use the bounds we obtained for $\mathcal{L}_q(t,k)$, $\mathcal{L}_\pi(t,k)$, $\|v_t^i+v_t^{-i}\|_\infty$, and $\|v_t^i-v_*^i\|_\infty$ in Lemma \ref{le:Nash_Gap}. For simplicity, we use $a\lesssim b$ to mean that there exists a \textit{numerical} constant $c$ such that $a\leq  cb$. Now, we have by the previous inequality, Eq. (\ref{eq:cici}), and Eq. (\ref{eq:cipa}) that
\begin{align*}
    &\mathbb{E}\left[\left\|v^i_{*,\pi_{T,K}^{-i}}-v^i_{\pi_{T,K}^i,\pi_{T,K}^{-i}}\right\|_\infty\right]\\
    \leq \;&\frac{2}{1-\gamma}\left(2\mathbb{E}[\|v_T^i+v_T^{-i}\|_\infty]+2\mathbb{E}[\|v^i_T-v^i_{*}\|_\infty]+\mathcal{L}_\pi(T,K)+2\tau \log(A_{\max})\right)\\
	\lesssim \;&\frac{\tau \log(A_{\max})}{(1-\gamma)^2}+\frac{|\mathcal{S}|A_{\max}T}{\tau(1-\gamma)^3}\left(\frac{1+\gamma}{2}\right)^{T-1}\\
    &+\frac{|\mathcal{S}|^{3/2}A_{\max}^{3/2}(K-z_\beta)^{1/2}}{\tau (1-\gamma)^5}\left(1-\frac{c_{\alpha,\beta}\alpha}{2}\right)^{\frac{K-z_\beta-1}{2}}+\frac{|\mathcal{S}|^2A_{\max}^2\hat{L}_\tau}{c_{\alpha,\beta}(1-\gamma)^5}z_\beta^2\alpha^{1/2},
\end{align*}
Finally, using the previous inequality in Lemma \ref{le:Nash_to_v} and we have
\begin{align*}
	\mathbb{E}[\textit{NG}(\pi_{T,K}^i,\pi_{T,K}^{-i})]\lesssim \;&\frac{\tau \log(A_{\max})}{(1-\gamma)^2}+\frac{|\mathcal{S}|A_{\max}T}{\tau(1-\gamma)^3}\left(\frac{1+\gamma}{2}\right)^{T-1}\\
    &+\frac{|\mathcal{S}|^{3/2}A_{\max}^{3/2}(K-z_\beta)^{1/2}}{\tau (1-\gamma)^5}\left(1-\frac{c_{\alpha,\beta}\alpha}{2}\right)^{\frac{K-z_\beta-1}{2}}\\
    &+\frac{|\mathcal{S}|^2A_{\max}^2\hat{L}_\tau}{c_{\alpha,\beta}(1-\gamma)^5}z_\beta^2\alpha^{1/2}.
\end{align*}
The proof of Theorem \ref{thm:tabular} is complete.

\subsubsection{Diminishing Stepsizes}

Consider using linearly diminishing stepsizes, i.e., $\alpha_k=\frac{\alpha}{k+h}$, $\beta_k=\frac{\beta}{k+h}$, and $\beta=c_{\alpha,\beta}\alpha$. Repeatedly using Eq. (\ref{eq:before_stepsize}) and we have for all $k\geq k_0$ that
\begin{align*}
	\mathbb{E}_t[\mathcal{L}_\pi(t,k)+\mathcal{L}_q(t,k)]
	\lesssim  \;&\frac{4|\mathcal{S}|A_{\max}}{(1-\gamma)^2}\underbrace{\prod_{m=k_0}^{k-1}\left(1-\frac{c_{\alpha,\beta}\alpha_m}{2}\right)}_{\hat{\mathcal{E}}_1}\nonumber\\
	&+\frac{|\mathcal{S}|^{3/2}A_{\max}^2\hat{L}_\tau}{(1-\gamma)^2}\underbrace{\sum_{n=k_0}^{k-1}z_n^2\alpha_n^2\prod_{m=n+1}^{k-1}\left(1-\frac{c_{\alpha,\beta}\alpha_m}{2}\right)}_{\hat{\mathcal{E}}_2}\\
	&+\frac{|\mathcal{S}|A_{\max} c_{\alpha,\beta}}{\tau}\|v_t^i+v_t^{-i}\|_\infty^2\underbrace{\sum_{n=k_0}^{k-1}\alpha_n\prod_{m=n+1}^{k-1}\left(1-\frac{c_{\alpha,\beta}\alpha_m}{2}\right)}_{\hat{\mathcal{E}}_3}.
\end{align*}
We next provide estimates for the terms $\{\hat{\mathcal{E}}_j\}_{1\leq j\leq 3}$. Bounds of terms like $\{\hat{\mathcal{E}}_j\}_{1\leq j\leq 3}$ are well-established in existing work studying the convergence rate of iterative algorithms \citep{srikant2019finite,lan2020first,chen2021finite}. Specifically, we have from \cite[Appendix A.2.]{chen2021finite} that
\begin{align*}
	\hat{\mathcal{E}}_1\leq
	\left(\frac{k_0+h}{k+h}\right)^{c_{\alpha,\beta}\alpha/2},\quad 
	\hat{\mathcal{E}}_2\leq
	\frac{4ez_k^2\alpha^2}{c_{\alpha,\beta}\alpha/2-1}\frac{1}{k+h},\;\text{ and }\;
	\hat{\mathcal{E}}_3\leq \frac{2}{c_{\alpha,\beta}}.
\end{align*}
It follows that 
\begin{align}\label{eq:ci_di}
	&\mathbb{E}_t[\mathcal{L}_\pi(t,k)+\mathcal{L}_q(t,k)]\nonumber\\
	\lesssim  \;&\frac{|\mathcal{S}|A_{\max}}{(1-\gamma)^2}\left(\frac{k_0+h}{k+h}\right)^{c_{\alpha,\beta}\alpha/2}+\frac{|\mathcal{S}|^{3/2}A_{\max}^2\hat{L}_\tau}{(1-\gamma)^2}\frac{z_k^2\alpha^2}{c_{\alpha,\beta}\alpha/2-1}\frac{1}{k+h}\nonumber\\
	&+\frac{|\mathcal{S}|A_{\max} }{\tau}\|v_t^i+v_t^{-i}\|_\infty^2\nonumber\\
	\lesssim  \;&\frac{|\mathcal{S}|A_{\max}}{(1-\gamma)^2}\left(\frac{\alpha_k}{\alpha_{k_0}}\right)^{c_{\alpha,\beta}\alpha/2}+\frac{|\mathcal{S}|^{3/2}A_{\max}^2\hat{L}_\tau}{(1-\gamma)^2}\frac{z_k^2\alpha^2}{c_{\alpha,\beta}\alpha/2-1}\frac{1}{k+h}\nonumber\\
	&+\frac{|\mathcal{S}|A_{\max} }{\tau}\|v_t^i+v_t^{-i}\|_\infty^2,
\end{align}
which implies
\begin{align*}
	\mathbb{E}_t[\mathcal{L}_\pi(t,k)]
	\lesssim  \;&\frac{|\mathcal{S}|A_{\max}}{(1-\gamma)^2}\left(\frac{\alpha_k}{\alpha_{k_0}}\right)^{c_{\alpha,\beta}\alpha/2}+\frac{|\mathcal{S}|^{3/2}A_{\max}^2\hat{L}_\tau}{(1-\gamma)^2}\frac{z_k^2\alpha^2}{c_{\alpha,\beta}\alpha/2-1}\frac{1}{k+h}\\
	&+\frac{|\mathcal{S}|A_{\max} }{\tau}\|v_t^i+v_t^{-i}\|_\infty^2.
\end{align*}
Using the previous bound on $\mathbb{E}_t[\mathcal{L}_\pi(t,k)]$ in Eq. (\ref{eq:Lyapunov_q}) and we have
\begin{align*}
	\mathbb{E}_t[\mathcal{L}_q(t,k+1)]\leq \;&\left(1-c_\tau\alpha_k+\frac{17A_{\max}^2\beta_k}{\tau(1-\gamma)^2}\right)\mathbb{E}_t[\mathcal{L}_q(t,k)]\\
	&+\frac{\beta_k}{16}\mathbb{E}_t[\mathcal{L}_\pi(t,k)]+\frac{352|\mathcal{S}|^{3/2}A_{\max}^{3/2}\hat{L}_\tau}{(1-\gamma)^2}z_k\alpha_k\alpha_{k-z_k,k-1}\\
	\lesssim \;&\left(1-\frac{c_\tau\alpha_k}{2}\right)\mathbb{E}_t[\mathcal{L}_q(t,k)]+\frac{|\mathcal{S}|^{3/2}A_{\max}^2}{\alpha_{k_0}(1-\gamma)^2}z_k^2\alpha_k^2\\
	&+\frac{|\mathcal{S}|A_{\max}c_{\alpha,\beta}\alpha_k }{\tau}\|v_t^i+v_t^{-i}\|_\infty^2.
\end{align*}
Repeatedly using the previous inequality starting from $k_0$ and we have
\begin{align*}
	\mathbb{E}_t[\mathcal{L}_q(t,k)]\lesssim\;&\frac{|\mathcal{S}|A_{\max}}{(1-\gamma)^2}\left(\frac{\alpha_k}{\alpha_{k_0}}\right)^{c_\tau\alpha/2}+\frac{|\mathcal{S}|^{3/2}A_{\max}^2}{\alpha_{k_0}(1-\gamma)^2}z_k^2\alpha_k\\
	&+\frac{|\mathcal{S}|A_{\max}c_{\alpha,\beta} }{c_\tau\tau}\|v_t^i+v_t^{-i}\|_\infty^2\\
	\lesssim\;&\frac{|\mathcal{S}|^{3/2}A_{\max}^2}{\alpha_{k_0}(1-\gamma)^2}z_k^2\alpha_k+\frac{|\mathcal{S}|A_{\max}c_{\alpha,\beta} }{c_\tau\tau}\|v_t^i+v_t^{-i}\|_\infty^2
\end{align*}
Since $\sum_{i=1,2}\mathbb{E}_t\left[\|q_{t,K}^i-\bar{q}_{t,K}^i\|_2\right]\lesssim \mathbb{E}_t[\mathcal{L}_q(t,K)]^{1/2}$, 
we have
\begin{align}\label{eeeq}
	\sum_{i=1,2}\mathbb{E}_t\left[\|q_{t,K}^i-\bar{q}_{t,K}^i\|_2\right]\leq\;&\frac{c_1'|\mathcal{S}|^{3/4}A_{\max}}{\alpha^{1/2}_{k_0}(1-\gamma)}z_k\alpha_k^{1/2}+\frac{c_2'\sqrt{|\mathcal{S}|A_{\max}}c^{1/2}_{\alpha,\beta} }{c_{\tau}^{1/2}\tau^{1/2}}\|v_t^i+v_t^{-i}\|_\infty,
\end{align}
where $c_1'$ and $c_2'$ are numerical constants.
Taking total expectation on boths sides of the previous inequality and then using the result in Eq. (\ref{eq:Lyapunov_v+}), and we have
\begin{align*}
	&\mathbb{E}[\|v_{t+1}^i+v_{t+1}^{-i}\|_\infty]\\
	\leq\;& \left(\gamma+\frac{c_2'\sqrt{|\mathcal{S}|A_{\max}}c^{1/2}_{\alpha,\beta} }{c_{\tau}^{1/2}\tau^{1/2}}\right)\mathbb{E}[\|v_t^i+v_t^{-i}\|_\infty]+\frac{c_1'|\mathcal{S}|^{3/4}A_{\max}}{\alpha^{1/2}_{k_0}(1-\gamma)}z_k\alpha_k^{1/2}\\
	\leq \;&\left(\frac{\gamma+1}{2}\right)\mathbb{E}[\|v_t^i+v_t^{-i}\|_\infty]+\frac{c_1'|\mathcal{S}|^{3/4}A_{\max}}{\alpha^{1/2}_{k_0}(1-\gamma)}z_k\alpha_k^{1/2},
\end{align*}
where the last line follows from Condition \ref{con:stepsize}. Repeatedly using the previous inequality starting from $0$ and we have
\begin{align}\label{eq:zls}
	\mathbb{E}[\|v_t^i+v_t^{-i}\|_\infty]\lesssim \frac{2}{1-\gamma}\left(\frac{\gamma+1}{2}\right)^t+\frac{|\mathcal{S}|^{3/4}A_{\max}}{\alpha^{1/2}_{k_0}(1-\gamma)^2}z_k\alpha_k^{1/2}.
\end{align}
The next step is to bound $\|v_t^i-v_*^i\|_\infty$. Recall from Eq. (\ref{eq:Lyapunov_v}) that
\begin{align*}
    \mathbb{E}[\|v_{t+1}^i-v_*^i\|_\infty]
	\leq \;&\gamma  \mathbb{E}[\|v_t^i-v_*^i\|_\infty]+2\mathbb{E}[\|v_t^i+v_t^{-i}\|_\infty]+4\tau \log(A_{\max})\\
	&+2\mathbb{E}[\mathcal{L}_\pi(t,K)]+2\mathbb{E}[\mathcal{L}_q(t,K)]^{1/2}.
\end{align*}
Since Eq. (\ref{eq:ci_di}) and Eq. (\ref{eeeq}) imply that
\begin{align*}
    &\mathbb{E}[\|v_t^i+v_t^{-i}\|_\infty]+\tau \log(A_{\max})+\mathbb{E}[\mathcal{L}_\pi(t,K)]+\mathbb{E}[\mathcal{L}_q(t,K)]^{1/2}\\
	\lesssim \;&\frac{|\mathcal{S}|A_{\max} }{\tau(1-\gamma)^2}\left(\frac{\gamma+1}{2}\right)^t+\tau \log(A_{\max})+\frac{|\mathcal{S}|^2A_{\max}^2\hat{L}_\tau}{\alpha_{k_0} c_{\alpha,\beta}(1-\gamma)^3}z_K^2\alpha_K^{1/2},
\end{align*}
we have 
\begin{align*}
    \mathbb{E}[\|v_{t+1}^i-v_*^i\|_\infty]\leq\;& \gamma  \mathbb{E}[\|v_t^i-v_*^i\|_\infty]\\
    &+c''\left[\frac{|\mathcal{S}|A_{\max} }{\tau(1-\gamma)^2}\left(\frac{\gamma+1}{2}\right)^t+\tau \log(A_{\max})+\frac{|\mathcal{S}|^2A_{\max}^2\hat{L}_\tau}{\alpha_{k_0} c_{\alpha,\beta}(1-\gamma)^3}z_K^2\alpha_K^{1/2}\right]
\end{align*}
for some numerical constant $c''$. Repeatedly using the previous inequality starting from $0$ to $T-1$ and we have
\begin{align*}
	\mathbb{E}[\|v_T^i-v_*^i\|_\infty]\lesssim \frac{|\mathcal{S}|A_{\max} T}{\tau(1-\gamma)^2}\left(\frac{\gamma+1}{2}\right)^{T-1}+\frac{\tau \log(A_{\max})}{(1-\gamma)}+\frac{|\mathcal{S}|^2A_{\max}^2\hat{L}_\tau}{\alpha_{k_0} c_{\alpha,\beta}(1-\gamma)^4}z_K^2\alpha_K^{1/2}
\end{align*}
Using the previous inequality, Eq. (\ref{eq:ci_di}), and Eq. (\ref{eq:zls}) 
in Lemma \ref{le:Nash_Gap}, and we obtain
\begin{align*}
	\mathbb{E}[\|v^i_{*,\pi_{T,K}^{-i}}-v^i_{\pi_{T,K}^i,\pi_{T,K}^{-i}}\|_\infty]\lesssim\;&\frac{|\mathcal{S}|A_{\max} T}{\tau(1-\gamma)^3}\left(\frac{\gamma+1}{2}\right)^{T-1}+\frac{\tau \log(A_{\max})}{(1-\gamma)^2}\\
	&+\frac{|\mathcal{S}|^2A_{\max}^2\hat{L}_\tau}{\alpha_{k_0} c_{\alpha,\beta}(1-\gamma)^5}z_K^2\alpha_K^{1/2}
\end{align*}
Finally, we have by the previous inequality and Lemma  \ref{le:Nash_to_v} that
\begin{align*}
	\mathbb{E}[\textit{NG}(\pi_{T,K}^i,\pi_{T,K}^{-i})]\lesssim\;&\frac{|\mathcal{S}|A_{\max} T}{\tau(1-\gamma)^3}\left(\frac{\gamma+1}{2}\right)^{T-1}+\frac{\tau \log(A_{\max})}{(1-\gamma)^2}\\
	&+\frac{|\mathcal{S}|^2A_{\max}^2\hat{L}_\tau}{\alpha_{k_0} c_{\alpha,\beta}(1-\gamma)^5}z_K^2\alpha_K^{1/2}.
\end{align*}
The proof of Theorem \ref{thm:diminishing} is complete.

\subsection{Proof of All Supporting Lemmas}

\subsubsection{Proof of Lemma \ref{le:boundedness}}\label{pf:le:boundedness}
We first show by induction that whenever $\|v_t^i\|_\infty\leq \frac{1}{1-\gamma}$, we have $\|q_{t,k}^i\|_\infty\leq \frac{1}{1-\gamma}$ for all $k\geq 0$. Note that $\|q_{t,0}^i\|_\infty\leq \frac{1}{1-\gamma}$ holds by our initialization. Suppose that $\|q_{t,k}^i\|_\infty\leq \frac{1}{1-\gamma}$ for some $k\geq 0$. Then we have for all $(s,a^i)$ that
\begin{align}
	&|q_{t,k+1}^i(s,a^i)|\nonumber\\
	=\;&|q_{t,k}^i(s,a^i)+\alpha_k\mathds{1}_{\{(s,a^i)=(S_k,A_k^i)\}}(\mathcal{R}^i(S_k,A_k^i,A_k^{-i})+\gamma v_t^i(S_{k+1})-q_{t,k}^i(S_k,A_k^i))|\nonumber\\
	\leq \;&(1-\alpha_k\mathds{1}_{\{(s,a^i)=(S_k,A_k^i)\}})|q_{t,k}^i(s,a^i)|\nonumber\\
	&+\alpha_k\mathds{1}_{\{(s,a^i)=(S_k,A_k^i)\}}|\mathcal{R}^i(S_k,A_k^i,A_k^{-i})+\gamma v_t^i(S_{k+1})|\nonumber\\
	\leq \;&(1-\alpha_k\mathds{1}_{\{(s,a^i)=(S_k,A_k^i)\}})\frac{1}{1-\gamma}+\alpha_k\mathds{1}_{\{(s,a^i)=(S_k,A_k^i)\}}\left(1+\frac{\gamma}{1-\gamma}\right)\label{eq:boundedness1}\\
	= \;&\frac{1}{1-\gamma},\nonumber
\end{align}
where Eq. (\ref{eq:boundedness1}) follows from the induction hypothesis $\|q_{t,0}^i\|_\infty\leq \frac{1}{1-\gamma}$, $\|v_t^i\|_\infty\leq \frac{1}{1-\gamma}$, and $\max_{s,a^i,a^{-i}}|\mathcal{R}^i(s,a^i,a^{-i})|\leq 1$. The induction is now complete and we have  $\|q_{t,k}^i\|_\infty\leq \frac{1}{1-\gamma}$ for all $k\geq 0$ whenever $\|v_t^i\|_\infty\leq \frac{1}{1-\gamma}$.

We next again use induction to show that $\|v_t^i\|_\infty\leq \frac{1}{1-\gamma}$ for all $t\geq 0$. Our initialization ensures that $\|v_0^i\|_\infty\leq \frac{1}{1-\gamma}$. Suppose that $\|v_t^i\|_\infty\leq \frac{1}{1-\gamma}$ for some $t\geq  0$. Using the update equation for $v_{t+1}^i$ (i.e., Algorithm \ref{algorithm:tabular} Line $8$) and the fact that $\|q_{t,k}^i\|_\infty\leq \frac{1}{1-\gamma}$ for all $k\geq 0$, we have for all $s\in\mathcal{S}$ that
\begin{align*}
	|v_{t+1}^i(s)|=\left|\sum_{a^i\in\mathcal{A}^i}\pi^i_{t,K}(a^i|s)q_{t,K}^i(s,a^i)\right|\leq \sum_{a^i\in\mathcal{A}^i}\pi^i_{t,K}(a^i|s)\|q_{t,K}^i\|_\infty\leq \frac{1}{1-\gamma}.
\end{align*}
The induction for $\{v_t^i\}$ is now complete and we have $\|v_t^i\|_\infty\leq \frac{1}{1-\gamma}$ for all $t\geq 0$.

\subsubsection{Proof of Lemma \ref{le:margin}}\label{pf:le:margin}
Observe that for any $x\in\mathbb{R}^d$ and $j\in \{1,2,\cdots,d\}$, we have
\begin{align*}
	\frac{\exp(x_j)}{\sum_{\ell=1}^d\exp(x_\ell)}=\;&\frac{\exp(x_j)}{\exp(x_j)+\sum_{\ell\neq j}\exp(x_\ell)}\\
	=\;&\frac{1}{1+\sum_{\ell\neq j}\exp(x_\ell-x_j)}\\
	\geq \;&\frac{1}{1+(d-1)\exp(2\|x\|_\infty)}.
\end{align*}
Therefore, since $\|q^i_{t,k}\|_\infty\leq 1/(1-\gamma)$ for all $t,k\geq 0$ (cf. Lemma \ref{le:boundedness}), we have for all $t,k\geq 0$ and $(s,a^i)$ that 
\begin{align}\label{eq:softmax_lemma}
	\sigma_\tau(q_{t,k}^i(s))(a^i)\geq\;&\frac{1}{1+(A_{\max}-1)\exp(2/[(1-\gamma)\tau])}=\ell_{\tau}.
\end{align}

We next use induction to show that $\pi_{t,k}^i(a^i|s)\geq \ell_{\tau}$ for all $t,k\geq 0$. Given any $t\geq 0$, our uniform initialization of $\pi_{t,0}^i$ ensures that $\pi_{t,0}^i(a^i|s)\geq \ell_{\tau}$ for all $(s,a^i)$. Now suppose that $\pi_{t,k}^i(a^i|s)\geq \ell_{\tau}$ for all $(s,a^i)$ for some $k\geq 0$. Then we have from Algorithm \ref{algorithm:tabular} Line $4$ that
\begin{align*}
	\pi_{t,k+1}^i(a^i|s)=\;& (1-\beta_k)\pi_{t,k}^i(a^i|s)+\beta_k \sigma_\tau(q_{t,k}^i(s))(a^i)\\
	\geq\;& (1-\beta_k)\ell_{\tau}+\beta_k\ell_{\tau}\\
	=\;&\ell_{\tau},
\end{align*}
where the inequality follows from Eq. (\ref{eq:softmax_lemma}) and the induction hypothesis. The induction is now complete and we have $\pi_{t,k}^i(a^i|s)\geq \ell_{\tau}$ for all $t,k\geq 0$ and $(s,a^i)$. Similarly, we also have $\pi_{t,k}^{-i}(a^{-i}|s)\geq \ell_{\tau}$ for all $t,k\geq 0$ and $(s,a^{-i})$.

\subsubsection{Proof of Lemma \ref{thm:exploration}}\label{pf:thm:exploration}
Lemma \ref{thm:exploration} (1), (3), and (4) are identical to \cite[Proposition 3]{zhang2022global}. We here only prove Lemma \ref{thm:exploration} (2).

Consider the Markov chain $\{S_k\}$ induced by $\pi_b$.
	Since $\{S_k\}$ is irreducible and aperiodic, there exists a positive integer $r_b$ such that $P_{\pi_b}^{r_b}$ has strictly positive entries \cite[Proposition 1.7]{levin2017markov}. Therefore, there exists $\delta_b\in (0,1)$ such that
	\begin{align*}
		P_{\pi_b}^{r_b}(s,s')\geq \delta_b \mu_b(s')
	\end{align*}
	for all $(s,s')$. In addition, the constant $\rho_b$ introduced after Assumption \ref{as:MC} is explicitly given as $\rho_b=\exp(-\delta_b/r_b)$.
	The previous two equations are from the proof of the Markov chain convergence theorem presented in \cite[Section 4.3]{levin2017markov}.
	
	Next we consider the Markov chain $\{S_k\}$ induced by an arbitrary $\pi\in \Pi_\delta$. Since
	\begin{align*}
		\frac{\pi_b(a|s)}{\pi(a|s)}=\frac{\pi_b^i(a^i|s)\pi_b^{-i}(a^i|s)}{\pi^i(a^i|s)\pi^{-i}(a^i|s)}\leq  \frac{1}{\delta_i\delta_{-i}},\quad \forall\;a=(a^i,a^{-i}) \text{ and }s,
	\end{align*}
	we have for any $s,s'\in\mathcal{S}$ and $k\geq 1$ that
	\begin{align*}
		P_{\pi_b}^k(s,s')=\;&\sum_{s_0}P^{k-1}_{\pi_b}(s,s_0)P_{\pi_b}(s_0,s')\\
		=\;&\sum_{s_0}P^{k-1}_{\pi_b}(s,s_0)\sum_{a\in\mathcal{A}}\pi_b(a|s_0)P_a(s_0,s')\\
		= \;&\sum_{s_0}P^{k-1}_{\pi_b}(s,s_0)\sum_{a\in\mathcal{A}}\frac{\pi_b(a|s_0)}{\pi(a|s_0)}\pi(a|s_0)P_a(s_0,s')\\
		\leq  \;&\frac{1}{\delta_i\delta_{-i}}\sum_{s_0}P^{k-1}_{\pi_b}(s,s_0)\sum_{a\in\mathcal{A}}\pi(a|s_0)P_a(s_0,s')\\
		\leq  \;&\frac{1}{\delta_i\delta_{-i}}\sum_{s_0}P^{k-1}_{\pi_b}(s,s_0)P_{\pi}(s_0,s')\\
		=  \;&\frac{1}{\delta_i\delta_{-i}}[P^{k-1}_{\pi_b} P_{\pi}](s,s').
	\end{align*}
	Since the previous inequality holds for all $s$ and $s'$, we in fact have $\delta_i\delta_{-i} P_{\pi_b}^k\leq P^{k-1}_{\pi_b} P_{\pi}$ (which is an entry-wise inequality). Repeatedly using the previous inequality and we obtain 
	\begin{align*}
		(\delta_i\delta_{-i})^kP_{\pi_b}^k\leq P_{\pi}^k,
	\end{align*}
	which implies
	\begin{align*}
		P_\pi^{r_b}(s,s')\geq\;& (\delta_i\delta_{-i})^{r_b}P_{\pi_b}^{r_b}(s,s')\\
		\geq\;& \delta_b(\delta_i\delta_{-i})^{r_b}\mu_b(s')\\
		\geq\;& \delta_b(\delta_i\delta_{-i})^{r_b}\frac{\mu_b(s')}{\mu_\pi(s')}\mu_\pi(s')\\
		\geq\;& \delta_b(\delta_i\delta_{-i})^{r_b} \mu_{b,\min}\mu_\pi(s').
	\end{align*}
	Following the proof of the Markov chain convergence theorem in \cite[Section 4.3]{levin2017markov} and we have
	\begin{align}\label{eq1:thm:exploration}
		\|P_\pi^k(s,\cdot)-\mu_\pi(\cdot)\|_{\text{TV}}\leq\;&  (1-\delta_b(\delta_i\delta_{-i})^{r_b}\mu_{b,\min})^{k/r_b-1},\quad \forall\;s\in\mathcal{S},\; \pi\in \Pi_\delta.
	\end{align}
	Since $A_{\max}\geq 2$ (otherwise there is no decision to make in this Markov game), we have $\delta_i\delta_{-i}\leq \frac{1}{2}$. It follows that  $1-\delta_b(\delta_i\delta_{-i})^{r_b}\mu_{b,\min}> 1/2$.
	Using the previous inequality in Eq. (\ref{eq1:thm:exploration}) and we have 
	\begin{align*}
		\sup_{\pi\in \Pi_\delta}\max_{s\in\mathcal{S}}\|P_\pi^k(s,\cdot)-\mu_\pi(\cdot)\|_{\text{TV}}\leq\;&  2(1-\delta_b(\delta_i\delta_{-i})^{r_b}\mu_{b,\min})^{k/r_b}\\
		\leq\;&  2\exp\left(-\delta_b(\delta_i\delta_{-i})^{r_b}\mu_{b,\min}k/r_b\right)\\
		=\;&  2\rho_b^{(\delta_i\delta_{-i})^{r_b}\mu_{b,\min}k}\tag{Recall that $\rho_b=\exp(-\delta_b/r_b)$}\\
		=\;&  2\rho_\delta^k.
	\end{align*}
	
	We next compute the mixing time. Using the previous inequality and the definition of the total variation distance, we have
	\begin{align*}
		\sup_{\pi\in \Pi_\delta}\max_{s\in\mathcal{S}}\|P_\pi^k(s,\cdot)-\mu_\pi(\cdot)\|_{\text{TV}}\leq \eta
	\end{align*}
	as long as
	\begin{align*}
		k\geq \frac{\log(2/\eta)}{\log(1/\rho_\delta)}=\frac{1}{(\delta_i\delta_{-i})^{r_b}\mu_{b,\min}} \frac{\log(2/\eta)}{\log(1/\rho_b)}\geq \frac{t_{\pi_b,\eta}}{(\delta_i\delta_{-i})^{r_b}\mu_{b,\min}}.
	\end{align*}

\subsubsection{Proof of Lemma \ref{le:Nash_to_v}}\label{pf:le:Nash_to_v}
Using the definition of utility functions and we have
\begin{align*}
	&\sum_{i=1,2}\left(\max_{\pi^i}U^i(\pi^i,\pi_{T,K}^{-i})-U^i(\pi_{T,K}^i,\pi_{T,K}^{-i})\right)\\
	= \;&\sum_{i=1,2}\left(\max_{\pi^i}\mathbb{E}_{S\sim p_o}\left[v^i_{\pi^i,\pi_{T,K}^{-i}}(S)-v^i_{\pi_{T,K}^i,\pi_{T,K}^{-i}}(S)\right]\right)\\
	\leq \;&\sum_{i=1,2}\left(\mathbb{E}_{S\sim p_o}\left[\max_{\pi^i}v^i_{\pi^i,\pi_{T,K}^{-i}}(S)-v^i_{\pi_{T,K}^i,\pi_{T,K}^{-i}}(S)\right]\right)\tag{Jensen's inequality}\\
	= \;&\sum_{i=1,2}\left(\mathbb{E}_{S\sim p_o}\left[v^i_{*,\pi_{T,K}^{-i}}(S)-v^i_{\pi_{T,K}^i,\pi_{T,K}^{-i}}(S)\right]\right)\\
	\leq  \;&\sum_{i=1,2}\left\|v^i_{*,\pi_{T,K}^{-i}}-v^i_{\pi_{T,K}^i,\pi_{T,K}^{-i}}\right\|_\infty.
\end{align*}

\subsubsection{Proof of Lemma \ref{le:Nash_Gap}}\label{pf:le:Nash_Gap}

For any $t\geq 0$, $s\in\mathcal{S}$, and $i\in \{1,2\}$, we have
\begin{align}
	0\leq \;&\left|v^i_{*,\pi_{t,K}^{-i}}(s)-v^i_{\pi_{t,K}^i,\pi_{t,K}^{-i}}(s)\right|\nonumber\\
	=\;&v^i_{*,\pi_{t,K}^{-i}}(s)-v^i_{\pi_{t,K}^i,\pi_{t,K}^{-i}}(s)\nonumber\\
	\leq  \;&v^i_{*,\pi_{t,K}^{-i}}(s)-v^i_{\pi_{t,K}^i,*}(s)\nonumber\\
	=\;&-v^{-i}_{\pi_{t,K}^{-i},*}(s)-v^i_{\pi_{t,K}^i,*}(s)\nonumber\\
	=\;&v^i_*(s)-v^{-i}_{\pi_{t,K}^{-i},*}(s)+v^{-i}_*(s)-v^i_{\pi_{t,K}^i,*}(s)\nonumber\\
	\leq \;&\|v^{-i}_*-v^{-i}_{\pi_{t,K}^{-i},*}\|_\infty+\|v^i_*-v^i_{\pi_{t,K}^i,*}\|_\infty.\label{eq:last_policy_bound}
\end{align}
It remains to bound the two terms on the RHS of the previous inequality. For the first term, note that for any $s\in\mathcal{S}$ and $t\geq 0$, we have
\begin{align}
	0\leq\;& v^{-i}_*(s)-v^{-i}_{\pi_{t,K}^{-i},*}(s)\nonumber\\
	=\;& v^i_{*,\pi_{t,K}^{-i}}(s)-v^i_*(s)\nonumber\\
	=\;&\max_{\mu^i}(\mu^i)^\top  \mathcal{T}^i(v^i_{*,\pi_{t,K}^{-i}})(s)\pi_{t,K}^{-i}(s)-\max_{\mu^i}\min_{\mu^{-i}}(\mu^i)^\top \mathcal{T}^i(v_*^i)(s)\mu^{-i}\nonumber\\
	=\;&|\max_{\mu^i}(\mu^i)^\top  \mathcal{T}^i(v^i_{*,\pi_{t,K}^{-i}})(s)\pi_{t,K}^{-i}(s)-\max_{\mu^i}(\mu^i)^\top  \mathcal{T}^i(v^i_{*})(s)\pi_{t,K}^{-i}(s)|\nonumber\\
	&+|\max_{\mu^i}(\mu^i)^\top  \mathcal{T}^i(v^i_{*})(s)\pi_{t,K}^{-i}(s)-\max_{\mu^i}\min_{\mu^{-i}}(\mu^i)^\top \mathcal{T}^i(v_*^i)(s)\mu^{-i}|\nonumber\\
	\leq \;&\max_{\mu^i}|(\mu^i)^\top  (\mathcal{T}^i(v^i_{*,\pi_{t,K}^{-i}})(s)-\mathcal{T}^i(v^i_{*})(s))\pi_{t,K}^{-i}(s)|\nonumber\\
	&+|\max_{\mu^i}(\mu^i)^\top  \mathcal{T}^i(v^i_{*})(s)\pi_{t,K}^{-i}(s)-\max_{\mu^i}\min_{\mu^{-i}}(\mu^i)^\top  \mathcal{T}^i(v^i_t)(s)\mu^{-i}|\nonumber\\
	&+|\max_{\mu^i}\min_{\mu^{-i}}(\mu^i)^\top  \mathcal{T}^i(v^i_t)(s)\mu^{-i}-\max_{\mu^i}\min_{\mu^{-i}}(\mu^i)^\top \mathcal{T}^i(v_*^i)(s)\mu^{-i}|\nonumber\\
	\leq \;&\underbrace{\max_{\mu^i}|(\mu^i)^\top  (\mathcal{T}^i(v^i_{*,\pi_{t,K}^{-i}})(s)-\mathcal{T}^i(v^i_{*})(s))\pi_{t,K}^{-i}(s)|}_{\hat{E}_1}\nonumber\\
	&+\underbrace{|\max_{\mu^i}(\mu^i)^\top  \mathcal{T}^i(v^i_{*})(s)\pi_{t,K}^{-i}(s)-\max_{\mu^i}(\mu^i)^\top  \mathcal{T}^i(v^i_t)(s)\pi_{t,K}^{-i}(s)|}_{\hat{E}_2}\nonumber\\
	&+\underbrace{\max_{\mu^i}(\mu^i)^\top  \mathcal{T}^i(v^i_t)(s)\pi_{t,K}^{-i}(s)-\max_{\mu^i}\min_{\mu^{-i}}(\mu^i)^\top  \mathcal{T}^i(v^i_t)(s)\mu^{-i}}_{\hat{E}_3}\nonumber\\
	&+\underbrace{|\max_{\mu^i}\min_{\mu^{-i}}(\mu^i)^\top  \mathcal{T}^i(v^i_t)(s)\mu^{-i}-\max_{\mu^i}\min_{\mu^{-i}}(\mu^i)^\top \mathcal{T}^i(v_*^i)(s)\mu^{-i}|}_{\hat{E}_4}.\label{eq:connect1}
\end{align}
We next bound the terms $\{\hat{E}_j\}_{1\leq j\leq 4}$. For any $v_1^i,v_2^i\in\mathbb{R}^{|\mathcal{S}|}$, we have for any $(s,a^i,a^{-i})$ that
\begin{align*}
	&|\mathcal{T}^i(v_1^i)(s,a^i,a^{-i})-\mathcal{T}^i(v_2^i)(s,a^i,a^{-i})|\\
	=\;&\gamma|\mathbb{E}[v_1^i(S_1)-v_2^i(S_1)\mid S_0=s,A_0^i=a^i,A_0^{-i}=a^{-i}]|\\
	\leq\;& \gamma\|v_1^i-v_2^i\|_\infty,
\end{align*}
which implies $\|\mathcal{T}^i(v_1^i)-\mathcal{T}^i(v_2^i)\|_\infty\leq \gamma\|v_1^i-v_2^i\|_\infty$.
As a result, we have
\begin{align*}
	\hat{E}_1\leq\;& \|\mathcal{T}^i(v^i_{*,\pi_{t,K}^{-i}})-\mathcal{T}^i(v^i_{*})\|_\infty\leq \gamma\|v^i_{*,\pi_{t,K}^{-i}}-v^i_{*}\|_\infty,\\
	\hat{E}_2\leq \;&\|\mathcal{T}^i(v_t^i)-\mathcal{T}^i(v^i_{*})\|_\infty\leq \gamma\|v^i_t-v^i_{*}\|_\infty,\\
	\hat{E}_4\leq\;& \|\mathcal{T}^i(v_t^i)-\mathcal{T}^i(v^i_{*})\|_\infty\leq \gamma\|v^i_t-v^i_{*}\|_\infty.
\end{align*}
Bounding the term $\hat{E}_3$ requires more effort. First observe that
\begin{align*}
	\hat{E}_3
	\leq \;&\left|\max_{\mu^i}(\mu^i)^\top \mathcal{T}^i(v_t^i)(s)\pi_{t,K}^{-i}(s)-\min_{\mu^{-i}}\pi_{t,K}^i(s)\mathcal{T}^i(v_t^i)(s)\mu^{-i}\right|\\
	\leq \;&\left|\max_{\mu^{-i}}(\mu^{-i})^\top  \mathcal{T}^{-i}(v_t^{-i})(s)\pi_{t,K}^i(s)+\min_{\mu^{-i}}(\mu^{-i})^\top \mathcal{T}^i(v_t^i)(s)^\top \pi_{t,K}^i(s)\right|\\
	&+\left|\sum_{i=1,2}\max_{\mu^i}(\mu^i)^\top \mathcal{T}^i(v_t^i)(s)\pi_{t,K}^{-i}(s)\right|\\
	\leq \;&\left|\max_{\mu^{-i}}(\mu^{-i})^\top  \mathcal{T}^{-i}(v_t^{-i})(s)\pi_{t,K}^i(s)-\max_{\mu^{-i}}(\mu^{-i})^\top [-\mathcal{T}^i(v_t^i)(s)]^\top \pi_{t,K}^i(s)\right|\\
	&+\left|\sum_{i=1,2}\max_{\mu^i}(\mu^i)^\top \mathcal{T}^i(v_t^i)(s)\pi_{t,K}^{-i}(s)\right|\\
	\leq \;&\max_{\mu^{-i}}\left|(\mu^{-i})^\top  (\mathcal{T}^{-i}(v_t^{-i})(s)+[\mathcal{T}^i(v_t^i)(s)]^\top) \pi_{t,K}^i(s)\right|\\
	&+\left|\sum_{i=1,2}\max_{\mu^i}(\mu^i)^\top \mathcal{T}^i(v_t^i)(s)\pi_{t,K}^{-i}(s)\right|\\
	\leq \;&\max_{a^i,a^{-i}}\left|\mathcal{T}^i(v_t^i)(s,a^i,a^{-i})+\mathcal{T}^{-i}(v_t^{-i})(s,a^i,a^{-i})\right|\\
	&+\left|\sum_{i=1,2}\max_{\mu^i}(\mu^i)^\top \mathcal{T}^i(v_t^i)(s)\pi_{t,K}^{-i}(s)\right|\\
	\leq \;&\gamma\|v_t^i+v_t^{-i}\|_\infty+\left|\sum_{i=1,2}\max_{\mu^i}(\mu^i)^\top \mathcal{T}^i(v_t^i)(s)\pi_{t,K}^{-i}(s)\right|,
\end{align*}
where the last line follows from
\begin{align*}
    &|\mathcal{T}^i(v_t^i)(s,a^i,a^{-i})+\mathcal{T}^{-i}(v_t^{-i})(s,a^i,a^{-i})|\\
    =\;&\gamma |\mathbb{E}[v_t^i(S_1)+v_t^i(S_1)\mid S_0=s,A_0^i=a^i,A_0^{-i}=a^{-i}]|\\
    \leq \;&\gamma\|v_t^i+v_t^{-i}\|_\infty.
\end{align*}
In addition, we have
\begin{align*}
	&\left|\sum_{i=1,2}\max_{\mu^i}(\mu^i)^\top \mathcal{T}^i(v_t^i)(s)\pi_{t,K}^{-i}(s)\right|\\
	= \;&\left|\sum_{i=1,2}\left\{\max_{\mu^i}(\mu^i)^\top \mathcal{T}^i(v^i_t)(s) \pi_{t,K}^{-i}(s)-(\pi_{t,K}^i(s))^\top \mathcal{T}^i(v^i_t)(s) \pi_{t,K}^{-i}(s)\right\}\right|\\
	&+\left|\sum_{i=1,2}(\pi_{t,K}^i(s))^\top \mathcal{T}^i(v^i_t)(s) \pi_{t,K}^{-i}(s)\right|\\
	= \;&\sum_{i=1,2}\left\{\max_{\mu^i}(\mu^i)^\top \mathcal{T}^i(v^i_t)(s) \pi_{t,K}^{-i}(s)-(\pi_{t,K}^i(s))^\top \mathcal{T}^i(v^i_t)(s) \pi_{t,K}^{-i}(s)\right\}\\
	&+\left|\sum_{i=1,2}(\pi_{t,K}^i(s))^\top \mathcal{T}^i(v^i_t)(s) \pi_{t,K}^{-i}(s)\right|\\
	\leq \;&\sum_{i=1,2}\max_{\mu^i}\left\{(\mu^i-\pi_{t,K}^i(s))^\top \mathcal{T}^i(v^i_t)(s) \pi_{t,K}^{-i}(s)+\tau \nu(\mu^i)-\tau \nu(\pi_{t,K}^i(s))\right\}+2\tau \log(A_{\max})\\
	&+\left|\sum_{i=1,2}(\pi_{t,K}^i(s))^\top \mathcal{T}^i(v^i_t)(s) \pi_{t,K}^{-i}(s)\right|\\
	=  \;&V_{v_t,s}(\pi_{t,K}^i(s),\pi_{t,K}^{-i}(s))+2\tau \log(A_{\max})+\left|\sum_{i=1,2}(\pi_{t,K}^i(s))^\top \mathcal{T}^i(v^i_t)(s) \pi_{t,K}^{-i}(s)\right|\\
	\leq  \;&V_{v_t,s}(\pi_{t,K}^i(s),\pi_{t,K}^{-i}(s))+2\tau \log(A_{\max})\\
	&+\max_{a^i,a^{-i}}\left|\mathcal{T}^i(v_t^i)(s,a^i,a^{-i})+\mathcal{T}^{-i}(v_t^{-i})(s,a^i,a^{-i})\right|\\
	\leq  \;&V_{v_t,s}(\pi_{t,K}^i(s),\pi_{t,K}^{-i}(s))+2\tau \log(A_{\max})+\gamma\|v_t^i+v_t^{-i}\|_\infty.
\end{align*}
It follows that
\begin{align}\label{eq:355}
	\hat{E}_3\leq 2\gamma\|v_t^i+v_t^{-i}\|_\infty+\max_{s}V_{v_t,s}(\pi_{t,K}^i(s),\pi_{t,K}^{-i}(s))+2\tau \log(A_{\max}).
\end{align}

Substituting the upper bounds we obtained for the terms $\{E_j\}_{1\leq j\leq 4}$ into Eq. (\ref{eq:connect1}) and we have
\begin{align*}
	\|v^{-i}_*-v^{-i}_{\pi_{t,K}^{-i},*}\|_\infty\leq\;&\gamma\|v^i_{*,\pi_{t,K}^{-i}}-v^i_{*}\|_\infty+2\gamma\|v_t^i+v_t^{-i}\|_\infty+2\gamma\|v^i_t-v^i_{*}\|_\infty\\
	&+\max_{s}V_{v_t,s}(\pi_{t,K}^i(s),\pi_{t,K}^{-i}(s))+2\tau \log(A_{\max})\\
	=\;&\gamma\|v^{-i}_*-v^{-i}_{\pi_{t,K}^{-i},*}\|_\infty+2\gamma\|v_t^i+v_t^{-i}\|_\infty+2\gamma\|v^i_t-v^i_{*}\|_\infty\\
	&+\max_{s}V_{v_t,s}(\pi_{t,K}^i(s),\pi_{t,K}^{-i}(s))+2\tau \log(A_{\max}),
\end{align*}
which after rearranging terms implies
\begin{align*}
	\|v^{-i}_*-v^{-i}_{\pi_{t,K}^{-i},*}\|_\infty\leq\;&\frac{1}{1-\gamma}\bigg(2\|v_t^i+v_t^{-i}\|_\infty+2\|v^i_t-v^i_{*}\|_\infty\\
	&+\max_{s}V_{v_t,s}(\pi_{t,K}^i(s),\pi_{t,K}^{-i}(s))+2\tau \log(A_{\max})\bigg).
\end{align*}
Similarly, we also have
\begin{align*}
	\|v^i_*-v^i_{\pi_{t,K}^i,*}\|_\infty\leq\;&\frac{1}{1-\gamma}\bigg(2\|v_t^i+v_t^{-i}\|_\infty+2\|v^i_t-v^i_{*}\|_\infty\\
	&+\max_{s}V_{v_t,s}(\pi_{t,K}^i(s),\pi_{t,K}^{-i}(s))+2\tau \log(A_{\max})\bigg).
\end{align*}
Substituting the previous two inequalities into Eq. (\ref{eq:last_policy_bound}) and we finally obtain
\begin{align*}
	\|v^i_{*,\pi_{t,K}^{-i}}-v^i_{\pi_{t,K}^i,\pi_{t,K}^{-i}}\|_\infty\leq \;&\frac{2}{1-\gamma}\bigg(2\|v_t^i+v_t^{-i}\|_\infty+2\|v^i_t-v^i_{*}\|_\infty\\
	&+\max_{s}V_{v_t,s}(\pi_{t,K}^i(s),\pi_{t,K}^{-i}(s))+2\tau \log(A_{\max})\bigg).
\end{align*}

\subsubsection{Proof of Lemma \ref{le:outer-loop}}\label{pf:le:outer-loop}
For any $i\in \{1,2\}$, we have by the outer-loop update equation (cf. Line $8$) of Algorithm \ref{algorithm:tabular} that
\begin{align*}
	v_{t+1}^i(s)=\;&\pi_{t,K}^i(s)^\top q_{t,K}^i(s)\\
	=\;&\textit{val}^i(\mathcal{T}^i(v^i_t)(s))+\pi_{t,K}^i(s)^\top q_{t,K}^i(s)-\textit{val}^i(\mathcal{T}^i(v^i_t)(s))
\end{align*}
Since $\textit{val}^i(\mathcal{T}^i(v_*^i)(s))=\mathcal{B}^i(v_*^i)(s)=v_*^i(s)$, we have
\begin{align}
	|v_{t+1}^i(s)-v_*^i(s)|
	=\;&|\textit{val}^i(\mathcal{T}^i(v^i_t)(s))-\textit{val}^i(\mathcal{T}^i(v_*^i)(s))|\nonumber\\
	&+|\pi_{t,K}^i(s)^\top q_{t,K}^i(s)-\textit{val}^i(\mathcal{T}^i(v^i_t)(s))|.\label{eq1:prop:outer}
\end{align}
For the first term on the RHS of Eq. (\ref{eq1:prop:outer}), we have by the contraction property of the minimax Bellman operator that
\begin{align*}
	\left|\textit{val}^i(\mathcal{T}^i(v^i_t)(s))-\textit{val}^i(\mathcal{T}^i(v_*^i)(s))\right|=\;&\left|\mathcal{B}^i(v_t^i)(s)-\mathcal{B}^i(v_*^i)(s)\right|\\
	\leq\;& \gamma \|v^i_t-v_*^i\|_\infty.
\end{align*}
For the second term on the RHS of Eq. (\ref{eq1:prop:outer}), we have
\begin{align*}
	\left|\pi_{t,K}^i(s)^\top q_{t,K}^i(s)-\textit{val}^i(\mathcal{T}^i(v^i_t)(s))\right|\leq \;&\underbrace{\left|\max_{\mu^i}(\mu^i)^\top \mathcal{T}^i(v^i_t)(s) \pi_{t,K}^{-i}(s)-\pi_{t,K}^i(s)^\top q_{t,K}^i(s)\right|}_{T_1}\\
	&+\underbrace{\left|\max_{\mu^i}(\mu^i)^\top \mathcal{T}^i(v^i_t)(s) \pi_{t,K}^{-i}(s)-\textit{val}^i(\mathcal{T}^i(v^i_t)(s))\right|}_{T_2}
\end{align*}
For the term $T_1$, we have
\begin{align*}
	T_1
	\leq \;&\left|\max_{\mu^i}(\mu^i)^\top \mathcal{T}^i(v^i_t)(s) \pi_{t,K}^{-i}(s)-(\pi_{t,K}^i(s))^\top \mathcal{T}^i(v^i_t)(s) \pi_{t,K}^{-i}(s)\right|\\
	&+\left|(\pi_{t,K}^i(s))^\top \mathcal{T}^i(v^i_t)(s) \pi_{t,K}^{-i}(s)-\pi_{t,K}^i(s)^\top q_{t,K}^i(s)\right|\\
	\leq \;&\max_{\mu^i}(\mu^i)^\top \mathcal{T}^i(v^i_t)(s) \pi_{t,K}^{-i}(s)-(\pi_{t,K}^i(s))^\top \mathcal{T}^i(v^i_t)(s) \pi_{t,K}^{-i}(s)\\
	&+\|\mathcal{T}^i(v^i_t)(s) \pi_{t,K}^{-i}(s)- q_{t,K}^i(s)\|_\infty\\
	\leq \;&\sum_{i=1,2}\left\{\max_{\mu^i}(\mu^i-\pi_{t,K}^i(s))^\top \mathcal{T}^i(v^i_t)(s) \pi_{t,K}^{-i}(s)\right\}\\
	&+\|\mathcal{T}^i(v^i_t)(s) \pi_{t,K}^{-i}(s)- q_{t,K}^i(s)\|_\infty\\
	\leq \;&\sum_{i=1,2}\left\{\max_{\mu^i}(\mu^i-\pi_{t,K}^i(s))^\top \mathcal{T}^i(v^i_t)(s) \pi_{t,K}^{-i}(s)+\tau \nu(\mu^i)-\tau\nu(\pi_{t,K}^i(s))\right\}\\
	&+2\tau\log(A_{\max})+\|\mathcal{T}^i(v^i_t)(s) \pi_{t,K}^{-i}(s)- q_{t,K}^i(s)\|_\infty\\
	\leq \;&V_{v_t,s}(\pi_{t,K}^i(s),\pi_{t,K}^{-i}(s))+2\tau\log(A_{\max})+\|\mathcal{T}^i(v^i_t)(s) \pi_{t,K}^{-i}(s)- q_{t,K}^i(s)\|_\infty.
\end{align*}
Note that $T_2$ is exactly the term $\hat{E}_3$ we analyzed in proving Lemma \ref{le:Nash_Gap}. Therefore, we have from Eq. (\ref{eq:355}) that
\begin{align*}
	T_2\leq 2\gamma\|v_t^i+v_t^{-i}\|_\infty+\max_{s}V_{v_t,s}(\pi_{t,K}^i(s),\pi_{t,K}^{-i}(s))+2\tau \log(A_{\max}).
\end{align*}
It follows that
\begin{align*}
	&\left|\pi_{t,K}^i(s)^\top q_{t,K}^i(s)-\textit{val}^i(\mathcal{T}^i(v^i_t)(s))\right|\\
	\leq \;&T_1+T_2\\
	\leq \;&2\max_{s}V(\pi_{t,K}^i(s),\pi_{t,K}^{-i}(s))+\max_{s}\|\mathcal{T}^i(v^i_t)(s) \pi_{t,K}^{-i}(s)- q_{t,K}^i(s)\|_\infty\\
	&+2\gamma\|v_t^i+v_t^{-i}\|_\infty+4\tau \log(A_{\max}).
\end{align*}
Using the upper bounds we obtained for the two terms on the RHS of Eq. (\ref{eq1:prop:outer}) and we have
\begin{align*}
	|v_{t+1}^i(s)-v_*^i(s)|
	\leq \;&\gamma  \|v_t^i-v_*^i\|_\infty+2\max_{s\in\mathcal{S}}V(\pi_{t,K}^i(s),\pi_{t,K}^{-i}(s))+4\tau \log(A_{\max})\\
	&+\max_{s\in\mathcal{S}}\|\mathcal{T}^i(v^i_t)(s) \pi_{t,K}^{-i}(s)- q_{t,K}^i(s)\|_\infty+2\gamma\|v_t^i+v_t^{-i}\|_\infty.
\end{align*}
Since the RHS of the previous inequality does not depend on $s$, we have for any $i\in \{1,2\}$ that
\begin{align*}
	\|v_{t+1}^i-v_*^i\|_\infty
	\leq \;&\gamma  \|v_t^i-v_*^i\|_\infty+2\max_{s\in\mathcal{S}}V(\pi_{t,K}^i(s),\pi_{t,K}^{-i}(s))+4\tau \log(A_{\max})\\
	&+\max_{s\in\mathcal{S}}\|\mathcal{T}^i(v^i_t)(s) \pi_{t,K}^{-i}(s)- q_{t,K}^i(s)\|_\infty+2\gamma\|v_t^i+v_t^{-i}\|_\infty.
\end{align*}

\subsubsection{Proof of Lemma \ref{le:outer-sum}}\label{pf:le:outer-sum}
Using the outer-loop update equation (cf. Algorithm \ref{algorithm:tabular} Line $8$) and we have
\begin{align*}
	\left|\sum_{i=1,2}v_{t+1}^i(s)\right|
	=\;&\sum_{i=1,2}\pi_{t,K}^i(s)^\top q_{t,K}^i(s)\\
	=\;&\left|\sum_{i=1,2}\pi_{t,K}^i(s)^\top (q_{t,K}^i(s)-\mathcal{T}^i(v_t^i)(s)\pi_{t,K}^{-i}(s))\right|\\
	&+\left|\sum_{i=1,2}\pi_{t,K}^i(s)\mathcal{T}^i(v_t^i)(s)\pi_{t,K}^{-i}(s)\right|\\
	\leq \;&\sum_{i=1,2}\max_{s\in\mathcal{S}}\| q_{t,K}^i(s)-\mathcal{T}^i(v_t^i)(s)\pi_{t,K}^{-i}(s)\|_\infty\\
	&+\max_{(s,a^i,a^{-i})}\left|\mathcal{T}^i(v_t^i)(s,a^i,a^{-i})+\mathcal{T}^{-i}(v_t^{-i})(s,a^i,a^{-i})\right|\\
	\leq \;&\sum_{i=1,2}\max_{s\in\mathcal{S}}\| q_{t,K}^i(s)-\mathcal{T}^i(v_t^i)(s)\pi_{t,K}^{-i}(s)\|_\infty+\gamma\|v_t^i+v_t^{-i}\|_\infty.
\end{align*}
Since the RHS of the previous inequality does not depend on $s$, we in fact have
\begin{align*}
	\|v_{t+1}^i+v_{t+1}^{-i}\|_\infty\leq \gamma\|v_t^i+v_t^{-i}\|_\infty+\sum_{i=1,2}\max_{s\in\mathcal{S}}\| q_{t,K}^i(s)-\mathcal{T}^i(v_t^i)(s)\pi_{t,K}^{-i}(s)\|_\infty.
\end{align*}

\subsubsection{Proof of Lemma \ref{le:properties_Lyapunov}}\label{pf:le:properties_Lyapunov}
To begin with, observe that
\begin{align*}
	{\arg\max}_{\hat{\mu}^i\in\Delta^{|\mathcal{A}^i|}}\left\{(\hat{\mu}^i)^\top X_i\mu^{-i}+\tau \nu(\hat{\mu}^i)\right\}=\sigma_\tau(X_i\mu^{-i}).
\end{align*}
Therefore, the function $V_X(\cdot,\cdot)$ can be equivalently written as
\begin{align*}
	V_X(\mu^i,\mu^{-i})=\sum_{i=1,2}\left[(\sigma_\tau(X_i\mu^{-i}))^\top X_i\mu^{-i}+\tau \nu(\sigma_\tau(X_i\mu^{-i}))-(\mu^i)^\top X_i \mu^{-i} -\tau \nu(\mu^i)\right],
\end{align*}
which will frequently used in our analysis.
\begin{enumerate}[(1)]
	\item It is clear that the function $V_X(\cdot,\cdot)$ is by definition non-negative. The strong convexity follows from the following two observations.
	\begin{enumerate}[(i)]
		\item The negative entropy $-\nu(\cdot)$ is $1$ -- strongly convex with respect to $\|\cdot\|_2$ \cite[Example 5.27]{beck2017first}.
		\item The following function of $\mu^i$ 
		\begin{align*}
			(\sigma_\tau(X_{-i}\mu^i))^\top X_{-i}\mu^i+\tau \nu(\sigma_\tau(X_{-i}\mu^i))=\max_{\hat{\mu}^{-i}\in\Delta^{|\mathcal{A}^{-i}|}}\left\{(\hat{\mu}^{-i})^\top X_{-i}\mu^i+\tau \nu(\hat{\mu}^{-i})\right\}
		\end{align*}
		is the maximum of linear functions in $\mu^i$, and hence is convex.
	\end{enumerate}
	Therefore, the function $V_X(\mu^i,\mu^{-i})$ is $\tau$ -- strongly convex in $\mu^i$ with respect to $\|\cdot\|_2$ uniformly for all $\mu^{-i}$.
	\item The smoothness follows from the following two results.
	\begin{enumerate}[(i)]
		\item Since the Hessian matrix of the negative entropy function $-\nu(\mu^i)$ satisfies
		\begin{align*}
			H_{-\nu}(\mu^i)=\text{diag}\left(\frac{1}{\mu^i(a_1^i)},\cdots,\frac{1}{\mu^i(a_n^i)}\right)\leq  \frac{I_{|\mathcal{A}^i|}}{\delta_i}
		\end{align*}
		for all $\mu^i\in\Delta^{|\mathcal{A}^i|}$ satisfying $\min_{a^i\in\mathcal{A}^i}\mu^i(a^i)\geq \delta_i$, we have from the second order characterization of smoothness that $-\nu(\mu^i)$ is a $\frac{1}{\delta_i}$ -- smooth function on $\{\mu^i\in\Delta^{|\mathcal{A}^i|}\mid \mu^i(a^i)\geq \delta_i,\forall\;a^i\in\mathcal{A}^i\}$ with respect to $\|\cdot\|_2$ .
		\item 
		Using the optimality condition and we have
		\begin{align*}
			&\nabla_{\mu^i}(\sigma_\tau(X_{-i}\mu^i))^\top X_{-i}\mu^i+\tau \nu(\sigma_\tau(X_{-i}\mu^i))\\
			=\;&
			\nabla_{\mu^i}\max_{\hat{\mu}^{-i}\in\Delta^{|\mathcal{A}^{-i}|}}\left\{(\hat{\mu}^{-i})^\top X_{-i}\mu^i+\tau \nu(\hat{\mu}^{-i})\right\}\\
			=\;&X_{-i}^\top \sigma_\tau(X_{-i}\mu^i).
		\end{align*}
		Therefore, using the formula for the gradient of the softmax function \citep{gao2017properties}, the Hessian $H(\mu^i)$ of the function $(\sigma_\tau(X_{-i}\mu^i))^\top X_{-i}\mu^i+\tau \nu(\sigma_\tau(X_{-i}\mu^i))$ satisfies
		\begin{align*}
			H(\mu^i)=\;&\frac{1}{\tau} X_{-i}^\top (\text{diag}(\sigma_\tau(X_{-i}\mu^i))-\sigma_\tau(X_{-i}\mu^i)\sigma_\tau(X_{-i}\mu^i)^\top )X_{-i}\\
			\leq \;&\frac{1}{\tau} X_{-i}^\top \text{diag}(\sigma_\tau(X_{-i}\mu^i))X_{-i}\\
			\leq \;&\frac{1}{\tau} X_{-i}^\top X_{-i}\\
		    \leq \;&\frac{\sigma^2_{\max}(X_{-i})}{\tau} I_{|\mathcal{A}^i|}
		\end{align*}
		Using the second order characterization of smoothness and we conclude that the function  $(\sigma_\tau(X_{-i}\mu^i))^\top X_{-i}\mu^i+\tau \nu(\sigma_\tau(X_{-i}\mu^i))$ is $\frac{\sigma^2_{\max}(X_{-i})}{\tau}$ -- smooth with respect to $\|\cdot\|_2$.
	\end{enumerate}
	Combining (i) and (ii) and we conclude that the function $V_X(\mu^i,\mu^{-i})$ is a $(\frac{\sigma_{\max}^2(X_{-i})}{\tau}+\frac{\tau}{\delta_i})$ -- smooth function on $\{\mu^i\in\Delta^{|\mathcal{A}^i|}\mid \mu^i(a^i)\geq \delta_i,\forall\;a^i\in\mathcal{A}^i\}$ with respect to $\|\cdot\|_2$ uniformly for all $\mu^{-i}$.
	\item  
	We first compute the gradient $\nabla_1 V_X(\mu^i,\mu^{-i})$ using Danskin's theorem:
	\begin{align}\label{eq:V_gradient}
		\nabla_1 V_X(\mu^i,\mu^{-i})= -(X_i+X_{-i}^\top)\mu^{-i}-\tau \nabla \nu(\mu^i)+X_{-i}^\top \sigma_\tau(X_{-i} \mu^i).
	\end{align}
	It follows that
	\begin{align}
		&\langle \nabla_1V_X(\mu^i,\mu^{-i}),\sigma_\tau(X_i\mu^{-i})-\mu^i \rangle\nonumber\\
		=\;&\langle -(X_i+X_{-i}^\top)\mu^{-i}-\tau \nabla \nu(\mu^i)+X_{-i}^\top \sigma_\tau(X_{-i} \mu^i), \sigma_\tau(X_i\mu^{-i})-\mu^i \rangle\nonumber\\
		=\;&\langle -(X_i+X_{-i}^\top)\mu^{-i}-\tau \nabla \nu(\mu^i)+X_{-i}^\top \sigma_\tau(X_{-i} \mu^i), \sigma_\tau(X_i\mu^{-i})-\mu^i \rangle\nonumber\\
		&+\langle X_i\mu^{-i}+\tau \nabla \nu(\sigma_\tau(X_i \mu^{-i})), \sigma_\tau(X_i\mu^{-i})-\mu^i \rangle\label{eq:oc}\\
		=\;&\tau\langle  \nabla \nu(\sigma_\tau(X_i \mu^{-i}))- \nabla \nu(\mu^i), \sigma_\tau(X_i\mu^{-i})-\mu^i \rangle\nonumber\\
		&+( \sigma_\tau(X_{-i} \mu^i)-\mu^{-i})^\top X_{-i}( \sigma_\tau(X_i\mu^{-i})-\mu^i )\nonumber.
	\end{align}
	where Eq. (\ref{eq:oc}) is due to the optimality condition $X_i\mu^{-i}+\tau \nabla \nu(\sigma_\tau(X_i \mu^{-i}))=0$. To proceed, observe that the concavity of $\nu(\cdot)$ and the optimality condition $X_i\mu^{-i}+\tau \nabla \nu(\sigma_\tau(X_i \mu^{-i}))=0$ together imply that
	\begin{align*}
		&\langle  \nabla \nu(\sigma_\tau(X_i \mu^{-i}))- \nabla \nu(\mu^i), \sigma_\tau(X_i\mu^{-i})-\mu^i \rangle\\
		=\;&\langle   \nabla \nu(\mu^i)-\nabla \nu(\sigma_\tau(X_i \mu^{-i})), \mu^i-\sigma_\tau(X_i\mu^{-i}) \rangle\\
		=\;&\langle   \nabla \nu(\mu^i), \mu^i-\sigma_\tau(X_i\mu^{-i}) \rangle-\langle\nabla \nu(\sigma_\tau(X_i \mu^{-i})), \mu^i-\sigma_\tau(X_i\mu^{-i}) \rangle\\
		\leq \;&\nu(\mu^i)-\nu(\sigma_\tau(X_i\mu^{-i}))-\langle\nabla \nu(\sigma_\tau(X_i \mu^{-i})), \mu^i-\sigma_\tau(X_i\mu^{-i}) \rangle\\
		=\;&\nu(\mu^i)-\nu(\sigma_\tau(X_i\mu^{-i}))+\frac{1}{\tau}\langle X_i\mu^{-i}, \mu^i-\sigma_\tau(X_i\mu^{-i}) \rangle\\
		=\;&\frac{1}{\tau}\left[(\mu^i)^\top X_i \mu^{-i} +\tau \nu(\mu^i)-\max_{\hat{\mu}^i\in\Delta^{|\mathcal{A}^i|}}\left\{(\hat{\mu}^i)^\top X_i\mu^{-i}+\tau \nu(\hat{\mu}^i)\right\}\right].
	\end{align*}
	Therefore, we have
	\begin{align*}
		&\langle \nabla_1V_X(\mu^i,\mu^{-i}),\sigma_\tau(X_i\mu^{-i})-\mu^i \rangle\\
		\leq \;&\left[(\mu^i)^\top X_i \mu^{-i} +\tau \nu(\mu^i)-\max_{\hat{\mu}^i\in\Delta^{|\mathcal{A}^i|}}\left\{(\hat{\mu}^i)^\top X_i\mu^{-i}+\tau \nu(\hat{\mu}^i)\right\}\right]\\
		&+( \sigma_\tau(X_{-i} \mu^i)-\mu^{-i})^\top X_{-i}( \sigma_\tau(X_i\mu^{-i})-\mu^i )
	\end{align*}
	Similarly, we also have 
	\begin{align*}
		&\langle \nabla_2V_X(\mu^i,\mu^{-i}),\sigma_\tau(X_{-i}\mu^i)-\mu^{-i} \rangle\\
		\leq \;&\left[(\mu^{-i})^\top X_{-i} \mu^i +\tau \nu(\mu^{-i})-\max_{\hat{\mu}^{-i}\in\Delta^{|\mathcal{A}^{-i}|}}\left\{(\hat{\mu}^{-i})^\top X_{-i}\mu^i+\tau \nu(\hat{\mu}^{-i})\right\}\right]\\
		&+( \sigma_\tau(X_i \mu^{-i})-\mu^i)^\top X_i( \sigma_\tau(X_{-i}\mu^i)-\mu^{-i} )
	\end{align*}
	Adding up the previous two inequalities and we obtain
	\begin{align}
		&\langle \nabla_1V_X(\mu^i,\mu^{-i}),\sigma_\tau(X_i\mu^{-i})-\mu^i \rangle+\langle \nabla_2V_X(\mu^i,\mu^{-i}),\sigma_\tau(X_{-i}\mu^i)-\mu^{-i} \rangle\nonumber\\
		\leq \;&-V_X(\mu^i,\mu^{-i})+( \sigma_\tau(X_i \mu^{-i})-\mu^i)^\top (X_i+X_{-i}^\top )( \sigma_\tau(X_{-i}\mu^i)-\mu^{-i} ).\label{eq:gradient_V_1}
	\end{align}
	To control the second term on the RHS of Eq. (\ref{eq:gradient_V_1}), observe that
	\begin{align}
		&( \sigma_\tau(X_i \mu^{-i})-\mu^i)^\top (X_i+X_{-i}^\top )( \sigma_\tau(X_{-i}\mu^i)-\mu^{-i})\nonumber\\
		\leq\;& \|\sigma_\tau(X_i \mu^{-i})-\mu^i\|_2 \|X_i+X_{-i}^\top \|_2\| \sigma_\tau(X_{-i}\mu^i)-\mu^{-i}\|_2\nonumber\\
		\leq\;& (\|\sigma_\tau(X_i \mu^{-i})\|_2+\|\mu^i\|_2) \|X_i+X_{-i}^\top \|_2\| \sigma_\tau(X_{-i}\mu^i)-\mu^{-i}\|_2\nonumber\\
		\leq \;&2\|X_i+X_{-i}^\top \|_2\| \sigma_\tau(X_{-i}\mu^i)-\mu^{-i}\|_2\nonumber\\
		\leq \;&c_1\|X_i+X_{-i}^\top \|_2^2+\frac{1}{c_1}\| \sigma_\tau(X_{-i}\mu^i)-\mu^{-i}\|_2^2\tag{This is true for all $c_1>0$}\nonumber\\
		\leq \;&c_1\|X_i+X_{-i}^\top \|_2^2+\frac{1}{c_1}(\| \sigma_\tau(X_{-i}\mu^i)-\mu^{-i}\|_2^2+\|\sigma_\tau(X_i \mu^{-i})-\mu^i\|_2^2).\label{eq:gradientV12}
	\end{align}
	To proceed, note that the function
	\begin{align*}
	    F_{X_i}(\mu^i,\mu^{-i}):=\max_{\hat{\mu}^i}\left\{(\hat{\mu}^i-\mu^i)^\top X_i\mu^{-i}+\tau \nu(\hat{\mu}^i)-\tau\nu(\mu^i)\right\}
	\end{align*}
	is a $\tau$-strongly convex function of $\mu^i$ uniformly for all $\mu^{-i}$. Therefore, we have
	\begin{align*}
	    F_{X_i}(\mu^i,\mu^{-i})=\;&F_{X_i}(\mu^i,\mu^{-i})-F_{X_i}(\sigma_\tau(X_i\mu^{-i}),\mu^{-i})\\
	    =\;&F_{X_i}(\mu^i,\mu^{-i})-\min_{\mu^i}F_{X_i}(\mu^i,\mu^{-i})\\
	    \geq  \;&\frac{\tau}{2}\|\sigma_\tau(X_i\mu^{-i})-\mu^i\|_2^2,
	\end{align*}
	which is called the quadratic growth property in optimization literature. It follows that
	\begin{align*}
	    \|\sigma_\tau(X_i\mu^{-i})-\mu^i\|_2^2\leq \frac{2}{\tau }\max_{\hat{\mu}^i}\left\{(\hat{\mu}^i-\mu^i)^\top X_i\mu^{-i}+\tau \nu(\hat{\mu}^i)-\tau\nu(\mu^i)\right\}.
	\end{align*}
	Similarly, we also have
	\begin{align*}
	    \|\sigma_\tau(X_{-i}\mu^i)-\mu^{-i}\|_2^2\leq \frac{2}{\tau }\max_{\hat{\mu}^{-i}}\left\{(\hat{\mu}^{-i}-\mu^{-i})^\top X_{-i}\mu^i+\tau \nu(\hat{\mu}^{-i})-\tau\nu(\mu^{-i})\right\}.
	\end{align*}
	Adding up the previous two inequalities and we have
	\begin{align*}
	    \|\sigma_\tau(X_{-i}\mu^i)-\mu^{-i}\|_2^2+\|\sigma_\tau(X_i\mu^{-i})-\mu^i\|_2^2\leq \frac{2}{\tau}V_X(\mu^i,\mu^{-i}).
	\end{align*}
	Using the previous inequality in Eq. (\ref{eq:gradientV12}) and we have
	\begin{align*}
		&( \sigma_\tau(X_i \mu^{-i})-\mu^i)^\top (X_i+X_{-i}^\top )( \sigma_\tau(X_{-i}\mu^i)-\mu^{-i})\\
		\leq \;&c_1\|X_i+X_{-i}^\top \|_2^2+\frac{1}{c_1}(\| \sigma_\tau(X_{-i}\mu^i)-\mu^{-i}\|_2^2+\|\sigma_\tau(X_i \mu^{-i})-\mu^i\|_2^2)\\
		\leq \;&c_1\|X_i+X_{-i}^\top \|_2^2+\frac{2}{c_1\tau}V_X(\mu^i,\mu^{-i})\\
		= \;&\frac{16}{\tau}\|X_i+X_{-i}^\top \|_2^2+\frac{1}{8}V_X(\mu^i,\mu^{-i}),
	\end{align*}
	where the last line follows from choosing $c_1=16/\tau$. Using the previous inequality in Eq. (\ref{eq:gradient_V_1}) and we obtain
	\begin{align*}
		&\langle \nabla_1V_X(\mu^i,\mu^{-i}),\sigma_\tau(X_i\mu^{-i})-\mu^i \rangle+\langle \nabla_2V_X(\mu^i,\mu^{-i}),\sigma_\tau(X_{-i}\mu^i)-\mu^{-i} \rangle\\
		\leq \;&-\frac{7}{8}V_X(\mu^i,\mu^{-i})+\frac{16}{\tau}\|X_i+X_{-i}^\top \|_2^2.
	\end{align*}
	\item For any $u^i\in\mathbb{R}^{|\mathcal{A}^i|}$, using the explicit expression of the gradient of $V_X(\cdot,\cdot)$ from Eq. (\ref{eq:V_gradient}) and we have
	\begin{align*}
		&\langle \nabla_1V_X(\mu^i,\mu^{-i}),\sigma_\tau(u^i)-\sigma_\tau(X_i\mu^{-i})\rangle\\
		=\;&\langle -(X_i+X_{-i}^\top)\mu^{-i}-\tau \nabla \nu(\mu^i)+X_{-i}^\top \sigma_\tau(X_{-i} \mu^i),\sigma_\tau(u^i)-\sigma_\tau(X_i\mu^{-i})\rangle\\
		=\;&\langle -(X_i+X_{-i}^\top)\mu^{-i}-\tau \nabla \nu(\mu^i)+X_{-i}^\top \sigma_\tau(X_{-i} \mu^i), \sigma_\tau(u^i)-\sigma_\tau(X_i\mu^{-i}) \rangle\\
		&+\langle X_i\mu^{-i}+\tau \nabla \nu(\sigma_\tau(X_i \mu^{-i})), \sigma_\tau(u^i)-\sigma_\tau(X_i\mu^{-i}) \rangle\\
		=\;&\tau\langle  \nabla \nu(\sigma_\tau(X_i \mu^{-i}))- \nabla \nu(\mu^i), \sigma_\tau(u^i)-\sigma_\tau(X_i\mu^{-i}) \rangle\\
		&+( \sigma_\tau(X_{-i} \mu^i)-\mu^{-i})^\top X_{-i}( \sigma_\tau(u^i)-\sigma_\tau(X_i\mu^{-i}) )\\
		\leq \;&\tau\|\nabla \nu(\sigma_\tau(X_i \mu^{-i}))- \nabla \nu(\mu^i)\|_2 \|\sigma_\tau(u^i)-\sigma_\tau(X_i\mu^{-i}) \|_2\\
		&+\|\sigma_\tau(X_{-i} \mu^i)-\mu^{-i}\|_2 \|X_{-i}\|_2\| \sigma_\tau(u^i)-\sigma_\tau(X_i\mu^{-i})\|_2 \\
		\leq \;&\frac{\tau}{\delta_i}\|\sigma_\tau(X_i \mu^{-i})- \mu^i\|_2 \|\sigma_\tau(u^i)-\sigma_\tau(X_i\mu^{-i}) \|_2\\
		&+ \|X_{-i}\|_2\|\sigma_\tau(X_{-i} \mu^i)-\mu^{-i}\|_2\| \sigma_\tau(u^i)-\sigma_\tau(X_i\mu^{-i})\|_2,
	\end{align*}
	where the last inequality follows from the smoothness of $\nu(\cdot)$ in Lemma \ref{le:properties_Lyapunov} (2).
	Similarly, we also have for any $u^{-i}\in\mathbb{R}^{|\mathcal{A}^{-i}|}$ that 
	\begin{align*}
		&\langle \nabla_2V_X(\mu^i,\mu^{-i}),\sigma_\tau(u^{-i})-\sigma_\tau(X_{-i}\mu^i) \rangle\\
		\leq \;&\frac{\tau}{\delta_{-i}} \|\sigma_\tau(X_{-i} \mu^i)- \mu^{-i}\|_2 \|\sigma_\tau(u^{-i})-\sigma_\tau(X_{-i}\mu^i) \|_2\\
		&+ \|X_i\|_2\|\sigma_\tau(X_i \mu^{-i})-\mu^i\|_2\| \sigma_\tau(u^{-i})-\sigma_\tau(X_i\mu^i)\|_2.
	\end{align*}
	Adding up the previous two inequalities and we have
	\begin{align*}
		&\langle \nabla_1V_X(\mu^i,\mu^{-i}),\sigma_\tau(u^i)-\sigma_\tau(X_i\mu^{-i})\rangle+\langle \nabla_2V_X(\mu^i,\mu^{-i}),\sigma_\tau(u^{-i})-\sigma_\tau(X_{-i}\mu^i)\rangle\\
		\leq \;&\left(\frac{\tau}{\delta_i}+\frac{\tau}{\delta_{-i}}+\|X_i\|_2+\|X_{-i}\|_2\right)\left(\|\sigma_\tau(X_i \mu^{-i})- \mu^i\|_2+\|\sigma_\tau(X_{-i} \mu^i)-\mu^{-i}\|_2\right)\\
		&\times \left(\|\sigma_\tau(u^i)-\sigma_\tau(X_i\mu^{-i}) \|_2+\| \sigma_\tau(u^{-i})-\sigma_\tau(X_i\mu^i)\|_2\right)\\
		\leq \;&\frac{1}{2}\left(\frac{\tau}{\delta_i}+\frac{\tau}{\delta_{-i}}+\|X_i\|_2+\|X_{-i}\|_2\right)\bigg[\Bar{c}\left(\|\sigma_\tau(X_i \mu^{-i})- \mu^i\|_2+\|\sigma_\tau(X_{-i} \mu^i)-\mu^{-i}\|_2\right)^2\\
		&+\frac{1}{\Bar{c}}\left(\|\sigma_\tau(u^i)-\sigma_\tau(X_i\mu^{-i}) \|_2+\| \sigma_\tau(u^{-i})-\sigma_\tau(X_i\mu^i)\|_2\right)^2\bigg]\tag{This is true for all $\Bar{c}>0$}\\
		\leq \;&\left(\frac{\tau}{\delta_i}+\frac{\tau}{\delta_{-i}}+\|X_i\|_2+\|X_{-i}\|_2\right)\bigg[\Bar{c}\|\sigma_\tau(X_i \mu^{-i})- \mu^i\|_2^2+\Bar{c}\|\sigma_\tau(X_{-i} \mu^i)-\mu^{-i}\|_2^2\\
		&+\frac{1}{\Bar{c}}\|\sigma_\tau(u^i)-\sigma_\tau(X_i\mu^{-i}) \|_2^2+\frac{1}{\Bar{c}}\| \sigma_\tau(u^{-i})-\sigma_\tau(X_i\mu^i)\|_2^2\bigg]\tag{This is true because $(a+b)^2\leq 2(a^2+b^2)$ for all $a,b\in\mathbb{R}$}\\
		\leq \;&\left(\frac{\tau}{\delta_i}+\frac{\tau}{\delta_{-i}}+\|X_i\|_2+\|X_{-i}\|_2\right)\bigg[\frac{2\Bar{c}}{\tau}V_X(\mu^i,\mu^{-i})+\frac{1}{\Bar{c}\tau^2}\|u^i-X_i\mu^{-i} \|_2^2\\
		&+\frac{1}{\Bar{c}\tau^2}\| u^{-i}-X_i\mu^i\|_2^2\bigg],
	\end{align*}
	where the last line follows from the quadratic growth property of strongly convex functions and the Lipschitz continuity of the softmax function \citep{gao2017properties}.
\end{enumerate}

\subsubsection{Proof of Lemma \ref{le:policy_drift}}\label{pf:le:policy_drift}

Since $\min_{i=1,2}\min_{s,a^i}\pi_k^i(a^i|s)\geq \ell_\tau$ (cf. Lemma \ref{le:margin}), Lemma \ref{le:properties_Lyapunov} (2) implies that the function $V_{v,s}(\mu^i,\mu^{-i})$ as a function of $\mu^i$ is $L_{\tau,i}$ -- smooth on $\{\mu^i\in\Delta^{|\mathcal{A}^i|}\mid \min_{a^i}\mu^i(a^i)\geq \ell_\tau\}$ uniformly for all $\mu^{-i}$, where
\begin{align*}
    L_{\tau,i}:= \;&\frac{\sigma_{\max}^2(\mathcal{T}^{-i}(v^{-i})(s))}{\tau}+\frac{\tau}{\ell_{\tau}}.
\end{align*}
We next bound $L_{\tau,i}$ from above.
Since $\|v^i\|_\infty\leq 1/(1-\gamma)$ and $\|v^{-i}\|_\infty\leq 1/(1-\gamma)$, we have for any $(s,a^i,a^{-i})$ that
\begin{align*}
    |\mathcal{T}^{-i}(v^{-i})(s,a^{-i},a^i)|\leq \;&|\mathcal{R}^{-i}(s,a^{-i},a^i)|+\gamma \mathbb{E}[|v^{-i}(S_1)|\mid S_0=s,A_0^i=a^i,A_0^{-i}=a^{-i}]\\
    \leq \;&1+\frac{\gamma}{1-\gamma}\\
    =\;&\frac{1}{1-\gamma},
\end{align*}
which implies
\begin{align}\label{eqeq:cici}
    \sigma_{\max}(\mathcal{T}^{-i}(v^{-i})(s))=\;&\|\mathcal{T}^{-i}(v^{-i})(s))\|_2\leq \frac{\sqrt{|\mathcal{A}^i||\mathcal{A}^{-i}|}}{1-\gamma}\leq \frac{A_{\max}}{1-\gamma}.
\end{align}
As a result, we have by $\tau\leq 1$ and $\ell_\tau\leq 1$ that
\begin{align*}
    L_{\tau,i}= \frac{\sigma_{\max}^2(\mathcal{T}^{-i}(v^{-i})(s))}{\tau}+\frac{\tau}{\ell_{\tau}}\leq \frac{A_{\max}^2}{\tau(1-\gamma)^2}+\frac{\tau}{\ell_{\tau}}\leq \frac{2A_{\max}^2}{\ell_{\tau}(1-\gamma)^2}:=L_\tau.
\end{align*}
Similarly, $V_{v,s}(\mu^i,\mu^{-i})$ as a function of $\mu^{-i}$ is also $L_\tau$ -- smooth on the set $\{\mu^{-i}\in\Delta^{|\mathcal{A}^{-i}|}\mid \min_{a^{-i}}\mu^i(a^{-i})\geq \ell_\tau\}$ uniformly for all $\mu^i$.

Using the smoothness of $V_{v,s}(\cdot,\cdot)$ established above, for any $s\in\mathcal{S}$, we have by the policy update equation (cf. Algorithm \ref{algorithm:inner-loop} Line $3$) that  
\begin{align}
	&V_{v,s}(\pi_{k+1}^i(s),\pi_{k+1}^{-i}(s))\nonumber\\
	\leq \;&V_{v,s}(\pi_k^i(s),\pi_k^{-i}(s))+\beta_k\langle \nabla_2V_{v,s}(\pi_k^i(s),\pi_k^{-i}(s)),\sigma_\tau(q_k^{-i}(s))-\pi_k^{-i}(s) \rangle\nonumber\\
	&+\beta_k\langle \nabla_1V_{v,s}(\pi_k^i(s),\pi_{k+1}^{-i}(s)),\sigma_\tau(q_k^i(s))-\pi_k^i(s) \rangle\nonumber\\
	&+\frac{L_\tau\beta_k^2}{2}\|\sigma_\tau(q_k^i(s))-\pi_k^i(s)\|_2^2+\frac{L_\tau\beta_k^2}{2}\|\sigma_\tau(q_k^{-i}(s))-\pi_k^{-i}(s)\|_2^2\nonumber\\
	\leq \;&V_{v,s}(\pi_k^i(s),\pi_k^{-i}(s))+\underbrace{\beta_k\langle \nabla_2V_{v,s}(\pi_k^i(s),\pi_k^{-i}(s)),\sigma_\tau(\mathcal{T}^{-i}(v^{-i})(s)\pi_k^i(s))-\pi_k^{-i}(s) \rangle}_{\hat{N}_1}\nonumber\\
	&+\underbrace{\beta_k\langle \nabla_1V_{v,s}(\pi_k^i(s),\pi_{k+1}^{-i}(s)),\sigma_\tau(\mathcal{T}^i(v^i)(s)\pi_k^{-i}(s))-\pi_k^i(s) \rangle}_{\hat{N}_2}\nonumber\\
	&+\underbrace{\beta_k\langle \nabla_2V_{v,s}(\pi_k^i(s),\pi_k^{-i}(s)),\sigma_\tau(q_k^{-i}(s))-\sigma_\tau(\mathcal{T}^{-i}(v^{-i})(s)\pi_k^i(s)) \rangle}_{\hat{N}_3}\nonumber\\
	&+\underbrace{\beta_k\langle \nabla_1V_{v,s}(\pi_k^i(s),\pi_{k+1}^{-i}(s)),\sigma_\tau(q_k^i(s))-\sigma_\tau(\mathcal{T}^i(v^i)(s)\pi_k^{-i}(s)) \rangle}_{\hat{N}_4}\nonumber\\
	&+2L_\tau\beta_k^2.\label{eq:policy_overall}
\end{align}
We next bound the terms $\{\hat{N}_j\}_{1\leq j\leq 4}$ on the RHS of Eq. (\ref{eq:policy_overall}) using Lemma \ref{le:properties_Lyapunov} (3) and (4). 

First consider $\hat{N}_1+\hat{N}_2$. We have by Lemma \ref{le:properties_Lyapunov} (3) that
\begin{align*}
	\hat{N}_1+\hat{N}_2
	\leq \;&-\frac{7\beta_k}{8} V_{v,s}(\pi_k^i(s),\pi_k^{-i}(s))
	+\frac{16 \beta_k}{\tau}\| \mathcal{T}^i(v^i)(s)+\mathcal{T}^{-i}(v^{-i}(s)^\top \|_2^2.
\end{align*}
To proceed, note that for any $\mu^{-i}\in \mathbb{R}^{|\mathcal{A}^{-i}|}$ satisfying $\|\mu^{-i}\|_2=1$, we have
\begin{align*}
	&\| (\mathcal{T}^i(v^i)(s)+\mathcal{T}^{-i}(v^{-i}(s)^\top)\mu^i \|_2^2\\
	=\;&\gamma^2 \sum_{a^i}\left[\sum_{a^{-i}} \mathbb{E}[v^i(S_1)+v^{-i}(S_1)\mid S_0=s,A_0^i=a^i,A_0^{-i}=a^{-i}]\mu^{-i}(a^{-i})\right]^2\\
	\leq \;&\gamma^2\|v^i+v^{-i}\|_\infty^2 \sum_{a^i}\left[\sum_{a^{-i}} \mu^{-i}(a^{-i})\right]^2\\
	\leq \;&\gamma^2A_{\max}\|v^i+v^{-i}\|_\infty^2,
\end{align*}
which implies 
\begin{align*}
	\| \mathcal{T}^i(v^i)(s)+\mathcal{T}^{-i}(v^{-i}(s)^\top \|_2^2\leq \gamma^2A_{\max}\|v^i+v^{-i}\|_\infty^2.
\end{align*}
It follows that
\begin{align*}
	\hat{N}_1+\hat{N}_2
	\leq \;&-\frac{7\beta_k}{8} V_{v,s}(\pi_k^i(s),\pi_k^{-i}(s))
	+\frac{16A_{\max} \beta_k}{\tau}\|v^i+v^{-i}\|_\infty^2.
\end{align*}

We next consider $\hat{N}_3+\hat{N}_4$. Since
\begin{align*}
	\max(\|\mathcal{T}^i(v^i)(s)\|_2 ,\|\mathcal{T}^{-i}(v^{-i})(s)\|_2)\leq \frac{A_{\max}}{1-\gamma},\tag{See Eq. (\ref{eqeq:cici})}
\end{align*}
we have by Lemma \ref{le:properties_Lyapunov} (4) that
\begin{align*}
	&\hat{N}_3+\hat{N}_4\\
	\leq\;& 2\beta_k\left(\frac{\tau}{\ell_{\tau}}+\frac{A_{\max}}{1-\gamma}\right)\bigg[\frac{2\Bar{c}}{\tau}V_{v,s}(\pi_k^i(s),\pi_k^{-i}(s))\\
	&+\frac{1}{\Bar{c}\tau^2}\|q_k^i(s)-\mathcal{T}^i(v^i)(s)\pi_k^{-i}(s) \|_2^2+\frac{1}{\Bar{c}\tau^2}\| q_k^{-i}(s)-\mathcal{T}^{-i}(v^{-i})(s)\pi_k^i(s)\|_2^2\bigg]
\end{align*}
for any $\Bar{c}>0$.
By choosing $\Bar{c}=\frac{\tau}{32}(\frac{\tau}{\ell_{\tau}}+\frac{A_{\max}}{1-\gamma})^{-1}$,
we have from the previous inequality that
\begin{align*}
	\hat{N}_3+\hat{N}_4
	\leq\;& \frac{\beta_k}{8} V_{v,s}(\pi_k^i(s),\pi_k^{-i}(s))+\frac{64\beta_k\left(\frac{\tau}{\ell_{\tau}}+\frac{A_{\max}}{1-\gamma}\right)^2}{\tau^3}\|q_k^i(s)-\mathcal{T}^i(v^i)(s)\pi_k^{-i}(s) \|_2^2\\
	&+\frac{64\beta_k\left(\frac{\tau}{\ell_{\tau}}+\frac{A_{\max}}{1-\gamma}\right)^2}{\tau^3}\| q_k^{-i}(s)-\mathcal{T}^{-i}(v^{-i})(s)\pi_k^i(s)\|_2^2\\
	\leq\;& \frac{\beta_k}{8} V_{v,s}(\pi_k^i(s),\pi_k^{-i}(s))+\frac{256A_{\max}^2\beta_k}{\ell_{\tau}^2\tau^3(1-\gamma)^2}\|q_k^i(s)-\mathcal{T}^i(v^i)(s)\pi_k^{-i}(s) \|_2^2\\
	&+\frac{256A_{\max}^2\beta_k}{\ell_{\tau}^2\tau^3(1-\gamma)^2}\| q_k^{-i}(s)-\mathcal{T}^{-i}(v^{-i})(s)\pi_k^i(s)\|_2^2.
\end{align*}

Finally, using the upper bounds we obtained for the terms $\hat{N}_1+\hat{N}_2$ and $\hat{N}_3+\hat{N}_4$ in Eq. (\ref{eq:policy_overall}) and we have
\begin{align*}
	&V_{v,s}(\pi_{k+1}^i(s),\pi_{k+1}^{-i}(s))\\
	\leq \;&\left(1-\frac{3\beta_k}{4}\right)V_{v,s}(\pi_k^i(s),\pi_k^{-i}(s))+\frac{16A_{\max} \beta_k}{\tau}\|v^i+v^{-i}\|_\infty^2.\\
	&+\frac{256A_{\max}^2\beta_k}{\ell_{\tau}^2\tau^3(1-\gamma)^2}\|q_k^i(s)-\mathcal{T}^i(v^i)(s)\pi_k^{-i}(s) \|_2^2\\
	&+\frac{256A_{\max}^2\beta_k}{\ell_{\tau}^2\tau^3(1-\gamma)^2}\| q_k^{-i}(s)-\mathcal{T}^{-i}(v^{-i})(s)\pi_k^i(s)\|_2^2\\
	&+\frac{4A_{\max}^2}{\ell_{\tau}(1-\gamma)^2}\beta_k^2.
\end{align*}
Summing up both sides of the previous inequality for all $s$ and then taking expectation, and we have the desired result.

\subsubsection{Proof of Lemma \ref{le:operators}}\label{pf:le:operators}
\begin{enumerate}[(1)]
	\item For any $(q_1^i,q_2^i)$ and $(s_0,a_0^i,a_0^{-i},s_1)$, we have
	\begin{align*}
		&\|F^i(q_1^i,s_0,a_0^i,a_0^{-i},s_1)-F^i(q_2^i,s_0,a_0^i,a_0^{-i},s_1)\|_2^2\\
		=\;&\sum_{(s,a^i)}([F^i(q_1^i,s_0,a_0^i,a_0^{-i},s_1)](s,a^i)-[F^i(q_2^i,s_0,a_0^i,a_0^{-i},s_1)](s,a^i))^2\\
		= \;&\sum_{(s,a^i)}\mathds{1}_{\{(s,a^i)=(s_0,a_0^i)\}}\left(q_1^i(s_0,a_0^i)-q_2^i(s_0,a_0^i)\right)^2\\
		\leq \;&\|q_1^i-q_2^i\|_2^2.
	\end{align*}
	\item 
	For any $(s_0,a_0^i,a_0^{-i},s_1)$, we have
	\begin{align*}
		\|F^i(\bm{0},s_0,a_0^i,a_0^{-i},s_1)\|_2^2
		=\;&\sum_{(s,a^i)}([F^i(\bm{0},s_0,a_0^i,a_0^{-i},s_1)](s,a^i))^2\\
		=\;&\sum_{(s,a^i)}\mathds{1}_{\{(s,a^i)=(s_0,a_0^i)\}}\left(\mathcal{R}^i(s_0,a_0^i,a_0^{-i})+\gamma v^i(s_1)\right)^2\\
		\leq \;&\frac{1}{(1-\gamma)^2},
	\end{align*}
	where the last line follows from $\|v^i\|_\infty\leq 1/(1-\gamma)$ and $|\mathcal{R}^i(s_0,a_0^i,a_0^{-i})|\leq 1$.
	\item We first write down the explicit expression of $\Bar{F}_k^i(\cdot)$. Using the definition of $\mathcal{T}^i(\cdot)$ and we have
	\begin{align*}
		\Bar{F}_k^i(q^i)(s)=\mu_k(s)\Pi_k^i(s)\left(\mathcal{T}^i(v^i)(s)\pi_k^{-i}(s)-q^i(s)\right),\;\forall\;s\in\mathcal{S},
	\end{align*}
	where $\Pi_k^i(s):=\text{diag}(\pi_k^i(s))$. Since $\min_{0\leq k\leq K-1}\min_{s\in\mathcal{S}}\mu_k(s)>0$ (cf. Lemma \ref{le:margin} and Lemma \ref{thm:exploration} (4)) and $\Pi_k^i(s)$ has strictly positive diagonal entries for all $s$ and $k$ (cf. Lemma \ref{le:margin}),
	the equation $\bar{F}_k^i(q^i)=0$ has a unique solution $\Bar{q}_k^i\in\mathbb{R}^{|\mathcal{S}||\mathcal{A}^i|}$, which is explicitly given by
	\begin{align*}
		\bar{q}_k^i(s)=\mathcal{T}^i(v^i)(s)\pi_k^{-i}(s),\;\forall\;s\in\mathcal{S}.
	\end{align*}
	\item 
	Using the explicit expression of $\bar{F}_k^i(\cdot)$ and we have for any $q_1^i$ and $q_2^i$ that
	\begin{align*}
		\langle \Bar{F}_k^i(q_1^i)-\Bar{F}_k^i(q_2^i),q_1^i-q_2^i\rangle=\;&-\sum_{s,a^i}\mu_k(s)\pi_k^i(a^i|s)(q_1^i(s,a^i)-q_2^i(s,a^i))^2\\
		\leq  \;&-\min_{s,a^i}\mu_k(s)\pi_k^i(a^i|s)\|q_1^i-q_2^i\|_2^2\\
		\leq  \;&-\mu_\tau\ell_{\tau}\|q_1^i-q_2^i\|_2^2\tag{Lemma \ref{thm:exploration} (4) and Lemma \ref{le:margin}}\\
		= \;&-c_\tau\|q_1^i-q_2^i\|_2^2,
	\end{align*}
	where we recall that $c_\tau=\mu_\tau \ell_\tau$.
	
\end{enumerate}

\subsubsection{Proof of Lemma \ref{le:noise}}\label{pf:le:noise}

For any $k\geq 0$, we have
\begin{align*}
	N_2=\;&\mathbb{E}[\langle F^i(q_k^i,S_k,A_k^i,A_k^{-i},S_{k+1})-\bar{F}_k^i(q_k^i),q_k^i-\bar{q}_k^i\rangle]\\
	= \;&\underbrace{\mathbb{E}[\langle F^i(q_{k-z_k}^i,S_k,A_k^i,A_k^{-i},S_{k+1})-\bar{F}_{k-z_k}^i(q_{k-z_k}^i),q_{k-z_k}^i-\bar{q}_{k-z_k}^i\rangle]}_{N_{2,1}}\\
	&+\underbrace{\mathbb{E}[\langle F^i(q_{k-z_k}^i,S_k,A_k^i,A_k^{-i},S_{k+1})-\bar{F}_{k-z_k}^i(q_{k-z_k}^i),q_k^i-q_{k-z_k}^i\rangle]}_{N_{2,2}}\\
	&+\underbrace{\mathbb{E}[\langle F^i(q_{k-z_k}^i,S_k,A_k^i,A_k^{-i},S_{k+1})-\bar{F}_{k-z_k}^i(q_{k-z_k}^i),\bar{q}_{k-z_k}^i-\bar{q}_k^i\rangle]}_{N_{2,3}}\\
	&+\underbrace{\mathbb{E}[\langle F^i(q_k^i,S_k,A_k^i,A_k^{-i},S_{k+1})-F^i(q_{k-z_k}^i,S_k,A_k^i,A_k^{-i},S_{k+1}),q_k^i-\bar{q}_k^i\rangle]}_{N_{2,4}}\\
	&+\underbrace{\mathbb{E}[\langle \bar{F}_{k-z_k}^i(q_{k-z_k}^i)-\bar{F}_k^i(q_k^i),q_k^i-\bar{q}_k^i\rangle]}_{N_{2,5}}\\
\end{align*}
We next control the terms $\{N_{2,j}\}_{1\leq j \leq 5}$ on the RHS of the previous inequality. Before that, the following two lemmas are needed. The proof of Lemma \ref{le:difference} follows from that of \cite[Lemma 3]{srikant2019finite} and \cite[Lemma A.3]{chen2021finite}. Lemma \ref{le:difference_pi} is the policy-counterpart of Lemma \ref{le:difference}.

\begin{lemma}[Proof in Appendix \ref{pf:le:difference}]\label{le:difference}
	Given positive integers $k_1\leq k_2$ satisfying $\alpha_{k_1,k_2-1}\leq 1/4$, we have for any $k\in \{k_1,k_1+1,\cdots,k_2\}$ that
	\begin{align*}
		\|q_k^i-q_{k_1}^i\|_2\leq \;& \min(2\alpha_{k_1,k_2-1},1/2)\left(\|q_{k_1}^i\|_2+\frac{1}{1-\gamma}\right),\\
		\|q_k^i-q_{k_1}^i\|_2\leq \;& \min(4\alpha_{k_1,k_2-1},1)\left(\|q_{k_2}^i\|_2+\frac{1}{1-\gamma}\right).
	\end{align*}
\end{lemma}

\begin{lemma}[Proof in Appendix \ref{pf:le:difference_pi}]\label{le:difference_pi}
	Given positive integers $k_1\leq k_2$ satisfying $\beta_{k_1,k_2-1}\leq 1/4$, we have for any $s\in\mathcal{S}$ and $k\in \{k_1,k_1+1,\cdots,k_2\}$ that
	\begin{align*}
		\|\pi_k^i(s)-\pi_{k_1}^i(s)\|_2\leq \;& \min(2\beta_{k_1,k_2-1},1/2)\left(\|\pi_{k_1}^i(s)\|_2+1\right),\\
		\|\pi_k^i(s)-\pi_{k_1}^i(s)\|_2\leq \;& \min(4\beta_{k_1,k_2-1},1)\left(\|\pi_{k_2}^i(s)\|_2+1\right).
	\end{align*}
\end{lemma}

We next bound the terms $\{N_{2,j}\}_{1\leq j\leq 5}$. Let $\mathcal{F}_k$ be the $\sigma$-algebra generated the sequence of random variables $\{S_0,A_0^i,A_0^{-i},\cdots,S_{k-1},A_{k-1}^i,A_{k-1}^{-i},S_k\}$.

\paragraph{The Term $N_{2,1}$.}
Using the tower property of conditional expectations and we have
\begin{align*}
	&N_{2,1}\\
	=\;&\mathbb{E}[\langle F^i(q_{k-z_k}^i,S_k,A_k^i,A_k^{-i},S_{k+1})-\bar{F}_{k-z_k}^i(q_{k-z_k}^i),q_{k-z_k}^i-\bar{q}_{k-z_k}^i\rangle]\\
	=\;&\mathbb{E}[\langle \mathbb{E}[F^i(q_{k-z_k}^i,S_k,A_k^i,A_k^{-i},S_{k+1})\mid \mathcal{F}_{k-z_k}]-\bar{F}_{k-z_k}^i(q_{k-z_k}^i),q_{k-z_k}^i-\bar{q}_{k-z_k}^i\rangle]\\
	\leq \;&\mathbb{E}[\| \mathbb{E}[F^i(q_{k-z_k}^i,S_k,A_k^i,A_k^{-i},S_{k+1})\mid \mathcal{F}_{k-z_k}]-\bar{F}_{k-z_k}^i(q_{k-z_k}^i)\|_2\|q_{k-z_k}^i-\bar{q}_{k-z_k}^i\|_2]\\
	\leq \;&\frac{2\sqrt{|\mathcal{S}|A_{\max}}}{1-\gamma}\mathbb{E}[\underbrace{\|F^i(q_{k-z_k}^i,S_k,A_k^i,A_k^{-i},S_{k+1})\mid \mathcal{F}_{k-z_k}]-\bar{F}_{k-z_k}^i(q_{k-z_k}^i)\|_2}_{N_{2,1,1}}],
\end{align*}
where the last line follows from $\|q_{k-z_k}^i\|_2\leq \sqrt{|\mathcal{S}|A_{\max}}\|q_{k-z_k}^i\|_\infty\leq \sqrt{|\mathcal{S}|A_{\max}}/(1-\gamma)$ and similarly $\|\Bar{q}_{k-z_k}^i\|_2\leq \sqrt{|\mathcal{S}|A_{\max}}/(1-\gamma)$.
For the term $N_{2,1,1}$, using triangle inequality and we have
\begin{align}
	&\|F^i(q_{k-z_k}^i,S_k,A_k^i,A_k^{-i},S_{k+1})\mid \mathcal{F}_{k-z_k}]-\bar{F}_{k-z_k}^i(q_{k-z_k}^i)\|_2\nonumber\\
	\leq \;&\|\bar{F}_k^i(q_{k-z_k}^i)\!-\!\bar{F}_{k-z_k}^i(q_{k-z_k}^i)\|_2\nonumber\\
	&+\|F^i(q_{k-z_k}^i,S_k,A_k^i,A_k^{-i},S_{k+1})\mid \mathcal{F}_{k-z_k}]\!-\!\bar{F}_k^i(q_{k-z_k}^i)\|_2.\label{eq:N211}
\end{align}
To control the first term on the RHS of Eq. (\ref{eq:N211}), recall that
\begin{align*}
	\Bar{F}_k^i(q^i)(s)=\mu_k(s)\Pi_k^i(s) \left(\mathcal{T}^i(v^i)(s)\pi_k^{-i}(s)-q^i(s)\right),\;\forall\;s\in\mathcal{S},\;q^i\in\mathbb{R}^{|\mathcal{S}| |\mathcal{A}^i|},
\end{align*}
where $\Pi_k^i(s)=\text{diag}(\pi_k^i(s))$.
Therefore, we have for any $s\in\mathcal{S}$ and $q^i\in\mathbb{R}^{|\mathcal{S}| |\mathcal{A}^i|}$ that
\begin{align}
	&\|\bar{F}_k^i(q^i)(s)-\bar{F}_{k-z_k}^i(q^i)(s)\|_2\nonumber\\
	\leq \;&\|\mu_k(s)\Pi_k^i(s) \mathcal{T}^i(v^i)(s)\pi_k^{-i}(s)-\mu_{k-z_k}(s)\Pi_{k-z_k}^i(s) \mathcal{T}^i(v^i)(s)\pi_{k-z_k}^{-i}(s)\|_2\nonumber\\
	&+\|(\mu_k(s)\Pi_k^i(s) -\mu_{k-z_k}(s)\Pi_{k-z_k}^i(s)) q^i(s)\|_2.\label{eq:N2112}
\end{align}
We next bound the two terms on the RHS of Eq. (\ref{eq:N2112}). For the first term, we have
\begin{align*}
	&\|(\mu_k(s)\Pi_k^i(s) -\mu_{k-z_k}(s)\Pi_{k-z_k}^i(s)) q^i(s)\|_2\\
	\leq \;&\frac{A_{\max}^{1/2}}{1-\gamma}\|\mu_k(s)\Pi_k^i(s) -\mu_{k-z_k}(s)\Pi_{k-z_k}^i(s)\|_2\tag{$\|q^i(s)\|_2\leq \frac{A_{\max}^{1/2}}{1-\gamma}$}\\
	=\;&\frac{A_{\max}^{1/2}}{1-\gamma}\left(\|\mu_k(s)(\Pi_k^i(s)-\Pi_{k-z_k}^i(s))\|_2+\|(\mu_k(s) -\mu_{k-z_k}(s))\Pi_{k-z_k}^i(s)\|_2\right)\\
	\leq \;&\frac{A_{\max}^{1/2}}{1-\gamma}\left(\mu_k(s)\|\Pi_k^i(s)-\Pi_{k-z_k}^i(s)\|_2+|\mu_k(s) -\mu_{k-z_k}(s)|\right)\tag{$\|\Pi_{k-z_k}^i(s)\|_2\leq 1$}\\
	=  \;&\frac{A_{\max}^{1/2}}{1-\gamma}\left(\mu_k(s)\|\pi_k^i(s)-\pi_{k-z_k}^i(s)\|_\infty+|\mu_k(s) -\mu_{k-z_k}(s)|\right).
\end{align*}
For the second term on the RHS of Eq. (\ref{eq:N2112}), we have
\begin{align*}
	&\|\mu_k(s)\Pi_k^i(s) \mathcal{T}^i(v^i)(s)\pi_k^{-i}(s)-\mu_{k-z_k}(s)\Pi_{k-z_k}^i(s) \mathcal{T}^i(v^i)(s)\pi_{k-z_k}^{-i}(s)\|_2\\
	\leq \;&\|(\mu_k(s)\Pi_k^i(s)-\mu_{k-z_k}(s)\Pi_{k-z_k}^i(s)) \mathcal{T}^i(v^i)(s)\pi_k^{-i}(s)\|_2\\
	&+\|\mu_{k-z_k}(s)\Pi_{k-z_k}^i(s) \mathcal{T}^i(v^i)(s)(\pi_k^{-i}(s)-\pi_{k-z_k}^{-i}(s))\|_2\\
	\leq \;&\frac{A_{\max}^{1/2}}{1-\gamma}\|\mu_k(s)\Pi_k^i(s)-\mu_{k-z_k}(s)\Pi_{k-z_k}^i(s)\|_2 +\frac{1}{1-\gamma}\mu_{k-z_k}(s)\|\pi_k^{-i}(s)-\pi_{k-z_k}^{-i}(s)\|_\infty\\
	\leq \;&\frac{A_{\max}^{1/2}}{1-\gamma}\left(\mu_k(s)\|\Pi_k^i(s)-\Pi_{k-z_k}^i(s)\|_2 +|\mu_k(s)-\mu_{k-z_k}(s)|\right)\\
	&+\frac{\mu_{k-z_k}(s)}{1-\gamma}\|\pi_k^{-i}(s)-\pi_{k-z_k}^{-i}(s)\|_\infty\\
	\leq \;&\frac{A_{\max}^{1/2}}{1-\gamma}\left(|\mu_k(s)-\mu_{k-z_k}(s)|+(\mu_k(s)+\mu_{k-z_k}(s))\|\pi_k^{-i}(s)-\pi_{k-z_k}^{-i}(s)\|_\infty\right).
\end{align*}
Using the previous two inequalities in Eq. (\ref{eq:N2112}) and we have
\begin{align*}
	&\|\bar{F}_k^i(q^i)(s)-\bar{F}_{k-z_k}^i(q^i)(s)\|_2\\
	\leq\;& \frac{A_{\max}^{1/2}}{1-\gamma}\left(2|\mu_k(s)-\mu_{k-z_k}(s)|+(\mu_k(s)+\mu_{k-z_k}(s))\|\pi_k^{-i}(s)-\pi_{k-z_k}^{-i}(s)\|_\infty\right)\\
	&+\frac{A_{\max}^{1/2}}{1-\gamma}\mu_k(s)\|\pi_k^i(s)-\pi_{k-z_k}^i(s)\|_\infty,
\end{align*}
which implies
\begin{align*}
	&\|\bar{F}_k^i(q^i)(s)-\bar{F}_{k-z_k}^i(q^i)(s)\|_2^2\\
	\leq\;& \frac{3A_{\max}}{(1-\gamma)^2}\left(4|\mu_k(s)-\mu_{k-z_k}(s)|^2+2(\mu_k(s)^2+\mu_{k-z_k}(s)^2)\|\pi_k^{-i}(s)-\pi_{k-z_k}^{-i}(s)\|_\infty^2\right)\\
	&+\frac{3A_{\max}}{(1-\gamma)^2}\mu_k(s)^2\|\pi_k^i(s)-\pi_{k-z_k}^i(s)\|_\infty^2.
\end{align*}
It follows that
\begin{align*}
	\|\bar{F}_k^i(q^i)-\bar{F}_{k-z_k}^i(q^i)\|_2^2=\;&\sum_{s}\|\bar{F}_k^i(q^i)(s)-\bar{F}_{k-z_k}^i(q^i)(s)\|_2^2\\
	\leq \;&\frac{3A_{\max}}{(1-\gamma)^2}\left(4\|\mu_k-\mu_{k-z_k}\|_2^2+4\max_{s}\|\pi_k^{-i}(s)-\pi_{k-z_k}^{-i}(s)\|_\infty^2\right)\\
	&+\frac{3A_{\max}}{(1-\gamma)^2}\max_{s}\|\pi_k^i(s)-\pi_{k-z_k}^i(s)\|_\infty^2.
\end{align*}
Since
\begin{align*}
	\|\mu_k-\mu_{k-z_k}\|_2^2\leq\;& |\mathcal{S}| \|\mu_k-\mu_{k-z_k}\|_\infty^2\\
	\leq \;&|\mathcal{S}| \hat{L}_\tau^2 \|\pi_k^i-\pi_{k-z_k}^i\|_\infty^2\tag{Lemma \ref{thm:exploration} (3)}\\
	= \;&|\mathcal{S}| \hat{L}_\tau^2 \max_{s\in\mathcal{S}} \|\pi_k^i(s)-\pi_{k-z_k}^i(s)\|_1^2\\
	\leq  \;&|\mathcal{S}|^2 \hat{L}_\tau^2 \max_{s\in\mathcal{S}} \|\pi_k^i(s)-\pi_{k-z_k}^i(s)\|_2^2\\
	\leq  \;&16|\mathcal{S}|^2 \hat{L}_\tau^2\beta_{k-z_k,k-1}^2 \max_{s\in\mathcal{S}} (\|\pi_k^i(s)\|_2+1)^2\tag{Lemma \ref{le:difference_pi}}\\
	\leq \;&64|\mathcal{S}|^2 \hat{L}_\tau^2\beta_{k-z_k,k-1}^2
\end{align*}
and
\begin{align*}
	\max_{s}\|\pi_k^i(s)-\pi_{k-z_k}^i(s)\|_\infty^2\leq \;&\max_{s}\|\pi_k^i(s)-\pi_{k-z_k}^i(s)\|_2^2\\
	\leq \;&16\beta_{k-z_k,k-1}^2\max_{s}(\|\pi_k^i(s)\|_2+1)^2\tag{Lemma \ref{le:difference_pi}}\\
	\leq \;&64\beta_{k-z_k,k-1}^2,\quad i\in \{1,2\},
\end{align*}
we have
\begin{align*}
	\|\bar{F}_k^i(q^i)-\bar{F}_{k-z_k}^i(q^i)\|_2^2
	\leq \;&\frac{3A_{\max}}{(1-\gamma)^2}\left(256|\mathcal{S}|^2 \hat{L}_\tau^2\beta_{k-z_k,k-1}^2+320\beta_{k-z_k,k-1}^2\right)\\
	\leq \;&\frac{1728A_{\max}|\mathcal{S}|^2 \hat{L}_\tau^2\beta_{k-z_k,k-1}^2}{(1-\gamma)^2},
\end{align*}
which implies
\begin{align}\label{eqeqeq:N2111}
	\|\bar{F}_k^i(q^i)-\bar{F}_{k-z_k}^i(q^i)\|_2\leq \frac{42|\mathcal{S}|A_{\max}^{1/2}\hat{L}_\tau\beta_{k-z_k,k-1}}{1-\gamma}
\end{align}
for all $q^i\in\mathbb{R}^{|\mathcal{S}||\mathcal{A}^i|}$.

We next move on to bound the second term on the RHS of Eq. (\ref{eq:N211}). Recall that we denote $P_\pi\in\mathbb{R}^{|\mathcal{S}|\times |\mathcal{S}|}$ as the transition probability matrix of the Markov chain $\{S_k\}$ induced by the joint policy $\pi$.
Using the definition of conditional expectation and we have
\begin{align}
	&\|\mathbb{E}[F^i(q_{k-z_k}^i,S_k,A_k^i,A_k^{-i},S_{k+1})\mid \mathcal{F}_{k-z_k}]-\bar{F}_k^i(q_{k-z_k}^i)\|_2\nonumber\\
	=\;&\bigg\|\sum_{s}\left[\left(\prod_{j=k+1}^{k+z_k}P_{\pi_{j-z_k}}\right)(S_{k-z_k},s)-\mu_k(s)\right]\sum_{a^i}\pi_k^i(a^i|s)\sum_{a^{-i}}\pi_k^{-i}(a^{-i}|s)\nonumber\\
	&\times \sum_{s'}p(s'|s,a^i,a^{-i})F^i(q_{k-z_k}^i,s,a^i,a^{-i},s')\bigg\|_2\nonumber\\
	\leq \;&\left|\sum_{s}\left[\left(\prod_{j=k+1}^{k+z_k}P_{\pi_{j-z_k}}\right)(S_{k-z_k},s)-\mu_k(s)\right]\right|\left(\|q_{k-z_k}^i\|_2+\frac{1}{1-\gamma}\right)\tag{Lemma \ref{le:operators}}\nonumber\\
	\leq \;&\sum_{s}\left|\left(\prod_{j=k+1}^{k+z_k}P_{\pi_{j-z_k}}\right)(S_{k-z_k},s)-\mu_k(s)\right|\left(\|q_{k-z_k}^i\|_2+\frac{1}{1-\gamma}\right)\nonumber\\
	\leq  \;&\left\{\sum_{s}\left|\left(\prod_{j=k+1}^{k+z_k}P_{\pi_{j-z_k}}\right)(S_{k-z_k},s)-P_{\pi_k}^{z_k}(S_{k-z_k},s)\right|+\sum_{s}\left|P_{\pi_k}^{z_k}(S_{k-z_k},s)-\mu_k(s)\right|\right\}\nonumber\\
	&\times \left(\|q_{k-z_k}^i\|_2+\frac{1}{1-\gamma}\right)\nonumber\\
	\leq  \;&\left\{\left\|\prod_{j=k+1}^{k+z_k}P_{\pi_{j-z_k}}-P_{\pi_k}^{z_k}\right\|_\infty+2\rho_\tau^{z_k}\right\}\left(\|q_{k-z_k}^i\|_2+\frac{1}{1-\gamma}\right),\label{eq:probability}
\end{align}
where the last line follows from Lemma \ref{thm:exploration} (2) and Lemma \ref{le:margin}. Observe that 
\begin{align*}
	\left\|\prod_{j=k+1}^{k+z_k}P_{\pi_{j-z_k}}-P_{\pi_k}^{z_k}\right\|_\infty
	=\;&\left\|\sum_{\ell=1}^{z_k}\left(\prod_{j=k+1}^{k-\ell+1+z_k}P_{\pi_{j-z_k}}P_{\pi_k}^{\ell-1}-\prod_{j=k+1}^{k-\ell+z_k}P_{\pi_{j-z_k}}P_{\pi_k}^{\ell}\right)\right\|_\infty\\
	=\;&\left\|\sum_{\ell=1}^{z_k}\left(\prod_{j=k+1}^{k-\ell+z_k}P_{\pi_{j-z_k}}(P_{\pi_{k-\ell+1}}-P_{\pi_k})P_{\pi_k}^{\ell-1}\right)\right\|_\infty\\
	\leq \;&\sum_{\ell=1}^{z_k}\left\|\prod_{j=k+1}^{k-\ell+z_k}P_{\pi_{j-z_k}}\right\|_\infty\|P_{\pi_{k-\ell+1}}-P_{\pi_k}\|_\infty\|P_{\pi_k}^{\ell-1}\|_\infty.
\end{align*}
Since the induced $\ell_\infty$-norm for any stochastic matrix is $1$ and $P_\pi$ as a function of $\pi$ is $1$-Lipschitz continuous with respect to the $\ell_\infty$-norm, we have
\begin{align*}
	&\left\|\prod_{j=k+1}^{k+z_k}P_{\pi_{j-z_k}}-P_{\pi_k}^{z_k}\right\|_\infty\\
	\leq\;&\sum_{\ell=1}^{z_k}\|\pi_{k-\ell+1}-\pi_k\|_\infty\\
	=\;&\sum_{\ell=1}^{z_k}\max_{s\in\mathcal{S}}\sum_{a^i,a^{-i}}|\pi_{k-\ell+1}^i(a^i|s)\pi_{k-\ell+1}^{-i}(a^{-i}|s)-\pi_k^i(a^i|s)\pi_k^{-i}(a^{-i}|s)|\\
	\leq \;&\sum_{\ell=1}^{z_k}\max_{s\in\mathcal{S}}\sum_{a^i,a^{-i}}\pi_{k-\ell+1}^i(a^i|s)|\pi_{k-\ell+1}^{-i}(a^{-i}|s)-\pi_k^{-i}(a^{-i}|s)|\\
	&+\sum_{\ell=1}^{z_k}\max_{s\in\mathcal{S}}\sum_{a^i,a^{-i}}|\pi_{k-\ell+1}^i(a^i|s)-\pi_k^i(a^i|s)|\pi_k^{-i}(a^{-i}|s)\\
	= \;&\sum_{\ell=1}^{z_k}\max_{s\in\mathcal{S}}\left(\sum_{a^{-i}}|\pi_{k-\ell+1}^{-i}(a^{-i}|s)-\pi_k^{-i}(a^{-i}|s)|+\sum_{a^i}|\pi_{k-\ell+1}^i(a^i|s)-\pi_k^i(a^i|s)|\right)\\
	= \;&\sum_{\ell=1}^{z_k}\max_{s\in\mathcal{S}}\left(\|\pi_{k-\ell+1}^{-i}(s)-\pi_k^{-i}(s)\|_1+\|\pi_{k-\ell+1}^i(s)-\pi_k^i(s)\|_1\right)\\
	\leq \;&A_{\max}^{1/2}\sum_{\ell=1}^{z_k}\max_{s\in\mathcal{S}}\left(\|\pi_{k-\ell+1}^{-i}(s)-\pi_k^{-i}(s)\|_2+\|\pi_{k-\ell+1}^i(s)-\pi_k^i(s)\|_2\right)\\
	\leq \;&4z_k\beta_{k-z_k,k-1}A_{\max}^{1/2}\max_{s\in\mathcal{S}}\left(\|\pi_k^{-i}(s)\|_2+\|\pi_k^i(s)\|_2+2\right)\tag{Lemma \ref{le:difference_pi}}\\
	\leq \;&16z_k\beta_{k-z_k,k-1}A_{\max}^{1/2}\\
	\leq \;&16z_k\beta_{k-z_k,k-1}A_{\max}^{1/2}.
\end{align*}
It then follows from the previous inequality that
\begin{align*}
    &\|\mathbb{E}[F^i(q_{k-z_k}^i,S_k,A_k^i,A_k^{-i},S_{k+1})\mid \mathcal{F}_{k-z_k}]-\bar{F}_k^i(q_{k-z_k}^i)\|_2\\
    \leq \;&\left\{\left\|\prod_{j=k+1}^{k+z_k}P_{\pi_{j-z_k}}-P_{\pi_k}^{z_k}\right\|_\infty+2\rho_\tau^{z_k}\right\}\left(\|q_{k-z_k}^i\|_2+\frac{1}{1-\gamma}\right)\\
    \leq \;&\left(16A_{\max}^{1/2}z_k\beta_{k-z_k,k-1}+2\rho_\tau^{z_k}\right)\left(\|q_{k-z_k}^i\|_2+\frac{1}{1-\gamma}\right)\\
    \leq \;&\frac{2\sqrt{|\mathcal{S}|A_{\max}}}{1-\gamma}\left(16A_{\max}^{1/2}z_k\beta_{k-z_k,k-1}+2\rho_\tau^{z_k}\right)\\
    \leq \;&\frac{2\sqrt{|\mathcal{S}|A_{\max}}}{1-\gamma}\left(16A_{\max}^{1/2}z_k\beta_{k-z_k,k-1}+2\beta_k\right)\tag{Definition of $z_k$}\\
    \leq \;&\frac{36\sqrt{|\mathcal{S}|}A_{\max}z_k\beta_{k-z_k,k-1}}{1-\gamma}
\end{align*}
Substituting the previous inequality and the bound in Eq. (\ref{eqeqeq:N2111}) into Eq. (\ref{eq:N211}) and we have
\begin{align*}
    N_{2,1,1}\leq\;& \frac{42|\mathcal{S}|A_{\max}^{1/2}\hat{L}_\tau\beta_{k-z_k,k-1}}{1-\gamma}+\frac{36\sqrt{|\mathcal{S}|}A_{\max}z_k\beta_{k-z_k,k-1}}{1-\gamma}\\
    \leq \;&\frac{80|\mathcal{S}|A_{\max}\hat{L}_\tau z_k\beta_{k-z_k,k-1}}{1-\gamma}.
\end{align*}
It follows that
\begin{align*}
	N_{2,1}\leq \frac{2\sqrt{|\mathcal{S}|A_{\max}}}{1-\gamma}\mathbb{E}[N_{2,1,1}]\leq  \frac{160|\mathcal{S}|^{3/2}A_{\max}^{3/2}\hat{L}_\tau}{(1-\gamma)^2}z_k\beta_{k-z_k,k-1}.
\end{align*}

\paragraph{The Term $N_{2,2}$.}
For any $k\geq z_k$, we have
\begin{align*}
	N_{2,2}=\;&\mathbb{E}[\langle F^i(q_{k-z_k}^i,S_k,A_k^i,A_k^{-i},S_{k+1})-\bar{F}_{k-z_k}^i(q_{k-z_k}^i),q_k^i-q_{k-z_k}^i\rangle]\\
	\leq \;& \mathbb{E}[\underbrace{\|F^i(q_{k-z_k}^i,S_k,A_k^i,A_k^{-i},S_{k+1})-\bar{F}_{k-z_k}^i(q_{k-z_k}^i)\|_2}_{N_{2,2,1}}\underbrace{\|q_k^i-q_{k-z_k}^i\|_2}_{N_{2,2,2}}]
\end{align*}
Using Lemma \ref{le:operators} and we have
\begin{align*}
	&N_{2,2,1}\\
	=\;&\|F^i(q_{k-z_k}^i,S_k,A_k^i,A_k^{-i},S_{k+1})-\bar{F}_{k-z_k}^i(q_{k-z_k}^i)\|_2\\
	=\;&\|F^i(q_{k-z_k}^i,S_k,A_k^i,A_k^{-i},S_{k+1})\!-\!F^i(\bm{0},S_k,A_k^i,A_k^{-i},S_{k+1})\!+\!F^i(\bm{0},S_k,A_k^i,A_k^{-i},S_{k+1})\|_2\\
	&+\|\bar{F}_{k-z_k}^i(q_{k-z_k}^i)-\bar{F}_{k-z_k}^i(\bm{0})+\bar{F}_{k-z_k}^i(\bm{0})\|_2\\
	\leq \;&2\|q_{k-z_k}^i\|_2+\frac{2}{1-\gamma}\tag{Lemma \ref{le:boundedness} and Jensen's inequality}\\
	\leq \;& \frac{2|\mathcal{S}|^{1/2}A_{\max}^{1/2}}{1-\gamma}+\frac{2}{1-\gamma}\tag{Lemma \ref{le:boundedness}}\\
	\leq \;& \frac{4|\mathcal{S}|^{1/2}A_{\max}^{1/2}}{1-\gamma}.
\end{align*}
Moreover, we have by Lemma \ref{le:difference} and Lemma \ref{le:boundedness} that
\begin{align*}
	N_{2,2,2}\leq 4\alpha_{k-z_k,k-1}\left(\|q_k^i\|_2+\frac{1}{1-\gamma}\right)
	\leq \frac{8|\mathcal{S}|^{1/2}A_{\max}^{1/2}\alpha_{k-z_k,k-1}}{1-\gamma}.
\end{align*}
Therefore, we have
\begin{align*}
	N_{2,2}\leq \mathbb{E}[N_{2,2,1}\times N_{2,2,2}]\leq  \frac{32|\mathcal{S}|A_{\max}}{(1-\gamma)^2}\alpha_{k-z_k,k-1}.
\end{align*}

\paragraph{The Term $N_{2,3}$.}
For any $k\geq z_k$, we have
\begin{align*}
	N_{2,3}=\;&\mathbb{E}[\langle F^i(q_{k-z_k}^i,S_k,A_k^i,A_k^{-i},S_{k+1})-\bar{F}_{k-z_k}^i(q_{k-z_k}^i),\bar{q}_{k-z_k}^i-\bar{q}_k^i\rangle]\\
	\leq \;&\frac{c'}{2}\mathbb{E}[\underbrace{\|F^i(q_{k-z_k}^i,S_k,A_k^i,A_k^{-i},S_{k+1})-\bar{F}_{k-z_k}^i(q_{k-z_k}^i)\|_2^2}_{N_{2,3,1}}]+\frac{1}{2c'}\mathbb{E}[\underbrace{\|\bar{q}_{k-z_k}^i-\bar{q}_k^i\|_2^2}_{N_{2,3,2}}],
\end{align*}
where $c'>0$ is any positive real number. Since $N_{2,3,1}=N_{2,2,1}^2$, we have
\begin{align*}
	N_{2,3,1}\leq\frac{16|\mathcal{S}|A_{\max}}{(1-\gamma)^2}.
\end{align*}
To control the term $N_{2,3,2}$, using the explicit expression of $\bar{q}_k^i$ provided in Lemma \ref{le:operators} (3) and we have
\begin{align*}
	\|\bar{q}_{k-z_k}^i-\bar{q}_k^i\|_2^2=\;&\sum_{s}\|\mathcal{T}^i(v^i)(s)(\pi_{k-z_k}^{-i}(s)-\pi_k^{-i}(s))\|_2^2.
\end{align*}
Since
\begin{align*}
	\|\mathcal{T}^i(v^i)(s)(\pi_k^{-i}(s)-\pi_{k-z_k}^{-i}(s))\|_2
	\leq \;&\frac{A_{\max}}{1-\gamma}\|\pi_k^{-i}(s)-\pi_{k-z_k}^{-i}(s)\|_2\\
	\leq \;&\frac{4A_{\max}\beta_{k-z_k,k-1}}{1-\gamma}(\|\pi_k^{-i}(s)\|_2+1)\tag{Lemma \ref{le:difference_pi}}\\
	\leq \;&\frac{8A_{\max}\beta_{k-z_k,k-1}}{1-\gamma},
\end{align*}
we have
\begin{align*}
	N_{2,3,2}
	\leq \frac{64|\mathcal{S}|A_{\max}^2\beta_{k-z_k,k-1}^2}{(1-\gamma)^2}
\end{align*}
It follows that
\begin{align*}
	N_{2,3}=\;&\frac{c'}{2}\mathbb{E}[N_{2,3,1}]+\frac{1}{2c'}\mathbb{E}[N_{2,3,2}]\\
	\leq \;&\frac{8c'|\mathcal{S}|A_{\max}}{(1-\gamma)^2}+\frac{32|\mathcal{S}|A_{\max}^2\beta_{k-z_k,k-1}^2}{c'(1-\gamma)^2}\\
	\leq \;&\frac{8|\mathcal{S}|A_{\max}}{(1-\gamma)^2}\left(c'+\frac{4A_{\max}\beta_{k-z_k,k-1}^2}{c'}\right)\\
	=\;&\frac{32|\mathcal{S}|A_{\max}^{3/2}\beta_{k-z_k,k-1}}{(1-\gamma)^2},
\end{align*}
where the last line follows by choosing $c'=2A_{\max}^{1/2}\beta_{k-z_k,k-1}$.

\paragraph{The Term $N_{2,4}$.}
For any $k\geq 0$, we have
\begin{align*}
	N_{2,4}=\;&\mathbb{E}[\langle F^i(q_k^i,S_k,A_k^i,A_k^{-i},S_{k+1})-F^i(q_{k-z_k}^i,S_k,A_k^i,A_k^{-i},S_{k+1}),q_k^i-\bar{q}_k^i\rangle]\\
	\leq \;&\mathbb{E}[\| F^i(q_k^i,S_k,A_k^i,A_k^{-i},S_{k+1})-F^i(q_{k-z_k}^i,S_k,A_k^i,A_k^{-i},S_{k+1})\|_2\|q_k^i-\bar{q}_k^i\|_2]\\
	\leq \;&\mathbb{E}[\| q_k^i-q_{k-z_k}^i\|_2\|q_k^i-\bar{q}_k^i\|_2]\tag{Lemma \ref{le:operators}}\\
	\leq \;&4\alpha_{k-z_k,k-1}\mathbb{E}\left[\left(\| q_k^i\|_2+\frac{1}{1-\gamma}\right)\|q_k^i-\bar{q}_k^i\|_2\right]\tag{Lemma \ref{le:difference}}\\
	\leq \;&\frac{16|\mathcal{S}|A_{\max}\alpha_{k-z_k,k-1}}{(1-\gamma)^2},
\end{align*}
where the last line follows from $\|q_k^i\|_\infty\leq \frac{1}{1-\gamma}$ and $\|\Bar{q}_k^i\|_\infty\leq \frac{1}{1-\gamma}$.

\paragraph{The Term $N_{2,5}$.}
For any $k\geq 0$, we have
\begin{align*}
	N_{2,5}=\;&\mathbb{E}[\langle \bar{F}_k^i(q_k^i)-\bar{F}_{k-z_k}^i(q_{k-z_k}^i),q_k^i-\bar{q}_k^i\rangle]\\
	\leq \;&\mathbb{E}[\| \bar{F}_k^i(q_k^i)-\bar{F}_{k-z_k}^i(q_{k-z_k}^i)\|_2\|q_k^i-\bar{q}_k^i\|_2]\\
	\leq \;&\mathbb{E}[(\| \bar{F}_k^i(q_k^i)-\bar{F}_{k-z_k}^i(q_k^i)\|_2+\|\bar{F}_{k-z_k}^i(q_k^i)-\bar{F}_{k-z_k}^i(q_{k-z_k}^i)\|_2)\|q_k^i-\bar{q}_k^i\|_2]\\
	\leq \;&\mathbb{E}[\| q_k^i-q_{k-z_k}^i\|_2\|q_k^i-\bar{q}_k^i\|_2]+\frac{42|\mathcal{S}|A_{\max}^{1/2}\hat{L}_\tau\beta_{k-z_k,k-1}}{1-\gamma}\mathbb{E}[\|q_k^i-\bar{q}_k^i\|_2]\tag{Lemma \ref{le:operators} and Eq. (\ref{eqeqeq:N2111})}\\
	\leq \;&4\alpha_{k-z_k,k-1}\mathbb{E}\left[\left(\| q_k^i\|_2+\frac{1}{1-\gamma}\right)\|q_k^i-\bar{q}_k^i\|_2\right]\tag{Lemma \ref{le:difference}}\\
	&+\frac{42|\mathcal{S}|A_{\max}^{1/2}\hat{L}_\tau\beta_{k-z_k,k-1}}{1-\gamma}\mathbb{E}[\|q_k^i-\bar{q}_k^i\|_2]\\
	\leq \;&\frac{16|\mathcal{S}|A_{\max}\alpha_{k-z_k,k-1}}{(1-\gamma)^2}+\frac{84|\mathcal{S}|^{3/2}A_{\max}\hat{L}_\tau\beta_{k-z_k,k-1}}{(1-\gamma)^2}\\
	\leq \;&\frac{100|\mathcal{S}|^{3/2}A_{\max}\hat{L}_\tau\alpha_{k-z_k,k-1}}{(1-\gamma)^2}.
\end{align*}

Finally, combining the upper bounds we derived for the terms $\{N_{2,j}\}_{1\leq j\leq 5}$ and we have 
\begin{align*}
    N_2\leq \;&\sum_{j=1}^5N_{2,j}\\
    \leq \;&\frac{160|\mathcal{S}|^{3/2}A_{\max}^{3/2}\hat{L}_\tau}{(1-\gamma)^2}z_k\beta_{k-z_k,k-1}+\frac{32|\mathcal{S}|A_{\max}}{(1-\gamma)^2}\alpha_{k-z_k,k-1}\\
    &+\frac{32|\mathcal{S}|A_{\max}^{3/2}\beta_{k-z_k,k-1}}{(1-\gamma)^2}+\frac{16|\mathcal{S}|A_{\max}\alpha_{k-z_k,k-1}}{(1-\gamma)^2}\\
    &+\frac{100|\mathcal{S}|^{3/2}A_{\max}\hat{L}_\tau\alpha_{k-z_k,k-1}}{(1-\gamma)^2}\\
    \leq \;&\frac{340|\mathcal{S}|^{3/2}A_{\max}^{3/2}\hat{L}_\tau}{(1-\gamma)^2}z_k\alpha_{k-z_k,k-1}.
\end{align*}

\subsubsection{Proof of Lemma \ref{le:other_terms}}\label{pf:le:other_terms}
\begin{enumerate}[(1)]
	\item For any $k\geq 0$, using Lemma \ref{le:operators} and we have
	\begin{align*}
		\|q_{k+1}^i-q_k^i\|_2^2
		=\;&\alpha_k^2\|F^i(q_k^i,S_k,A_k^i,A_k^{-i},S_{k+1})\|_2^2\\
		=\;&\alpha_k^2\|F^i(q_k^i,S_k,A_k^i,A_k^{-i},S_{k+1})-F^i(\bm{0},S_k,A_k^i,A_k^{-i},S_{k+1})\\
		&+F^i(\bm{0},S_k,A_k^i,A_k^{-i},S_{k+1})\|_2^2\\
		\leq \;&\alpha_k^2\left(\|q_k^i\|_2+\frac{1}{1-\gamma}\right)^2\\
		\leq \;&\alpha_k^2\left(\frac{\sqrt{|\mathcal{S}|A_{\max}}}{1-\gamma}+\frac{1}{1-\gamma}\right)^2\tag{$\|q_k^i\|_\infty\leq \frac{1}{1-\gamma}$}\\
		\leq \;&\frac{4|\mathcal{S}|A_{\max}\alpha_k^2}{(1-\gamma)^2}.
	\end{align*}
	The result follows by taking expectation on both sides of the previous inequality.
	\item 
	For any $k\geq 0$, we have by Lemma \ref{le:operators} that
	\begin{align*}
		\|\bar{q}_k^i-\bar{q}_{k+1}^i\|_2^2=\;&\sum_{s}\|\mathcal{T}^i(v^i)(s)(\pi_{k+1}^{-i}(s)-\pi_k^{-i}(s))\|_2^2\\
		=\;&\beta_k^2\sum_{s}\|\mathcal{T}^i(v^i)(s)(\sigma_\tau(q_k^{-i}(s))-\pi_k^{-i}(s))\|_2^2\\
		\leq \;&\beta_k^2\sum_{s}(\|\mathcal{T}^i(v^i)(s)\sigma_\tau(q_k^{-i}(s))\|_2+\|\mathcal{T}^i(v^i)(s)\pi_k^{-i}(s)\|_2)^2\\
		\leq \;&\frac{4|\mathcal{S}|A_{\max}\beta_k^2}{(1-\gamma)^2}.
	\end{align*}
	The result follows by taking expectation on both sides of the previous inequality.
	\item 
	For any $k\geq 0$, we have
	\begin{align*}
		\langle q_{k+1}^i-q_k^i,\bar{q}_k^i-\bar{q}_{k+1}^i\rangle
		\leq \|q_{k+1}^i-q_k^i\|_2\|\bar{q}_k^i-\bar{q}_{k+1}^i\|_2
		\leq  \frac{4|\mathcal{S}|A_{\max}\alpha_k\beta_k}{(1-\gamma)^2},
	\end{align*}
	where the last inequality follows from Part (1) and Part (2) of this lemma.
	The result follows by taking expectation on both sides of the previous inequality.
	\item 
	For any $k\geq 0$, we have
	\begin{align}
		&\langle q_k^i-\bar{q}_k^i,\bar{q}_k^i-\bar{q}_{k+1}^i\rangle\nonumber\\
		=\;&\beta_k\sum_{s}\langle q_k^i(s)-\bar{q}_k^i(s),\mathcal{T}^i(v^i)(s)(\sigma_\tau(q_k^{-i}(s))-\pi_k^{-i}(s))\rangle\nonumber\\
		\leq \;&\beta_k\left(\frac{\hat{c}\| q_k^i-\bar{q}_k^i\|_2^2}{2}+\frac{\sum_{s}\|\mathcal{T}^i(v^i)(s)(\sigma_\tau(q_k^{-i}(s))-\pi_k^{-i}(s))\|_2^2}{2\hat{c}}\right),\label{eq:otherterms_2}
	\end{align}
	where $\hat{c}$ is an arbitrary positive real number.
	We next analyze the second term on the RHS of the previous inequality. For any $s\in\mathcal{S}$, we have
	\begin{align*}
		&\|\mathcal{T}^i(v^i)(s)(\sigma_\tau(q_k^{-i}(s))-\pi_k^{-i}(s))\|_2\\
		=\;&\|\mathcal{T}^i(v^i)(s)(\sigma_\tau(q_k^{-i}(s))-\sigma_\tau(\bar{q}_k^{-i}(s))+\sigma_\tau(\mathcal{T}^{-i}(v^{-i})(s)\pi_k^i(s))-\pi_k^{-i}(s))\|_2\\
		\leq \;&\underbrace{\|\mathcal{T}^i(v^i)(s)(\sigma_\tau(q_k^{-i}(s))-\sigma_\tau(\bar{q}_k^{-i}(s)))\|_2}_{B_1}\\
		&+\underbrace{\|\mathcal{T}^i(v^i)(s)(\sigma_\tau(\mathcal{T}^{-i}(v^{-i})(s)\pi_k^i(s))-\pi_k^{-i}(s))\|_2}_{B_2}.
	\end{align*}
	Since the softmax operator $\sigma_\tau(\cdot)$ is $\frac{1}{\tau}$ -- Lipschitz continuous with respect to $\|\cdot\|_2$ \cite[Proposition 4]{gao2017properties}, we have
	\begin{align*}
		B_1\leq \;&\|\mathcal{T}^i(v^i)(s)\|_2\|\sigma_\tau(q_k^{-i}(s))-\sigma_\tau(\bar{q}_k^{-i}(s))\|_2\\
		\leq \;&\frac{A_{\max}}{\tau(1-\gamma)}\|q_k^{-i}(s)-\bar{q}_k^{-i}(s)\|_2.
	\end{align*}
	We next analyze the term $B_2$. Using the quadratic growth property of strongly convex functions and we have
	\begin{align*}
		B_2=\;&\|\mathcal{T}^i(v^i)(s)(\sigma_\tau(\mathcal{T}^{-i}(v^{-i})(s)\pi_k^i(s))-\pi_k^{-i}(s))\|_2\\
		\leq \;&\|\mathcal{T}^i(v^i)(s)\|_2\|\sigma_\tau(\mathcal{T}^{-i}(v^{-i})(s)\pi_k^i(s))-\pi_k^{-i}(s)\|_2\\
		\leq \;&\frac{\sqrt{2}A_{\max}}{\sqrt{\tau}(1-\gamma)}V_{v,s}^{1/2}(\pi_k^i(s),\pi_k^{-i}(s)).
	\end{align*}
	Combine the upper bounds we obtained for the terms $B_1$ and $B_2$ and we obtain
	\begin{align*}
		&\sum_{s}\|\mathcal{T}^i(v^i)(s)(\sigma_\tau(q_k^{-i}(s))-\pi_k^{-i}(s))\|_2^2\\
		\leq \;&\sum_{s}(B_1+B_2)^2\\
		\leq \;&2\sum_{s}(B_1^2+B_2^2)\\
		\leq \;&2\sum_{s}\left(\frac{A_{\max}^2}{\tau^2(1-\gamma)^2}\|q_k^{-i}(s))-\bar{q}_k^{-i}(s)\|_2^2+\frac{2A_{\max}^2}{\tau(1-\gamma)^2}V_{v,s}(\pi_k^i(s),\pi_k^{-i}(s))\right)\\
		= \;&\frac{2A_{\max}^2}{\tau^2(1-\gamma)^2}\|q_k^{-i}-\bar{q}_k^{-i}\|_2^2+\frac{4A_{\max}^2}{\tau(1-\gamma)^2}\sum_{s}V_{v,s}(\pi_k^i(s),\pi_k^{-i}(s)).
	\end{align*}
	Coming back to Eq. (\ref{eq:otherterms_2}), using the previous inequality and we have
	\begin{align*}
		&\langle q_k^i-\bar{q}_k^i,\bar{q}_k^i-\bar{q}_{k+1}^i\rangle\\
		\leq\;&\beta_k\left(\frac{\hat{c}\| q_k^i-\bar{q}_k^i\|_2^2}{2}+\frac{\sum_{s}\|\mathcal{T}^i(v^i)(s)(\sigma_\tau(q_k^{-i}(s))-\pi_k^{-i}(s))\|_2^2}{2\hat{c}}\right)\\
		\leq \;&\beta_k\left(\frac{\hat{c}\| q_k^i-\bar{q}_k^i\|_2^2}{2}+\frac{A_{\max}^2}{\hat{c}\tau^2(1-\gamma)^2}\|q_k^{-i}-\bar{q}_k^{-i}\|_2^2\right.\\
		&\left.+\frac{2A_{\max}^2}{\hat{c}\tau(1-\gamma)^2}\sum_{s}V_{v,s}(\pi_k^i(s),\pi_k^{-i}(s))\right).
	\end{align*}
	Choosing $\hat{c}=\frac{32A_{\max}^2}{\tau(1-\gamma)^2}$ in the previous inequality and then taking total expectation, and we obtain
	\begin{align*}
		\mathbb{E}[\langle q_k^i-\bar{q}_k^i,\bar{q}_k^i-\bar{q}_{k+1}^i\rangle]
		\leq \frac{17A_{\max}^2\beta_k}{\tau(1-\gamma)^2}\mathbb{E}[\| q_k^i-\bar{q}_k^i\|_2^2]+\frac{\beta_k}{16}\sum_{s}\mathbb{E}[V_{v,s}(\pi_k^i(s),\pi_k^{-i}(s))].
	\end{align*}
\end{enumerate}

\subsubsection{Proof of Lemma \ref{le:q-function-drift}}\label{pf:le:q-function-drift}

For $i\in \{1,2\}$, we have from Eq. (\ref{eq:all_terms}), Eq. (\ref{eq:q_drift}), Lemma \ref{le:noise}, and Lemma \ref{le:other_terms} that
\begin{align*}
	&\mathbb{E}[\|q_{k+1}^i-\bar{q}_{k+1}^i\|_2^2]\\
	\leq \;&\mathbb{E}[\|q_k^i-\bar{q}_k^i\|_2^2]-\alpha_kc_\tau\mathbb{E}[\|q_k^i-\bar{q}_k^i\|_2^2]\\
	&+\frac{4|\mathcal{S}|A_{\max}}{(1-\gamma)^2}(\alpha_k^2+\alpha_k\beta_k+\beta_k^2)\\
	&+\frac{340|\mathcal{S}|^{3/2}A_{\max}^{3/2}\hat{L}_\tau}{(1-\gamma)^2}z_k\alpha_k\alpha_{k-z_k,k-1}\\
	&+\frac{17A_{\max}^2\beta_k}{\tau(1-\gamma)^2}\mathbb{E}[\| q_k^i-\bar{q}_k^i\|_2^2]+\frac{\beta_k}{16}\sum_{s}\mathbb{E}[V_{v,s}(\pi_k^i(s),\pi_k^{-i}(s))]\\
	\leq \;&\left(1-\alpha_kc_\tau+\frac{17A_{\max}^2\beta_k}{\tau(1-\gamma)^2}\right)\mathbb{E}[\|q_k^i-\bar{q}_k^i\|_2^2]\\
	&+\frac{12|\mathcal{S}|A_{\max}}{(1-\gamma)^2}\alpha_k^2+\frac{340|\mathcal{S}|^{3/2}A_{\max}^{3/2}\hat{L}_\tau}{(1-\gamma)^2}z_k\alpha_k\alpha_{k-z_k,k-1}\\
	&+\frac{\beta_k}{16}\sum_{s}\mathbb{E}[V_{v,s}(\pi_k^i(s),\pi_k^{-i}(s))]\\
	\leq \;&\left(1-\alpha_kc_\tau+\frac{17A_{\max}^2\beta_k}{\tau(1-\gamma)^2}\right)\mathbb{E}[\|q_k^i-\bar{q}_k^i\|_2^2]\\
	&+\frac{352|\mathcal{S}|^{3/2}A_{\max}^{3/2}\hat{L}_\tau}{(1-\gamma)^2}z_k\alpha_k\alpha_{k-z_k,k-1}+\frac{\beta_k}{16}\sum_{s}\mathbb{E}[V_{v,s}(\pi_k^i(s),\pi_k^{-i}(s))],
\end{align*}
where the second inequality follows from $\beta_k=c_{\alpha,\beta}\alpha_k$ with $c_{\alpha,\beta}\leq 1$.

\subsubsection{Proof of Lemma \ref{le:difference}}\label{pf:le:difference}
For any $k\in [k_1,k_2-1]$, we have
\begin{align}
	\|q_{k+1}^i\|_2-\|q_k^i\|_2
	\leq\;& 	
	\|q_{k+1}^i-q_k^i\|_2\tag{triangle inequality}\\
	=\;&\alpha_k \|F^i(q_k^i,S_k,A_k^i,A_k^{-i},S_{k+1})\|_2\nonumber\\
	\leq \;&\alpha_k \|F^i(q_k^i,S_k,A_k^i,A_k^{-i},S_{k+1})-F^i(\bm{0},S_k,A_k^i,A_k^{-i},S_{k+1})\|_2\nonumber\\
	&+\alpha_k\|F^i(\bm{0},S_k,A_k^i,A_k^{-i},S_{k+1})\|_2\nonumber\\
	\leq \;&\alpha_k \left(\|q_k^i\|_2+\frac{1}{1-\gamma}\right)\label{eq:10},
\end{align}
where the last inequality follows from Lemma \ref{le:operators}.
Adding $1/(1-\gamma)$ to both sides of the previous inequality and we have
\begin{align*}
	\|q_{k+1}^i\|_2+\frac{1}{1-\gamma}\leq (1+\alpha_k)\left(\|q_k^i\|_2+\frac{1}{1-\gamma}\right).
\end{align*}
Repeatedly using the previous inequality and we have for all $k\in [k_1,k_2]$:
\begin{align*}
	\|q_k^i\|_2\leq \prod_{j=k_1}^{k-1}(1+\alpha_j)\left(\|q_{k_1}^i\|_2+\frac{1}{1-\gamma}\right)-\frac{1}{1-\gamma}.
\end{align*}
Since $1+x\leq  e^x\leq 1+2x$ for all $x\in [0,1/2]$ and $\alpha_{k_1,k_2-1}\leq 1/4$, we have
\begin{align*}
	\prod_{j=k_1}^{k-1}(1+\alpha_j)\leq \exp\left(\alpha_{k_1,k-1}\right)\leq 1+2\alpha_{k_1,k-1}.
\end{align*}
As a result, we have for all $k\in [k_1,k_2]$ that
\begin{align*}
	\|q_k^i\|_2\leq (1+2\alpha_{k_1,k-1})\|q_{k_1}^i\|_2+\frac{2\alpha_{k_1,k-1}}{1-\gamma}.
\end{align*}
Using the previous inequality in Eq. (\ref{eq:10}) and we have for any $k\in [k_1,k_2-1]$:
\begin{align*}
	\|q_{k+1}^i-q_k^i\|_2&\leq \alpha_k\left(\|q_k^i\|_2+\frac{1}{1-\gamma}\right)\\
	&\leq \alpha_k (1+2\alpha_{k_1,k-1})\|q_{k_1}^i\|_2+\frac{2\alpha_k\alpha_{k_1,k-1}}{1-\gamma}\\
	&\leq 2\alpha_k \left(\|q_{k_1}^i\|_2+\frac{1}{1-\gamma}\right),
\end{align*}
where the last line follows from $\alpha_{k_1,k-1}\leq1/4$.
Therefore, we have for any $k\in [k_1,k_2]$:
\begin{align*}
	\|q_k^i-q_{k_1}^i\|_2\leq\;& \sum_{j=k_1}^{k-1}\|q_{j+1}^i-q_j^i\|_2\\
	\leq \;&2\sum_{j=k_1}^{k-1}\alpha_j\left(\|q_{k_1}^i\|_2+\frac{1}{1-\gamma}\right)\\
	=\;& 2\alpha_{k_1,k-1}\left(\|q_{k_1}^i\|_2+\frac{1}{1-\gamma}\right)\\
	\leq\;& 2\alpha_{k_1,k_2-1}\left(\|q_{k_1}^i\|_2+\frac{1}{1-\gamma}\right),
\end{align*}
where the last line follows from  $\alpha_{k_1,k-1}\leq \alpha_{k_1,k_2-1}$. This proves the first claimed inequality.

To prove the second claimed inequality, note that
\begin{align*}
	\|q_{k_2}^i-q_{k_1}^i\|_2&\leq 2\alpha_{k_1,k_2-1}\left(\|q_{k_1}^i\|_2+\frac{1}{1-\gamma}\right)\\
	&\leq 2\alpha_{k_1,k_2-1}\left(\|q_{k_1}^i-q_{k_2}^i\|_2+\|q_{k_2}^i\|_2+\frac{1}{1-\gamma}\right)\\
	&\leq \frac{1}{2}\|q_{k_2}^i-q_{k_1}^i\|_2+ 2\alpha_{k_1,k_2-1}\left(\|q_{k_2}^i\|_2+\frac{1}{1-\gamma}\right),
\end{align*}
we have $\|q_{k_2}^i-q_{k_1}^i\|_2\leq   4\alpha_{k_1,k_2-1}(\|q_{k_2}^i\|_2+1/(1-\gamma))$. Therefore, we have for any $k\in [k_1,k_2]$:
\begin{align*}
	\|q_k^i-q_{k_1}^i\|_2&\leq 2\alpha_{k_1,k_2-1}\left(\|q_{k_1}^i\|_2+\frac{1}{1-\gamma}\right)\\
	&\leq 2\alpha_{k_1,k_2-1}\left(\|q_{k_1}^i-q_{k_2}^i\|_2+\|q_{k_2}^i\|_2+\frac{1}{1-\gamma}\right)\\
	&\leq 2\alpha_{k_1,k_2-1}\left(4\alpha_{k_1,k_2-1}\left(\|q_{k_2}^i\|_2+\frac{1}{1-\gamma}\right)+\|q_{k_2}^i\|_2+\frac{1}{1-\gamma}\right)\\\
	&\leq 4\alpha_{k_1,k_2-1}\left(\|q_{k_2}^i\|_2+\frac{1}{1-\gamma}\right),
\end{align*}
where the last inequality follows from $\alpha_{k_1,k_2-1}\leq 1/4$. The proof is now complete.

\subsubsection{Proof of Lemma \ref{le:difference_pi}}\label{pf:le:difference_pi}

For any $k\geq 0$ and $s\in\mathcal{S}$, we have
\begin{align*}
	\|\pi_{k+1}^i(s)\|_2-\|\pi_k^i(s)\|_2\leq \;&\|\pi_{k+1}^i(s)-\pi_k^i(s)\|_2\\
	=\;&\beta_k\|\sigma_\tau(q_k^i(s))-\pi_k^i(s)\|_2\\
	\leq \;&\beta_k(\|\pi_k^i(s)\|_2+1).
\end{align*}
The rest of the proof is identical to that of Lemma  \ref{le:difference} (after Eq. (\ref{eq:10})).

\section{Proof of Theorem \ref{thm:bandit}}\label{pf:thm:bandit}
Note that Algorithm \ref{algo:bandit} is a special case of Algorithm \ref{algorithm:inner-loop} when the Markov game has only one state, and the inputs $v^i$ and $v^{-i}$ satisfy $v^i=v^{-i}=\bm{0}$. Therefore, Lemma \ref{le:policy_drift} and Lemma \ref{le:q-function-drift} are both applicable, and are restated in the following.

\begin{lemma}
	The following inequality holds for all $k\geq 0$.
	\begin{align}\label{eq:pi_bandit_drift}
		\mathbb{E}[V_{\mathcal{R}}(\pi_{k+1}^i,\pi_{k+1}^{-i})]
		\leq &\left(1-\frac{3\beta_k}{4}\right)\mathbb{E}[V_{\mathcal{R}}(\pi_k^i,\pi_k^{-i})]+\frac{4A_{\max}^2\beta_k^2}{\ell_{\tau}}\nonumber\\
		&+\frac{256A_{\max}^2\beta_k}{\ell_{\tau}^2\tau^3}\sum_{i=1,2}\mathbb{E}[\|q_k^i-\mathcal{R}^i\pi_k^{-i}\|_2^2].
	\end{align}
\end{lemma}

\begin{lemma}
	The following inequality holds for all $k\geq 0$:
	\begin{align}\label{eq:q_bandit_drift}
		\sum_{i=1,2}\mathbb{E}[\|q_{k+1}^i-\mathcal{R}^i\pi_{k+1}^{-i}\|_2^2]\leq \;&\left(1-\ell_\tau\alpha_k+\frac{17A_{\max}^2\beta_k}{\tau}\right)\sum_{i=1,2}\mathbb{E}[\|q_k^i-\mathcal{R}^i\pi_k^{-i}\|_2^2]\nonumber\\
		&+352A_{\max}^{3/2}\alpha_k^2+\frac{\beta_k}{8}\mathbb{E}[V_{\mathcal{R}}(\pi_k^i,\pi_k^{-i})].
	\end{align}
\end{lemma}

Adding up Eqs. (\ref{eq:q_bandit_drift}) and (\ref{eq:pi_bandit_drift}) and we have for any $k\geq 0$ that 
\begin{align*}
	&\sum_{i=1,2}\mathbb{E}[\|q_{k+1}^i-\mathcal{R}^i\pi_{k+1}^{-i}\|_2^2]+\mathbb{E}[V_{\mathcal{R}}(\pi_{k+1}^i,\pi_{k+1}^{-i})]\\
	\leq \;&\left(1-\ell_\tau\alpha_k+\frac{280A_{\max}^2\beta_k}{\ell_{\tau}^2\tau^3}\right)\sum_{i=1,2}\mathbb{E}[\|q_k^i-\mathcal{R}^i\pi_k^{-i}\|_2^2]\nonumber\\
	&+\left(1-\frac{\beta_k}{2}\right)\mathbb{E}[V_{\mathcal{R}}(\pi_k^i,\pi_k^{-i})]+\frac{4A_{\max}^2\beta_k^2}{\ell_{\tau}}+352A_{\max}^{3/2}\alpha_k^2\\
	\leq 
	&\left(1-\frac{c_{\alpha,\beta}\alpha_k}{2}\right)(\mathbb{E}[\|q_k^i-\mathcal{R}^i\pi_k^{-i}\|_2^2]+\mathbb{E}[V_{\mathcal{R}}(\pi_k^i,\pi_k^{-i})])\\
	&+\frac{4A_{\max}^2c_{\alpha,\beta}^2\alpha_k^2}{\ell_{\tau}}+356A_{\max}^{3/2}\alpha_k^2,
\end{align*}
where the last line follows from Condition \ref{con:stepsize}.
Denote $M_k=\mathbb{E}[V_{\mathcal{R}}(\pi_k^i,\pi_k^{-i})]+\sum_{i=1,2}\mathbb{E}[\|q_k^i-\mathcal{R}^i\pi_k^{-i}\|_2^2]$. The previous inequality reads
\begin{align*}
	M_{k+1}\leq \left(1-\frac{c_{\alpha,\beta}\alpha_k}{2}\right)M_k+356A_{\max}^{3/2}\alpha_k^2.
\end{align*}
Repeatedly using the previous inequality and we have for all $k\geq 0$ that
\begin{align}\label{eq:recursion_bandit}
	M_k\leq \prod_{m=0}^{k-1}\left(1-\frac{c_{\alpha,\beta}\alpha_m}{2}\right)M_0+356A_{\max}^{3/2}\sum_{n=0}^{k-1}\alpha_n^2\prod_{m=n+1}^{k-1}\left(1-\frac{c_{\alpha,\beta}\alpha_m}{2}\right).
\end{align}
Due to our initialization, we have
\begin{align*}
	M_0=V_{\mathcal{R}}(\pi_0^i,\pi_0^{-i})+\sum_{i=1,2}\|q_0^i-\mathcal{R}^i\pi_0^{-i}\|_2^2\leq 3.
\end{align*}
It remains to bound $\prod_{m=0}^{k-1}\left(1-\frac{c_{\alpha,\beta}\alpha_m}{2}\right)$ and $\sum_{n=0}^{k-1}\alpha_n^2\prod_{m=n+1}^{k-1}\left(1-\frac{c_{\alpha,\beta}\alpha_m}{2}\right)$ when $\alpha_k$ is explicitly chosen. Using results in \cite[Appendix A.2]{chen2021finite} and we have the following inequalities:
\begin{enumerate}[(1)]
	\item When using constant stepsize, i.e., $\alpha_k\equiv \alpha$, we have for all $k\geq 0$ that
	\begin{align*}
		M_k\leq 3\left(1-\frac{c_{\alpha,\beta}\alpha}{2}\right)^k+\frac{712A_{\max}^{3/2}\alpha}{c_{\alpha,\beta}}.
	\end{align*}
	\item When $\alpha_k=\frac{\alpha}{k+h}$ with $\alpha c_{\alpha,\beta}=4$ and $h$ chosen such that $\alpha/h<1$, we have for all $k\geq 0$ that
	\begin{align*}
		M_k\leq 3\left(\frac{h}{k+h}\right)^2+356A_{\max}^{3/2}\frac{16e}{c_{\alpha,\beta}}\frac{\alpha}{k+h}.
	\end{align*}
	\item When $\alpha_k=\frac{\alpha}{(k+h)^z}$, where $\alpha>0$, $z\in (0,1)$, and $h\geq [4z/(c_{\alpha,\beta}\alpha)]^{1/(1-z)}$, we have for all $k\geq 0$ that
	\begin{align*}
		M_k\leq 3\exp\left(-\frac{\alpha}{2c_{\alpha,\beta}(1-z)}((k+h)^{1-z}-h^{1-z})\right)+356A_{\max}^{3/2}\frac{4}{c_{\alpha,\beta}}\frac{\alpha}{(k+h)^z}.
	\end{align*}
\end{enumerate}
The result follows by observing that
\begin{align*}
	&\mathbb{E}[\text{NG}(\pi_K^i,\pi_K^{-i})]\\
	=\;&\mathbb{E}\left[\sum_{i=1,2}\left(\max_{\hat{\pi}^i}(\hat{\pi}^i)^\top \mathcal{R}^i\pi_K^{-i}-(\pi_K^i)^\top \mathcal{R}^i\pi_K^{-i}\right)\right]\\
	\leq \;&\mathbb{E}\left[\sum_{i=1,2}\left(\max_{\hat{\pi}^i}\left\{(\hat{\pi}^i)^\top \mathcal{R}^i\pi_K^{-i}+\tau \nu(\hat{\pi}^i)\right\}-(\pi_K^i)^\top \mathcal{R}^i\pi_K^{-i}\right)\right]\\
	\leq \;&\mathbb{E}\left[\sum_{i=1,2}\left(\max_{\hat{\pi}^i}\left\{(\hat{\pi}^i)^\top \mathcal{R}^i\pi_K^{-i}+\tau \nu(\hat{\pi}^i)\right\}-(\pi_K^i)^\top \mathcal{R}^i\pi_K^{-i}-\tau \nu(\pi_K^i)\right)\right]+\tau \log(A_{\max}^2)\\
	= \;&\mathbb{E}[V_{\mathcal{R}}(\pi_K^i,\pi_K^{-i})]+\tau \log(A_{\max}^2)\\
	\leq \;&M_k+2\tau \log(A_{\max}).
\end{align*}

\section{Proof of Corollary \ref{co:convergent_opponent} and Corollary \ref{co:convergent_opponent_Markov}}

We first consider Corollary \ref{co:convergent_opponent}. The following proof idea was previous used in \cite{sayin2021decentralized} to show the rationality of their decentralized $Q$-learning algorithm.

Observe that Theorem \ref{thm:bandit} can be easily generalized to the case where the reward is corrupted by noise. Specifically, suppose that player $i$ takes action $a^i$ and player $-i$ takes action $a^{-i}$. Instead of assuming player $i$ receives a deterministic reward $\mathcal{R}^i(a^i,a^{-i})$, we assume that player $i$ receives a random reward $r^i(a^i,a^{-i},\xi)$, where $\xi\in \Xi$ (where $\Xi$ is a finite set) is a random variable with distribution $\mu_\xi$, and is independent of everything else. The proof is identical as long as $r^i+r^{-i}=0$, and the reward is uniformly bounded, i.e., $\max_{a^i,a^{-i},\xi}|r^i(a^i,a^{-i},\xi)|<\infty$. 

Now consider the case where player $i$'s opponent follows a stationary policy $\pi^{-i}$. We incorporate the randomness of player $-i$'s action into the model and introduce a fictitious opponent with only one action $a^*$. In particular, let $\hat{r}^i(a^i,a^*,a^{-i})=\mathcal{R}^i(a^i,a^{-i})$ for all $a^i$ and $a^{-i}$, and let $\hat{p}(s'\mid s,a^i,a^*)=\sum_{\pi^{-i}(a^{-i}|s)}p(s'\mid a^i,a^{-i},s)$. Now the problem can be reformulated as player $i$ playing against the fictitious player with a single action $a^*$, with reward function $\hat{r}^i$ ($i\in \{1,2\}$) and transition probabilities $\hat{p}$. Applying Theorem \ref{thm:bandit} and we have the $\mathcal{O}(1/\epsilon)$ sample complexity for player $i$ to find its best  response against $\pi^{-i}$, up to a smoothing bias.

The proof of Corollary \ref{co:convergent_opponent_Markov} is identical to that of Corollary \ref{co:convergent_opponent}.

\section{On the Mixing Time of MDPs with Almost Deterministic Policies}\label{ap:mixing_example}

Consider an MDP with two states $s_1,s_2$ and two actions $a_1,a_2$. The transition probability matrix $P_1$ of taking action $a_1$ is the identity matrix $I_2$, and the transition probability matrix $P_2$ of taking action $a_2$ is $P_2=[0,1;1,0]$. Given $\alpha\in (1/2,1)$, let $\pi_\alpha$ be a policy such that $\pi(a_1|s)=\alpha$ and $\pi(a_2|s)=1-\alpha$ for any $s\in \{s_1,s_2\}$. Denote $P_\alpha$ as the transition probability matrix under $\pi_\alpha$. It is easy to see that
\begin{align*}
	P_\alpha=\begin{bmatrix}
		\alpha&1-\alpha\\
		1-\alpha&\alpha
	\end{bmatrix}.
\end{align*}
Since $P_\alpha$ is a doubly stochastic matrix, and has strictly positive entries, it has a unique stationary distribution $\mu=\bm{1}^\top /2$.

We next compute a lower bound of the mixing time of the $\pi_\alpha$-induced Markov chain.
Let $e_1=[1,0]^\top $ be the initial distribution of the states, and denote $[x_k,1-x_k]^\top $ as the distribution of the states at time step $k$. Then we have
\begin{align*}
	x_{k+1}=\;&x_k\alpha+(1-x_k)(1-\alpha)\\
	=\;&(2\alpha-1)x_k+1-\alpha\\
	=\;&(2\alpha-1)^{k+1}x_0+\sum_{i=0}^k(1-\alpha)(2\alpha-1)^{k-i}\\
	=\;&\frac{1}{2}+\frac{(2\alpha-1)^{k+1}}{2}.
\end{align*}
It follows that
\begin{align*}
	t_{\pi_\alpha,\eta}=\;&\min_{k\geq 0}\left\{\max_{\mu_0\in\Delta^2}\left\|\mu_0^\top P_\alpha^k-\bm{1}^\top/2\right\|_{\text{TV}}\leq \eta\right\}\\
	\geq \;&\min_{k\geq 0}\left\{\left\|e_1^\top P_\alpha^k-\bm{1}^\top/2\right\|_{\text{TV}}\leq \eta\right\}\\
	= \;&\min_{k\geq 0}\left\{(2\alpha-1)^k\leq 2\eta\right\}\\
	\geq  \;&\frac{\log(1/2\eta)}{\log(1/(2\alpha-1))}-1,
\end{align*}
which implies $\lim_{\alpha\rightarrow 1}t_{\alpha,\eta}=\infty$. Therefore, as the policies become deterministic, the mixing time of the associated Markov chain can approach infinity.

\end{document}